\documentclass[12pt]{article}
\usepackage{amsfonts,amsmath}
\usepackage[mathscr]{eucal}
\usepackage{amssymb}
\usepackage{amsthm}
\usepackage{epsfig,epsf}
\usepackage{graphicx}
\usepackage{graphics}
\usepackage{bbold}
\theoremstyle{plain}

\def\theequation{\arabic{section}.\arabic{equation}}
\newcommand{\be}{\begin{eqnarray}}
\newcommand{\ee}{\end{eqnarray}}
\newcommand{\nn}{\nonumber \\}
\newcommand{\lb}{\label}
\newcommand{\p}[1]{(\ref{#1})}
\renewcommand{\th}{\theta}
\newcommand{\q}{\quad}

\newcommand\cN{{\cal N}}

\def\g{\gamma}

\def\d{\delta}

\def\ve{\varepsilon}

 \def\th{\theta}

\def\pa{\partial}

\newcommand{\pp}{{++}}
\newcommand{\m}{{--}}

\newcommand{\Dp}{D^{\pp}}
\newcommand{\Dm}{D^{\m}}

\newcommand{\dpp}{\partial^{\pp}}
\newcommand{\dm}{\partial^{\m}}

\newcommand{\Vp}{V^\pp}
\newcommand{\Vm}{V^\m}

\def\sfrac#1#2{{\textstyle\frac{#1}{#2}}}

\begin{document}
\begin{titlepage}

\renewcommand{\thefootnote}{\star}
\begin{center}
{\flushright {CPHT-RR036.0915}\\[15mm]}

\vspace{0.1cm}

{\LARGE\bf  Ultraviolet behavior of $6D$ supersymmetric
Yang-Mills theories and harmonic superspace}

\vspace{1.2cm}
\renewcommand{\thefootnote}{$\star$}

\quad {\large\bf Guillaume~Bossard}
 \vspace{0.3cm}

{\it Centre de Physique Th\'eorique, Ecole Polytechnique, CNRS\\
Universit\'e Paris-Saclay 91128 Palaiseau cedex, France}\\

\vspace{0.1cm}

{\tt bossard@cpht.polytechnique.fr}\\
\vspace{0.5cm}

\quad {\large\bf Evgeny~Ivanov}
 \vspace{0.3cm}

{\it Bogoliubov Laboratory of Theoretical Physics, JINR,}\\
{\it 6 Joliot-Curie str., 141980 Dubna, Moscow region, Russia} \\
\vspace{0.1cm}

{\tt eivanov@theor.jinr.ru}\\
\vspace{0.5cm}

{\large\bf Andrei~Smilga} \vspace{0.3cm}

{\it SUBATECH, Universit\'e de Nantes,}\\
{\it 4 rue Alfred Kastler, BP 20722, Nantes 44307, France;}\\
\vspace{0.1cm}

{\tt smilga@subatech.in2p3.fr}\\

\end{center}
\vspace{0.2cm} \vskip 0.6truecm \nopagebreak

\begin{abstract}
\noindent
 We revisit the issue of  higher-dimensional counterterms for the ${\cal N} = (1,1)$ supersymmetric Yang-Mills (SYM) theory in six dimensions
 using the off-shell ${\cal N}=(1,0)$ and on-shell ${\cal N}=(1,1)$ harmonic superspace approaches.
The second approach is developed
 in full generality and used to solve, for the first time, the ${\cal N} = (1,1)$ SYM constraints in terms of ${\cal N}=(1,0)$ superfields. This provides
a convenient tool to write explicit expressions for the candidate counterterms and other  ${\cal N} = (1,1)$ invariants and
may be conducive to proving non-renormalization theorems needed to explain the absence of certain logarithmic divergences
in higher-loop contributions to scattering amplitudes in ${\cal N} = (1,1)$ SYM.

\end{abstract}
\vspace{0.3cm}

\begin{center}
{\it Dedicated to the memory of Boris Zupnik}
\end{center}

\newpage

\end{titlepage}

\setcounter{footnote}{0}

\setcounter{equation}0

\tableofcontents
\section{Introduction}
Yang--Mills theory and its supersymmetric extensions have been studied extensively over the years,
and are of particular relevance in four dimensions, in which case they define
 renormalizable quantum field theories.
It is well known that these theories are not renormalizable by power counting
in higher dimensions, but they
nonetheless provide effective theory descriptions of some particular
low energy sectors of string theory,
such as D5-brane dynamics and open string theory compactifications. In
 this paper, we concentrate
on $6D$ supersymmetric Yang-Mills (SYM) theory.
 Only the maximally supersymmetric
 ${\cal N} = (1,1)$ theory,  involving both left-handed and right-handed supercharges, is anomaly free in six dimensions \cite{6Danomaly},
and is therefore physically relevant.
The effective action for coincident D5-branes defines a non-abelian
generalization of
Born--Infeld theory representing an infinite series that involves the standard
${\cal N} = (1,1)$ supersymmetric $6D$ Yang-Mills
Lagrangian and
higher-derivative corrections
\cite{Tseytlin:1997csa,Koerber:2001uu,DeFosse:2001mk,Koerber:2002zb,Drummond:2003ex}.

We wish to note right away that the individual terms in the effective
action need not  and do not  possess
the full extended supersymmetry that string theory enjoys. Only the whole infinite series has
this property. We will discuss this issue in detail later.

 The higher-derivative supersymmetric structures similar to those that appear
in the Born-Infeld action define also the candidate
 counterterms for the ultra-violet (UV)
 logarithmic divergences in  the $6D$ SYM theory.
The supersymmetric Ward identities for the  on-shell amplitudes
only require
 these counterterms (at least, the counterterms that are responsible for first logarithmic divergences
$\sim \ln \Lambda_{UV}$ in the amplitudes) to be invariant under extended supersymmetry
transformations {\it on mass shell}, {\it {\it i.e.}} modulo the equations of motion
 of the $6D$ SYM theory.
 Only when one can give a superspace formulation of the theory and  use  a
symmetry-preserving   regularization,
 the counterterms should possess the corresponding  supersymmetry off shell.

There is no off-shell ${\cal N} = (1,1)$ superspace formulation of the maximally
 supersymmetric $6D$ SYM
theory. Thus, we cannot expect the counterterms to enjoy the full
off-shell supersymmetry of the original action. The on-shell ${\cal N} = (1,1)$ supersymmetry
 should, however, be there. On the other hand, there exists
a ${\cal N} = (1,0)$
superspace formulation, and the relevant counterterms should
be  ${\cal N} = (1,0)$
off-shell [and ${\cal N} = (1,1)$ on-shell] supersymmetric.
A limited symmetry of relevant counterterms is a specific feature of  theories with extended supersymmetry.
In more simple  cases (think of the Euler-Heisenberg effective Lagrangian for QED or of higher-dimensional counterterms
in the effective chiral theory describing the low-energy sector of QCD), all individual terms in the effective Lagrangian possess
the same off-shell symmetries as the leading term.

The structure of higher-dimensional counterterms  was previously  studied  in the conventional superspace approach
\cite{HoweStelle} and in (on-shell) harmonic superspace in \cite{Bossard:2009sy}. A convenient way to
determine the structure of these counterterms
 is using the (off-shell) harmonic superspace  technique
developed in \cite{harm1,harm} and extended to six dimensions in
\cite{HSW,Zup1,Zup2}. That is what we do in the present paper.

We make use of the
 ${\cal N} = (1,0)$  off-shell harmonic superspace formalism of
refs. \cite{HSW,Zup1,Zup2},
to define in detail the ${\cal N} = (1,1)$  on-shell harmonic superspace invariants
introduced in \cite{Bossard:2009sy}. We rewrite the standard superspace
${\cal N} = (1,1)$ SYM constraints in ${\cal N} = (1,1)$ harmonic superspace.
The main new result here is to solve explicitly these constraints in
terms of ${\cal N} = (1,0)$ superfields. Since the constraints put the theory on shell,
the ${\cal N} = (1,0)$ superfields are necessarily subjected to their equations of motion.
Nevertheless,
while constructing the invariants from the constrained ${\cal N}=(1,1)$ superfield
strength as
integrals over superspaces involving the full ${\cal N}=(1,0)$ superspace as a
subspace,
these superfields can still be treated as off-shell ones.
The on-shell conditions are needed only while checking the hidden ${\cal N}=(0,1)$
supersymmetry   of these invariants. The ${\cal N} = (1,1)$ harmonic superspace formalism  allowed us
to write down explicit analytic expressions in  off-shell  ${\cal N} = (1,0)$ harmonic
superspace for the candidate counterterms.
Their analysis may help to
understand the reason why certain possible {\it a priori}
logarithmically divergent structures in the scattering amplitudes happen to be absent,
as was displayed in explicit 3-loop calculations \cite{Bern:2005iz,Bern:2010tq,Bern:2012uf}
\footnote{We are talking here only about {\it logarithmic} divergences; power UV divergences characteristic
of a non-renormalizable theory are present starting from the first loop in certain UV regularization
schemes --- Slavnov's higher-derivative scheme or lattice regularization.
These power divergences cannot be cared of by calculations in the papers just cited.}.

 The algebraic renormalization method \cite{Piguet:1995er} was generalized to non-renormalizable supersymmetric theories in \cite{Bossard:2009sy} as a tool
 to determine whether some counterterms could support logarithmic divergences. This allowed to explain the absence of 2-loop
divergences. But the same arguments fail to explain the absence of non-planar divergences
at the 3-loop level \cite{Bossard:2010pk}. Arguments
 using the pure spinor formalism \cite{Berkovits:2009aw:2009aw,Bjornsson:2010wm,Bjornsson:2010wu} allow to explain this result,
  but there is no direct quantum field theory understanding of
this non-renormalization theorem.
 We will see that the absence of 2-loop divergences
 can also be understood in the ${\cal N}=(1,0)$ harmonic
 superspace framework through the absence of
an ${\cal N} = (1,0)$ off-shell supersymmetric  and manifestly gauge-invariant counterterm
 of canonical dimension $d=8$ \footnote{Hereafter, we denote by $d$ the canonical dimension (in mass units) of the relevant component $6D$ Lagrangian.}.
 At the 3-loop level ($d=10$),
both  planar (or {\it single-trace}) and non-planar (or {\it double-trace})
 supersymmetric counterterms can be constructed.
 We shall argue, however, that   Ward identities combining the algebraic approach for non-linear hidden supersymmetry
 with the off-shell ${\cal N}=(1,0)$ harmonic superspace methods could potentially explain the 3-loop
 non-renormalization theorem.

The  $6D$ SYM theory also represents an interest as a toy model for
more complicated extended supergravity theories.
In particular, the absence of divergences for the double-trace structure in the 3-loop amplitude
in six dimensions obtained by explicit computations is somehow similar to the absence of divergences which was observed
for the  four-graviton amplitude in $\mathcal{N} = 4$  supergravity in four dimensions at
the  3-loop level and in $\mathcal{N} = 5$  supergravity  at the 4-loop level
\cite{Bern:2012cd,Tourkine:2012ip,Bern:2012gh,Bern:2014sna}.

Indeed, for pure ${\cal N} = 4$ supergravity (without matter), the first available supersymmetric
counterterm appears at the 3-loop level \cite{Deser:1977nt,Deser:1978br}.
The absence of anomaly for the Cremmer--Julia symmetry for
$\mathcal{N}$-extended supergravity with
$\mathcal{N}\ge5$ \cite{Bossard:2010dq} and inspection of the possible supersymmetry
invariants exhibit that the first available counterterm only appears at
$(\mathcal{N}-1)$-loop order \cite{Bossard:2010bd,Beisert:2010jx,Bossard:2011tq}.
This allows to understand the good ultra-violet behavior of $\mathcal{N}=8$
supergravity amplitudes which was observed in \cite{Bern:2007hh,Bern:2008pv,Bern:2009kd}
through four loops.
However, this symmetry principle fails to explain the absence of divergences
at three loops in pure $\mathcal{N}=4$ supergravity \cite{Bossard:2013rza},
as well as at four loops in $\mathcal{N}=5$ supergravity.
These unexplained cancellations suggest that, by the same currently unexplained reason,
maximal supergravity may not
diverge at seven loops,  in spite of the presence of a  counterterm satisfying all symmetries  \cite{Bossard:2011tq}.
On the other hand, the 4-loop amplitudes in  $\mathcal{N}=4$ supergravity are known to involve
logarithmic divergences, and one might think that the same is true for the 8-loop amplitudes in the
maximal $\mathcal{N}=8$ supergravity, as was predicted long time ago in \cite{Kallosh:1980fi}.

Although the non-renormalization theorems in $6D$ SYM  theory and in supergravity
were proven using different methods, one may hope that a future proof of the
non-renormalization theorem for the non-planar 3-loop logarithm divergence in
Yang-Mills theory could shed some light on possible generalizations to supergravity.

The structure of the paper is the following.

In Sect. 2, we attempt to give a pedagogical explanation of the above-mentioned fact that the individual terms in
the supersymmetric effective Lagrangian do not necessarily possess all the symmetries of the leading term.  We illustrate
this for the toy supersymmetric quantum mechanical model with only one bosonic degree of freedom.

In Sect. 3, we recall
 the basic notions and introduce notation
for 6-dimensional ${\cal N} = (1,0)$ harmonic superspace.
In Sect. 4, we use this formalism to construct the classical invariant actions  of canonical dimension 4 involving the ${\cal N} = (1,0)$ vector multiplet
and the hypermultiplet.  One of such actions enjoys the extended ${\cal N} = (1,1)$ supersymmetry,
with the ${\cal N} = (0,1)$ part of this symmetry being realized via the transformations of ${\cal N} = (1,0)$  harmonic superfields.

In Sect. 5, still working in the ${\cal N} = (1,0)$  superspace framework, we analyze higher-dimensional
${\cal N} = (1,1)$  supersymmetric Lagrangians. We show that
   \begin{itemize}
\item At the 1-loop level ($d=6$), all the candidate counterterms vanish on mass shell \cite{ISZ,IShyper}.
We  demonstrate in Sect. 6.1 and, in more details, in Appendix B that no $d=6$ off-shell
${\cal N} = (1,1)$  supersymmetric Lagrangian can be constructed.
\item At the 2-loop level ($d=8$), the candidate counterterms also vanish on mass shell, if we require them
to be ${\cal N} = (1,0)$  off-shell supersymmetric {\it and} gauge invariant \footnote{These requirements should be imposed
under the assumption that the perturbative calculations
can be done in a way that preserves at all steps the off-shell   ${\cal N} = (1,0)$ supersymmetry and gauge invariance,
both of them being kept by regularization. This assumption is very natural: the existence of ${\cal N} = (1,0)$ superspace
implies the existence of supergraph technique, and the higher-derivative ultraviolet regularization
keeps gauge invariance and ${\cal N} = (1,0)$ supersymmetry, but we are not aware of its formal rigorous proof.}.
\item On the other hand, one can construct an {\it on-shell} $d=8$ gauge-invariant Lagrangian involving
both the vector multiplet and hypermultiplet and possessing {\it both} ${\cal N} = (1,0)$ and ${\cal N} = (0,1)$
supersymmetries only {\it on shell}.
Its bosonic part starts with the structure $\sim F^4$. It does not appear as a counterterm for the
${\cal N} = (1,1), \ 6D$ SYM theory, but {\it is}
 present in the derivative expansion of the Born-Infeld action for  coincident D5-branes.
 \end{itemize}

The methods of Sect. 5 where the extra ${\cal N} = (0,1)$ supersymmetry is ``hidden" in the superfield transformations
proved not to be too efficient for constructing the 3-loop $d=10$ invariants; even the construction of the $d=8$
invariants by this ``brute force'' method is rather complicated technically. To perform such a study in a more systematic way,
we developed  in Sect. 6 and 7 the on-shell ${\cal N} = (1,1)$ harmonic superfield formalism. It involves a double set of harmonics, $u^\pm_i$ and
$u_A^{\hat{\pm}}$, as well as the extra $SU(2)$ doublet of the $(0, 1)$ chiral $6D$ fermionic superspace coordinates. We show that
\begin{itemize}

\item The known superspace constraints on the covariant spinor derivatives, which define the ${\cal N} = (1,1)$ SYM theory
\cite{HoweStelle, HST},
admit a compact rewriting in this bi-harmonic superspace as the conditions for the two types of covariant Grassmann harmonic analyticities.

\item These constraints are explicitly solved in Sect. 7 in terms of the ${\cal N}=(1,0)$ SYM gauge superfield and hypermultiplet,
simultaneously providing the ${\cal N}=(1,0)$ Grassmann  harmonic analyticity and the on-shell conditions for these superfields.

\item These ${\cal N}=(1,0)$ constituents are encompassed by the single double-analytic ${\cal N} = (1,1)$ superfield strength with
simple transformation properties under the ${\cal N}=(0,1)$ supersymmetry.

\end{itemize}

In Sect. 8, we write various invariant actions in terms of this ${\cal N} = (1,1)$ superfield strength as integrals over the full
${\cal N} = (1,1)$ superspace or its $1/2$ or $3/4$ analytic subspaces and further rewrite these invariants in the ${\cal N} = (1,0)$ superspaces.
\begin{itemize}

\item
We rederive in this way
the on-shell $d=8$ invariant obtained in Sect. 5 and also derive nontrivial expressions for the single-trace and double-trace  $d=10$ invariants
in terms of ${\cal N} = (1,0)$ superfields. We note that the  single-trace invariant can be represented as a full ${\cal N} = (1,1)$
superspace integral, whereas the double-trace invariant cannot. We suggest that, using the algebraic method in ${\cal N}=(1,0)$ harmonic superspace,
this may be enough for proving a non-renormalization theorem preventing  the appearance of the double trace as a counterterm.

\item We also present an alternative view of constructing higher-order invariants on the $d=8$ example. One can keep the off-shell
${\cal N} = (1,0)$ supersymmetry, but allow for the gauge invariance to be deformed, or modify the definition
of the Yang-Mills field strength curvature \cite{Bergshoeff:1986jm}.
We write the explicit expression for the $d=8$ action thus obtained.  This may provide
an alternative way to construct the supersymmetric Born--Infeld Lagrangian in ${\cal N}=(1,0)$ harmonic
superspace, although we do not investigate this issue in this paper.

\end{itemize}
There are three technical Appendices.  In Appendix A, we derive certain Bianchi identities relating different
${\cal N} = (1,0)$ superfields that are used in Sect. 5. In Appendix B, we describe a failed attempt to construct an off-shell
${\cal N} = (1,1)$ invariant $d=6$ action. We conclude that such an action in all probability  does not exist.
In Appendix C, we give an alternative derivation of the $d=8$ on-shell  ${\cal N} = (1,1)$ supersymmetric Lagrangian,
directly in the ${\cal N}=(1,0)$ superspace.

\section{Off-shell vs. on-shell}
\setcounter{equation}0

In this pedagogical section, we clarify  generic features of effective supersymmetric Lagrangians by studying in detail
two toy supersymmetric quantum mechanical models and recalling the familiar situation for $4D$ field theories.

\subsection{Witten's model}

The simplest possible example is Witten's supersymmetric quantum mechanical
system involving one bosonic degree of freedom
with the Lagrangian \cite{Witten81}
  \be
\lb{LSQMcom}
  L_0 \ =\ \frac {\dot{x}^2 - [V'(x)]^2}2  + \frac i2 \left(
\dot{\psi} \bar \psi - \psi \dot{\bar\psi} \right)
+  V''(x) \bar \psi \psi \, .
  \ee
The corresponding equations of motion are
 \be
 \lb{eqmotSQM}
\ddot{x} + V'(x) V''(x) - V'''(x) \bar \psi \psi &=& 0 \, , \nn
i \dot{\psi} - V''(x) \psi  &=& 0 \, , \nn
i \dot{\bar \psi} + V''(x) \bar \psi  &=& 0 \, .
  \ee

The Lagrangian \p{LSQMcom}
is invariant (up to a total time derivative) under the following nonlinear supersymmetry transformations
  \be
\lb{trans_bez_D}
\delta x \equiv \delta_\epsilon x +  \delta_{\bar \epsilon} x
\ = \ \epsilon \bar \psi + \psi \bar \epsilon , \nn
\delta \psi \equiv \delta_\epsilon \psi \ =\
-\epsilon [  i\dot{x} + V'(x)] , \nn
\delta \bar\psi \equiv \delta_{\bar \epsilon} \bar\psi
\ =\ \bar\epsilon [i\dot{x} - V'(x) ] \, .
    \ee
Note now that it is {\it impossible} to write a
Lagrangian depending on the fields $x,\psi, \bar\psi$
and involving their {\it higher} time derivatives
 which would be
invariant under the transformations \p{trans_bez_D}.
This is due to the well-known fact that the Lie
brackets of the transformations \p{trans_bez_D} do not close off mass shell,
 but only on mass shell. When acting on the variable $x(t)$, the Lie bracket
 $(\delta_{ \bar \epsilon} \delta_{\xi} - \delta_{\xi} \delta_{ \bar \epsilon})$
boils down to a total time derivative. But it is not so for the fermion variables.
For example,
   \be
\lb{Lie_bracket}
\left( \delta_{ \bar \epsilon} \delta_{\xi} - \delta_{\xi} \delta_{ \bar \epsilon}
\right) \psi =  \ \xi \bar \epsilon [i\dot{\psi} + V''(x) \psi] \ =\ 2i\xi \bar \epsilon \, \dot{\psi} +
\xi \bar \epsilon \frac {\partial L}{\partial \bar \psi} \, .
     \ee
 The presence of the  second term in \p{Lie_bracket} does not affect the
invariance of $L_0$ under
\p{trans_bez_D}. Indeed,
  \be
 \lb{Lie_L0}
 \left(\delta_{\bar \epsilon} \delta_\xi -  \delta_\xi \delta_{\bar \epsilon} \right) L_0 \ =\
  2i\xi \bar \epsilon \, \dot{L_0} + \xi \bar \epsilon \, \left( \frac {\partial L_0}{\partial \psi}
 \frac {\partial L_0}{\partial \bar\psi} +  \frac {\partial L_0}{\partial \bar \psi}
 \frac {\partial L_0}{\partial \psi} \right) = 2i\xi \bar \epsilon \, \dot{L_0} \, .
   \ee
But,  for $L \neq L_0$, the second term in the Lie bracket
$\left(\delta_{\bar \epsilon} \delta_\xi -  \delta_\xi \delta_{\bar \epsilon} \right) L$  vanishes
only on the mass shell of $L_0$.

The standard way to solve this problem and to construct fully supersymmetric actions
of any dimension
is to introduce a superfield
    \be
  \lb{X}
   X(t, \theta, \bar\theta) \ =\ x + \theta \bar \psi + \psi \bar \theta + F \theta \bar \theta\, .
     \ee
 The transformations of the superspace coordinates generate {\it linear} supersymmetry
transformations of the dynamic variables,
   \be
\lb{trans_with_D}
\delta x \ = \ \epsilon \bar \psi + \psi \bar \epsilon , \nn
\delta \psi \ =\ \epsilon (F - i\dot{x} ) , \nn
\delta \bar\psi \ =\ \bar\epsilon (F + i\dot{x} ) , \nn
 \delta F = i(\epsilon \dot{\bar \psi} - \dot{\psi} \bar \epsilon )\, .
    \ee
Any higher-derivative action of the form
\be
\lb{polynom}
 S \ =\ \int  dt\, d\bar\theta d\theta
\left(  \frac 12 \bar D X P\Bigl[ \frac{\partial\, }{\partial t}\Bigr] DX + V(X) \right),
  \ee
where $P(\partial/\partial t)$ is an arbitrary polynomial and
  \be
D = \frac {\partial }{\partial \theta } + i\bar \theta \frac {\partial}{\partial t}\, , \ \ \ \ \ \ \ \ \
\bar D = -\frac {\partial }{\partial \bar \theta } - i \theta \frac {\partial}{\partial t}
 \ee
are the supersymmetric covariant derivatives, is invariant under \p{trans_with_D}.

For a linear polynomial $P(z) = a + bz$, one obtains an interesting higher-derivative model whose
Hamiltonian is Hermitian in spite of the presence of the ghosts (no ground state in the
spectrum) \cite{Robert}. For higher-order polynomials, the Hermiticity
is lost, but we need not worry about it, we use this as a toy model displaying the structure
of the effective Wilsonian Lagrangian in a field theory of interest. We choose $P(z) = 1 - gz^2$.
One obtains then the following component Lagrangian,
  \be
     \lb{d2tcomp}
 L \ =\   \frac 12 (\dot{x}^2 +   F^2 ) +i \dot{\bar\psi}
    {\psi} + F V'(x) + V''(x) \bar\psi \psi  \ + g  \frac 12
\left( \ddot{x}^2 + \dot{F}^2 + 2i  \ddot{\bar\psi}
    \dot{\psi} \right).
       \ee
This Lagrangian is invariant under the transformations \p{trans_with_D}. On the other hand,
the formerly auxiliary field $F$ has become dynamical and cannot be algebraically eliminated
as it can in Witten's model with $g=0$. Still, one can integrate out the field $F$
perturbatively through the formal power series solution
   \be
F =- \sum_{n=0}^\infty g^n  \frac{d^{2n} V'(x) }{d t^{2n}}  \, .
  \ee

One obtains in this way the Lagrangian
   \be
\lb{LSQMeff}
  L  \ =\ \frac{1}{2} \bigl(  \dot{x}^2 + g \ddot{x}^2\bigr)  + i \dot{\bar\psi} \psi + i g \ddot{\bar\psi}
    \dot{\psi}
- \frac{1}{2} \sum_{n=0}^\infty (-g)^n \biggl(  \frac{d^{n}  V'(x) }{d t^{n}}  \biggr)^2 +  V''(x) \bar \psi \psi  \, ,
  \ee
which is by construction invariant with respect to the nonlinear supersymmetry transformations
    \be
\lb{trans_eff}
\delta x \ &=& \ \epsilon \bar \psi + \psi \bar \epsilon \, , \nn
\delta \psi \ &=&\ \epsilon \biggl(   -i\dot{x} -  \sum_{n=0}^\infty g^n  \frac{d^{2n} V'(x)  }{d t^{2n}}   \biggr) \,  , \qquad
\delta \bar\psi \ =\ \bar\epsilon  \biggl( i\dot{x} -  \sum_{n=0}^\infty g^n  \frac{d^{2n} V'(x) }{d t^{2n}}    \biggr),
    \ee
that close modulo the  equations of motion for the full Lagrangian \p{LSQMeff}. For example,
   \be
\lb{Lie_eff}
\left(\delta_{ \epsilon_1} \delta_{\epsilon_2} -
\delta_{\epsilon_2}  \delta_{\epsilon_1}
\right) \psi = \ -2\epsilon_1 \epsilon_2  \sum_{n=0}^\infty g^n  \frac{d^{2n} \, }{d t^{2n}}  \left(
\frac {\partial L}{\partial \psi} \right).
     \ee
The Lagrangian  \eqref{LSQMeff} represents a  perturbative series in $g$,
\be L = \sum_{n=0}^\infty g^n L_n = L_0 + g L_1 + g^2 L_2 + \dots \ee
and similarly for $\delta = \delta_0 + g \delta_1 + \dots $, where
$L_0$ is written in
 \eqref{LSQMcom} and $\delta_0$ in  \eqref{trans_bez_D}.
It follows by construction that the first-order correction,
   \be
\lb{L_1}
 L_1 =  \frac{1}{2} \ddot{x}^2  +  i \ddot{\bar\psi}
    \dot{\psi} + \frac{1}{2} \dot{x}^2 \bigl( V''(x) \bigr)^2 ,  \ee
is invariant under the action of $\delta_0$
modulo the classical equations of motion \eqref{eqmotSQM}
and a total time derivative,
      \be  \delta_0 L_1 + \delta_1 L_0  =
\frac{d \,  }{d t }  ( \cdots ) \, . \lb{two_delta} \ee
  In other words, the action $\int dt L_1$ is invariant with respect to nonlinear
supersymmetry transformations \p{trans_bez_D} {\it on shell}, but not off shell.

 On the contrary, the second-order correction
\be L_2 =  - \frac{1}{2} \bigl( \ddot{x} V''(x) + \dot{x}^2 V'''(x)\bigr)^2, \ee
is not invariant with respect to   $\delta_0$, but satisfies instead
    \be\delta_0 L_2 + \delta_1 L_1 +
\delta_2 L_0  =  \frac{d \,  }{d t }  ( \cdots ) \, . \lb{three_delta} \ee

The situation when the effective Lagrangian represents an infinite series of
higher-derivative terms, like in \p{LSQMeff}, and this Lagrangian is invariant under
modified supersymmetry transformations also representing an infinite series,
is quite general. One known example is the Born-Infeld effective Lagrangian mentioned in the introduction.

\subsection{$4D$ supersymmetric gauge theories}

 Consider first
the ${\cal N} =1, \ 4D$  supersymmetric SYM Lagrangian.
It involves the gauge fields and  gluinos and  is invariant
under certain nonlinear supersymmetry
transformations. One also  can write higher-derivative off-shell
supersymmetric Lagrangians of canonical
dimensions $d=6,8$, etc., but they necessarily include the auxiliary field $D$ of the vector
multiplet, which now becomes dynamical. In this case, supersymmetry is realized linearly.

The same is true for the  ${\cal N} =2$  supersymmetric SYM theory. We have a scalar superfield
${\cal W}$ involving a triplet of auxiliary fields $D^A$. Higher-derivatives supersymmetric
Lagrangians, like ${\cal L} \sim {\rm Tr} \int d^8\theta \, {\cal W}^2
{\bar {\cal W}}^2$,
can be written, and they involve  the derivatives of $D^A$. For the ``matter'' fields belonging
to the ${\cal N} =2$ hypermultiplet, the full set of the auxiliary fields is infinite. The latter can be
 presented as
components of a certain ${\cal N} =2$  harmonic
superfield. Higher--derivative off-shell invariant actions can also be written  in that case.

But for the ${\cal N} =4$ theory, the situation is different. Superfield formalism, with all
supersymmetries being manifest and off-shell,
is not developed, the full set of auxiliary fields is not known and probably does not exist.
Thus, one cannot write in this case an off-shell supersymmetric higher-derivative action.
On the other hand, nontrivial higher-derivative  actions enjoying on-shell ${\cal N} =4$
supersymmetry exist (see, e.g., \cite{Drummond:2003ex,BPS01,BuchI1,BuchI2}).

In four dimensions, these higher-derivative invariants are not relevant for perturbative
calculations --- they do not appear as counterterms for a renormalizable
(even finite for ${\cal N} =4$) theory \footnote{Though they can appear in the Wilsonian effective actions.}.
But such invariants are relevant in six dimensions.
As far as their structure is concerned, the situation is the same as in four dimensions. Using harmonic approach,
one can develop ${\cal N} = (1,0)$ harmonic superfield formalism and
write down many off-shell  ${\cal N} = (1,0)$
invariants. On the other hand, we have no off-shell ${\cal N} = (1,1)$ superfield formalism and off-shell
${\cal N} = (1,1)$ invariants probably do not exist. However, it is possible to write down
many on-shell ${\cal N} = (1,1)$ invariants, and we will do it explicitly for the canonical
dimensions $d=8$ and $d=10$.

\section{Harmonic superspace and harmonic superfields}
\setcounter{equation}0

 We give here some basic facts about the $6D$ spinor algebra, the
ordinary and harmonic
${\cal N}{=}(1,0)$ superspaces  and ${\cal N}{=}(1,0)$ superfields.
For more details, see refs. \cite{ISZ,IShyper}.

\subsection{Spinor algebra}

The group $Spin(5,1)$ possesses two different spinor representations,
the complex 4-component spinors $\lambda^a$ and the complex 4-component spinors
$\psi_a$.  In contrast to $Spin(6) \equiv SU(4)$, where two 4-dimensional
 representations are related to each other by complex conjugation, in  $Spin(5,1)$, they are completely
independent. The situation is opposite to that in $4D$ where the group  $Spin(3,1)$ involves two
complex conjugate spinor representations, while in $Spin(4) \equiv SU(2) \times SU(2)$ these representations
are independent.

For the vectors, it is convenient to introduce the notation
   \be
&&V_{ab}=\sfrac12(\gamma^M)_{ab}V_M,
\ee
where $ M = 0,\ldots, 5$ and
$(\gamma^M)_{ab}$ are  antisymmetric $6D$ matrices (the $6D$ analog of $\sigma_\mu$) satisfying
 \be
\gamma_M \tilde \gamma_N + \gamma_N \tilde \gamma_M = \ -2\eta_{MN}\,, \qquad  \eta_{MN}={\rm diag} (1,-1,-1,-1,-1,-1) \,,
 \ee
with
 \be
(\tilde \gamma_M)^{ab} \ =\ \frac 12 \varepsilon^{abcd}   (\gamma_M)_{cd} \, .
  \ee
One of the possible explicit representations of these matrices is
  \be
\gamma_0 = \tilde \gamma_0 = i\sigma_2 \otimes 1\!\!\!\!1 ; \ \ \ \
\gamma_1 = -\tilde \gamma_1 = i\sigma_2 \otimes \sigma_1 ; \ \ \ \
\gamma_2 = -\tilde \gamma_2 = i  1\!\!\!\!1 \otimes  \sigma_2; \nonumber \\
\gamma_3 = -\tilde \gamma_3 = i\sigma_2 \otimes \sigma_3; \ \ \ \
\gamma_4 = \tilde \gamma_4 =  \sigma_1 \otimes \sigma_2 ; \ \ \ \
\gamma_5 = \tilde \gamma_5 = \sigma_3 \otimes  \sigma_2 \, .
  \ee

Note the properties
    \be
\label{Trgam_epsilon}
{\rm Tr} \{\gamma_M \tilde \gamma_N \gamma_P \tilde \gamma_Q \gamma_R \tilde
\gamma_S \} = - {\rm Tr} \{\tilde \gamma_M  \gamma_N \tilde \gamma_P  \gamma_Q \tilde \gamma_R
\gamma_S \} \ =\ 4\,\varepsilon_{MNPQRS}  + {\rm symmetric\ part}
   \ee
($ \varepsilon_{012345} = 1$) and
   \be
\label{gamgam_epsilon}
(\gamma^A)_{ab} (\gamma_A)_{cd}  = 2\varepsilon_{abcd}, \ \ \ \
(\tilde \gamma^A)^{ab} (\tilde \gamma_A)^{cd}  = 2\varepsilon^{abcd} \, .
  \ee

The Dirac gamma-matrices $\Gamma_M$ satisfying the standard Clifford algebra
$$\Gamma_M \Gamma_N + \Gamma_N \Gamma_M = 2\eta_{MN}$$
can be chosen as
 \be
\lb{Gamma}
\Gamma_M \ =\ \left( \begin{array}{cc} 0 & {\tilde \gamma}_M \\ -\gamma_M & 0
\end{array} \right).
  \ee
One can also introduce
   \be
  \lb{Gamma7}
\Gamma_7 \ =\ \Gamma_0 \Gamma_1 \Gamma_2 \Gamma_3 \Gamma_4 \Gamma_5
      \ee
and observe that the spinors $\lambda^a$, $\psi_a$ are the chiral left-handed
and right-handed projections of
 a 8-component Dirac spinor, {\it {\it i.e.}} $\lambda, \psi  = (1\pm \Gamma_7)\Psi/2$.

The $Spin(5,1)$ generators are
 \be
(\sigma_{MN})^a_{\ b} \ = \ \frac 12 \left( \tilde \gamma_M \gamma_N -
 \tilde \gamma_N \gamma_M \right)^a_{\ b} =
\frac 12 \left(  \gamma_N \tilde \gamma_M -
  \gamma_M \tilde \gamma_N \right)^{\ a}_{b} \, .
 \ee
They are real\footnote{Thus, the algebra $spin(5,1)$ represents a real form of
$spin(6) \equiv su(4)$ and
is sometimes denoted  $su^*(4)$.}.
This makes it convenient to define, instead of a complex 4-component
spinor $\lambda^a\,$, a couple of spinors $\lambda^a_{j = 1,2}$ obeying the
 pseudoreality condition
\be
\overline{\lambda^a_j}\equiv -C_{\ b}^a(\lambda^b_j)^* =
\varepsilon^{jk} \lambda^a_k\, ,
\ee
where $C$ is the charge conjugation matrix with the properties
 $C = -C^T, \ C^2 = -1$. It can be chosen as $C = {\tilde \gamma}_0 \gamma_5$.

Similarly, instead of a generic complex $\psi_a$,
one can introduce a couple of right-handed
spinors $\psi^A_a$ related by the pseudoreality condition.

\subsection{Superspace}

The standard ${\cal N}{=}(1,0)$ superspace involves the coordinates
\be
z=(x^M, \th^a_i),  \label{Centr}
\ee
where $\th^a_i$ are Grassmann pseudoreal left-handed spinors.

Next we introduce the harmonics $u^{\pm i}$  ($u^-_i = (u^+_i)^* , u^{+i} u_i^- = 1$) , which describe the ``harmonic sphere'' $SU(2)_R/U(1)$,
where $SU(2)_R$ is R-symmetry group of the ${\cal N}=(1,0)$ Poincar\'e superalgebra \footnote{The constraint $u^{+i} u_i^- = 1$ leaves
in $u^\pm_i$ 3 degrees of freedom.  One more degree of freedom is neutralized due to the strict preservation of the harmonic $U(1)$ charge
in all invariant actions in the harmonic superspace.}.
The harmonic ${\cal N}{=}(1,0), \,6D$ superspace in the central basis is parametrized by the coordinates
\be
Z := (z, u) = (x^M, \theta^a_i, u^{\pm i})\,.\lb{hss6}
\ee
The harmonic superspace in the analytic basis involves the harmonics and
the coordinates $x_{({\rm an})}^M, \th^{\pm a}$:
\be
Z_{({\rm an})} := (x_{({\rm an})}^M, \th^{\pm a}, u^{\pm i})\,,\lb{hss6An}
\ee
\be
x^M_{({\rm an})}=x^M+ \frac i2 \th^a_k\g^M_{ab}\th^b_l u^{+k}u^{-l},\q \th^{\pm a}=
u^\pm_k\th^{ak}.
\ee
A very important property of the analytic basis is that the set of coordinates
\be
\zeta :=(x^M_{({\rm an})}, \th^{+a}, u^{\pm i}) \,, \label{AnalSS}
\ee
involving only a half of the original Grassmann coordinates
forms a subspace closed under the action of ${\cal N}{=}(1,0)\,, 6D$ supersymmetry.
The set \p{AnalSS} parametrizes what is called ``harmonic analytic superspace''.

It is convenient to define the  differential operators called
spinor and harmonic derivatives. In the analytic basis, they are expressed as
    \be
&& D^+_a=\pa_{-a}\,, \;
D^-_a=-\pa_{+ a}-2i\th^{-b}\pa_{ab}\,, \; D^0 = u^{+i} \frac {\pa}{ \pa u^{+i}} -
u^{-i} \frac {\pa}{ \pa u^{-i}} + \th^{+a} \pa_{+ a} -  \th^{-a} \pa_{- a}\,,
  \nn \\
&& \Dp=\dpp+i\th^{+a}\th^{+b}\pa_{ab}+\th^{+a}\pa_{-a}~,\q\Dm=\dm+i\th^{-a}
\th^{-b}\pa_{ab}+\th^{-a}\pa_{+a}~,
 \ee
where $\,\pa_{\pm a}\th^{\pm b} = \d^b_a\, $ and
$$ \dpp =  u^{+i} \frac {\pa }{ \pa u^{-i}} , \q \dm =  u^{-i} \frac {\pa }{ \pa u^{+i}}\ .
$$
The following commutation relations hold
 \be
\lb{commrel}
&& \{D^+_a,D^-_b\}=2i\pa_{ab}, \q [\Dp, \Dm] = D^0\,, \lb{1001} \\
&& \nn
&& [\Dp,D^+_a]=[\Dm,D^-_a]=0\, ,\q[\Dp,D^-_a]=D^+_a,\q[\Dm,D^+_a]=D^-_a.
\ee
We shall use the notation
\be
&&(D^\pm)^4=-\sfrac{1}{24}\ve^{abcd}D^\pm_a D^\pm_b D^\pm_c D^\pm_d\,,
\quad (D^\pm)^{3a} = -\sfrac16\ve^{abcd}D^\pm_b D^\pm_c D^\pm_d\,, \nn
&&(\theta^{\pm})^4 =-\sfrac{1}{24}\ve_{abcd}\theta^{\pm a}\theta^{\pm b}\theta^{\pm c}\theta^{\pm d}\,,
\quad (D^\pm)^4(\theta^{\mp})^4 = 1\,,
\ee
and the following conventions for the full and analytic superspace
integration measures:
\be
\label{Zizeta}
d Z_{({\rm an})}=d^6x_{({\rm an})}\,du (D^-)^4(D^+)^4,\q d\zeta^{(-4)}=d^6x_{({\rm an})} du\,(D^-)^4.
\ee
The measure $dZ_{({\rm an})}$ has canonical dimension $-2$ and $ d\zeta^{(-4)}$ --- dimension $-4$.
In what follows we will frequently suppress the subscript $``({\rm an})"$ of the analytic basis coordinate  $x$
and the integration measure.

The harmonic integrals $\int F \, du$ are nonzero only if the integrand $F$ has zero harmonic
charge, $D^0 F = 0$. They can be computed using the rules
\be
\label{harmint} \
\int d{u}  \, u^{{+}}_j  u^{{-}}_k &=& \frac 12 \epsilon_{jk}\, , \nn
\int d u \,   u^{+}_j   u^{+}_k  u^{-}_m   u^{-}_n
&=& \frac 16 \left( \epsilon_{jm} \epsilon_{kn} +  \epsilon_{jn} \epsilon_{km}  \right) \, ,
\nn
\int du  \,  u^{+}_j   u^{+}_k   u^{+}_m  u^{-}_n   u^{-}_l
 u^{-}_p &=&
\frac 1{24} (\epsilon_{jn} \epsilon_{kl} \epsilon_{mp} + {\rm 5\ more\ terms}) \, ,
\ee
etc.

\subsection{ Superfields }

A general $6D$ superfield depends on 8 odd coordinates $\theta^a_i$ (or $\theta^{\pm a}$),
which makes their component expansion
rather complicated. There is, however, an important class of superfields,
{ Grassmann-analytic} superfields, which are defined on the analytic superspace  \p{AnalSS}
and so depend only on the half of the original Grassmann coordinates.
The structure of Grassmann-analytic (G-analytic) superfields is much simpler than
that of a general superfield.

 A G-analytic superfield  $\phi(\zeta)$ satisfies the constraint
$D_a^+ \phi = 0\,$\footnote{It is quite analogous to the habitual
chirality constraint $D_\alpha \phi = 0$ in four dimensions}.
In the analytic basis,  $D_a^+$ is reduced to the partial derivative
$\partial/\partial \theta^{- a}$ and this constraint simply means that
$\phi$ lives in the superspace \p{AnalSS}.

The superfields can be classified according to their harmonic charge $q$,
the eigenvalue of $D^0$. The pure $6D$
SYM theory is formulated   in terms of the  G-analytic
anti-Hermitian
superfield gauge potential which has charge $+2$ and is denoted $V^{++}$.
It defines the covariant harmonic derivative
\be
\lb{nablapp}
\nabla^\pp=\Dp+\Vp\,, \quad \delta V^{++} = \nabla^{++}\Lambda\,,
\ee
where $\Lambda = \Lambda(\zeta)$ is an arbitrary analytic gauge parameter in the adjoint representation of the gauge group.
It is convenient to introduce also a non-analytic gauge connection $V^{--}$ which covariantizes the harmonic derivative $D^{--}$
\be
\nabla^{--} = D^{--} + V^{--}\,, \quad \delta V^{--} = \nabla^{--}\Lambda\,. \lb{nablamm}
\ee
Requiring $\nabla^{++}$ and $\nabla^{--}$ to satisfy the same algebra as their flat counterparts,
 \be
\lb{nabcomm} [\nabla^\pp, \nabla^\m] = D^0\,,
 \ee
implies the harmonic zero-curvature condition
\be
\lb{A2}
\Dp\Vm-\Dm\Vp+[\Vp,\Vm]=0\,.\lb{hzc}
\ee
It can be used to solve for $V^{--}$ in terms of $V^{++}$ as a series over products of $V^{++}$ taken at different
harmonic ``points'',
\be
\label{Vmmrjad}
\Vm(z,u)=\sum\limits^{\infty}_{n=1} (-1)^n \int du_1\ldots du_n\,
\frac{\Vp(z,u_1 )\ldots \Vp(z,u_n )}{(u^+ u^+_1)(u^+_1 u^+_2)\ldots (u^+_n u^+)}\,.
\ee
Here, the factors $(u^+ u^+_1)^{-1}$, etc are the harmonic
distributions \cite{harm} and the central basis coordinates $z$ are
defined in \p{Centr}.

For further use,  note the following tensor relation between arbitrary variations of harmonic
connections:
\be
\d\Vm &=&\sfrac12(\nabla^\m)^2\d\Vp
- \sfrac12 \nabla^{++}(\nabla^{--}\delta V^{--}) \,.
\label{Rel1}
\ee
It follows from
 \be
\lb{Rel2}
\nabla^\m\d\Vp=\nabla^\pp\d\Vm \,  ,
 \ee
which in turn follows from \p{A2}.

The connection $V^{--}$ can be used to build up spinor and vector superfield connections,
    \be
{\cal A}^-_a(V)=-D^+_a\Vm,\qquad {\cal A}_{ab}(V)= \sfrac{i}{2}D^+_aD^+_b\Vm,
    \ee
and the corresponding covariant spinor and vector derivatives,
   \be
&&{\nabla}^-_a = D^-_a + {\cal A}^-_a, \qquad \nabla_{ab} = \partial_{ab}
+ {\cal A}_{ab} \,, \lb{nablaspinvec} \\
&& \delta {\cal A}^-_a = {\nabla}^-_a\Lambda\,, \qquad \delta {\cal A}_{ab} = \nabla_{ab}\Lambda\,. \lb{AAgauge}
 \ee

The covariant derivatives \p{nablapp}, \p{nablamm}, \p{nablaspinvec} and $\nabla^+_a$ (in the G-analytic basis,
it keeps its flat form $D^+_a = \partial_{-a}$.) obey the same
(anti)commutation relations \p{commrel} as the flat ones,
\be
&& [\nabla^{--}, D^+_a] = \nabla^{-}_a\,, \quad [\nabla^{++}, \nabla^{-}_a] = D^+_a\,, \quad [\nabla^{++}, D^{+}_a]
= [\nabla^{--}, \nabla^{-}_a] = 0\,, \nn
&& [D^+_a, \nabla^-_b] = 2i \nabla_{ab}\,.
\ee
In addition,
$$[\nabla^\pp, \nabla_{ab}] = 0\,.$$

On the other hand, the commutators
of spinor covariant derivatives with $\nabla_{ab}$ do not vanish,
  \be
\lb{Da,nabcb}
 [D^+_a, \nabla_{bc}] \ =\ \sfrac{i}2\varepsilon_{abcd} W^{+d} \, , \ \ \ \ \ \ \
[\nabla^-_a, \nabla_{bc}] \ =\ \sfrac{i}{2}\varepsilon_{abcd} W^{-d}  \,,
 \ee
where $W^{\pm a}$ are the covariant (1,0) spinor superfield strengths,
\be
&& W^{+a}=-\sfrac{1}{6}\ve^{abcd}D^+_b D^+_c D^+_d\Vm\,,\lb{W+Def} \\
&& W^{- a} := \nabla^{--}W^{+a}\,.
\lb{W-Def}
\ee
One  can also define the G-analytic superfield
    \be
\label{Fpp}
F^\pp=\sfrac14 D^+_aW^{+a}=(D^+)^4 \Vm \, , \quad D^+_a F^{++} = 0\,.
    \ee
{}From the harmonic zero-curvature condition \p{hzc} and G-analyticity of $V^\pp$, the important properties follow
\be
&&\nabla^\pp W^{+a}  = \nabla^{--} W^{-a} \ =\ 0\,, \quad \nabla^{++}W^{-a} = W^{+ a}\,,  \lb{HarmW} \\
&& D^+_b W^{+ a} = \delta^a_b F^{++}\,,\lb{DW0} \\
&& \nabla^{++}F^{++} = 0\,.\label{Vazhnoe}
\ee
Note that all these objects are homogeneously transformed by the gauge group
\be
\delta W^{\pm a} = [W^{\pm a}, \Lambda]\,, \quad \delta F^{++} =[F^{++}, \Lambda]\,. \lb{TransfHom}
\ee
More details on the algebra of gauge-covariant derivatives and the relevant Bianchi
identities are collected in Appendix A.

The matter hypermultiplet is described by a pair of the G-analytic superfields $q^{+A}(\zeta)\,, A=1,2\, $.
If they belong to the real representation of
the gauge group (e.g. the adjoint one), they  can be subjected  to the reality condition $\widetilde{q^{+ A}} = \epsilon_{AB}q^{+ B},$
where the ${\,} \widetilde {\,}$ conjugation  is the product of the ordinary complex conjugation and
an antipodal map on the harmonic sphere $S^2 \sim SU(2)/U(1)$ (see \cite{harm} for details). Note that the gauge prepotentials $V^{\pm\pm}$ defined above, as well as
the connections ${\cal A}^-_a$, ${\cal A}_{ab}$ and covariant strengths $W^{\pm a}$, are anti-Hermitian
with respect to this generalized conjugation.
Since, in what follows, we will deal with the adjoint hypermultiplets, it is worth giving how such $q^{+ A}$ transform under
the gauge group action
\be
\delta q^{+ A} = [q^{+ A}, \Lambda]\,.
\ee

Finally, we note the useful {\underline{\bf Lemma}}:
\be
\nabla^{++} F^{-n} = 0 \quad \Rightarrow \quad F^{-n} = 0\;\;\; {\rm for}\; n \geq 1\,,\lb{lemma}
\ee
where the ${\cal N}=(1,0)$ superfield $F^{-n}$ transforms in some representation of the gauge group and we
suppressed the ``color'' indices.  This statement  can be proved by passing to the central basis of the harmonic superspace, where $D^{++} = \partial^{++}\,$,
and the so called $\tau$-frame for the gauge fields, where $\nabla^{++} = e^{-iV}\partial^{++} e^{iV}$ and $V$ is
a harmonic superfield taking values in the algebra of the gauge group generators in the given representation and called ``bridge''.
For the superfield $\tilde{F}^{-n} := e^{iV}F^{-n}$ the constraint in \p{lemma} implies $\partial^{++}\tilde{F}^{-n} = 0\,$, whence
$\tilde{F}^{-n} = 0$ (see eqs. (4.20), (4.21) in \cite{harm}) and ${F}^{-n} = 0\,$. Note also that the constraint $\nabla^{++} F^{+n} = 0\,, \;n \geq 0\,,$
implies $ F^{+n} = e^{-iV} F^{i_1\cdots i_n}u^+_{i_1} \cdots u^+_{i_n}\,$. This property will be widely used in Sect. 7.

\section{Invariant actions of the ${\cal N}=(1,0)$ vector multiplet and a hypermultiplet}
\setcounter{equation}0

In this section we present the actions of canonical dimension $d=4$
containing the standard  kinetic terms of the ${\cal N}=(1,0)$ vector gauge multiplet and a hypermultiplet.

\subsection{The dimension 4 Lagrangian  of the gauge multiplet}
The superfield action providing the supersymmetric extension of the standard $d=4$ Yang-Mills Lagrangian
for the $6D$ gauge fields $\sim {\rm Tr} (F^{MN}F_{MN})$ is given
by the following expression which is non-local in harmonics
\cite{Zup1},
   \be
S^{SYM} =\frac{1}{f^2}\sum\limits^{\infty}_{n=2} \frac{(-1)^{n}}{n} {\rm Tr} \int
d^6\!x\, d^8\theta\, du_1\ldots du_n \frac{\Vp(z,u_1 )
\ldots \Vp(z,u_n ) }{(u^+_1 u^+_2)\ldots (u^+_n u^+_1 )}\,,\lb{action1}
   \ee
where $f$ is a coupling  constant  carrying the dimension of inverse mass.
This action is invariant under the supergauge transformations (recall \p{nablapp}, \p{nablamm})
 \be
\lb{gaugetran1}
 \delta V^\pp = \nabla^\pp \Lambda \,.
  \ee

The gauge freedom \p{gaugetran1} allows one to bring the superfield $V^\pp$ in the  Wess-Zumino gauge,
 \be
 \lb{V++WZ}
V^{++} \ =\ \th^{+a} \th^{+b} A_{ab} + 2 (\th^+)^3_a \lambda^{-a} - 3 (\th^+)^4 {\cal D}^{--} \, ,
 \ee
where $A_{ab}$ is the gauge field, $\lambda^{-a} = \lambda^{ai} u_i^-$ is the gaugino and
${\cal D}^{--} = {\cal D}^{ik} u^-_i u^-_k$, where ${\cal D}^{ik} = {\cal D}^{ki}$ are the auxiliary fields.
The component fields entering \p{V++WZ}  depend only on the coordinates $x^M$, but not on the harmonic
variables.

The component Lagrangian derived from \p{action1} has a simple form,
\be
\lb{Lagr-comp-d4}
{\cal L} \ = \ \frac 1{2f^2} {\rm Tr}\left( - F_{MN}^2 + i\lambda^k \gamma^M \nabla_M \lambda_k - {\cal D}^{ik}
{\cal D}_{ik} \right),
 \ee
with $F_{MN} = \pa_M A_N - \pa_N A_M - i[A_M, A_N]$ and $\nabla_M = \pa_M - iA_M$. It
 gives rise to the standard equations of motion of the second order for the gauge fields
 and of the first order  for the fermions.

These equations can be derived from the superfield equation of motion following
from \p{action1} by using the general formula for variation of $S^{SYM}$,
\be
\delta S^{SYM} = -\frac{1}{f^2} {\rm Tr}\int dZ\, \delta V^{++} V^{--} = -\frac{1}{f^2} {\rm Tr}\int d\zeta^{(-4)}\delta V^{++} F^{++}\,.
\ee
This gives the extremely simple equation of motion
 \be
 F^\pp = 0 \, . \lb{Fpp=0}
  \ee

\subsection{The dimension 4 hypermultiplet Lagrangian}
The invariant action  for the hypermultiplet in the adjoint representation (being interested in ${\cal N}=(1,1)$ extension, we will deal only with
this assignment of the hypermultiplet), giving rise to the Lagrangian of the canonical dimension 4, is given by the following integral
over the analytic superspace
\be \lb{hypAct}
S^q = \frac{1}{2 f^2}{\rm Tr} \int d\zeta^{(-4)} q^{+ A}\nabla^{++} q^+_A\,, \quad \nabla^{++} q^+_A = D^{++}q^+_A + [V^{++}, q^+_A]\,.
\ee
The coupling constant can be chosen the same as in $S^{SYM}$, keeping in mind a freedom of rescaling of $q^{+ A}$. The corresponding
equation of motion is
\be
E^{+3} :=  \nabla^{++} q^+_A = 0\,.\lb{Urq1}
\ee
An equivalent form of the same equation is
\be
(\nabla^{--})^2 q^{+ A} = \nabla^{--}q^{-A} = 0\,, \quad q^{-A} := \nabla^{--} q^{+ A}\,.\lb{Urq2}
\ee
This can be proved by acting on the l.h.s. of \p{Urq2} by $\nabla^{++}$, observing that the result is zero
as a consequence of \p{Urq1}, and then applying the {\bf Lemma} of the previous section. Note also the useful relations
\be
D^+_aq^{-A} = -\nabla^-_aq^{+ A}\,, \quad \nabla^-_aq^{-A} = 0\,, \lb{onsh1}
\ee
which follow from the analyticity of $q^{+ A}$ and the equations of motion \p{Urq2}.

As an instructive example, we consider the superfield action of the {\it free} hypermultiplet
  \be
S_{\rm free}^q =  \int d\zeta^{(-4)}\,
q^{+ A}D^{++} q^+_A  \, . \lb{S0hyp}
\ee
The corresponding equation of motion is
 \be
\lb{eqmothyp}
 D^\pp q^{+A} \ =\ 0 \, .
 \ee
The on-shell constraint \p{eqmothyp} together with the G-analyticity condition
$D^+_a q^{+A} = 0$ can be resolved to find
  \be
\lb{q+free}
  q^{+A} \ =\ \varphi^{+A} - \theta^{+a} \psi_a^A - i \theta^{+a} \theta^{+b} \partial_{ab} \varphi^{-A} \, ,
 \ee
 where $\varphi^{\pm A} = \varphi^{jA} u^\pm_j$ are physical harmonic-independent
{\it on-shell} scalar fields. They satisfy the free equation of motion $\Box \varphi = 0$. And $\psi_a^A$ are
right-handed on-shell fermionic fields satisfying the free Dirac equation.

\subsection{The ${\cal N}\,{=}\,(1,1)$ SYM action and its hidden ${\cal N}\,{=}\,(0,1)$ \break supersymmetry}

  We now consider the actions \p{action1} and \p{hypAct}  together and write
\be
S^{Vq^+} = S^{SYM} + S^q = \frac{1}{f^2}\left(\int dZ {\cal L}^{\rm SYM} +
\frac12 {\rm Tr}\int d\zeta^{(-4)} q^{+ A}\nabla^{++} q^+_A \right).
\lb{Summ}
\ee
The sum \p{Summ} exhibits invariance under the extra hidden ${\cal N}=(0,1)$  supersymmetry,
\be
\delta_0 V^{++} = \epsilon^{+ A}q^+_A\,, \quad \delta_0 q^{+ A} = -(D^+)^4 (\epsilon^{-A} V^{--})\,, \quad
\epsilon^{\pm}_A = \epsilon_{aA}\theta^{\pm a}\,,\lb{Hidden}
\ee
which completes the manifest ${\cal N}=(1,0)$ supersymmetry to ${\cal N}=(1,1)\,$.
This means that \p{Summ} is in fact the ${\cal N}=(1,0)$ form of the ${\cal N}=(1,1)$ SYM theory action.
Note a useful representation for the variation $\delta_0q^{+ A}$ through the superfield strengths $F^{++}$ and $W^+_a$:
\be
\lb{deltaq}
\delta_0 q^+_A =- \epsilon_{aA}(\theta^{-a} F^{++} - W^{+ a}).
\ee
It is consistent with the analyticity of $q^{+A}$ because of the analyticity of $F^{++}$ and the relation \p{DW0}.

The invariance \p{Hidden} is quite analogous to the hidden ${\cal N}=2, \ 4D$ supersymmetry which completes
the manifest ${\cal N}=2, \ 4D$ supersymmetry
of the sum of the harmonic superspace actions for the ${\cal N}=2, \ 4D$ SYM field and the
adjoint  hypermultiplet
to 
 ${\cal N}=4$ supersymmetry \cite{harm}. This sum is thus nothing but a representation
of the ${\cal N}=4, \ 4D$ SYM action in terms of  ${\cal N}=2$ superfields.

Like in the ${\cal N}=4,\ 4D$ case, the transformations \p{Hidden}
have the correct closure with themselves and with the manifest ${\cal N}=(1,0)$ supersymmetry only on  mass shell,
when the equations of motion corresponding to the action \p{Summ},
\be
E^{++} := F^{++} + \frac12 [q^{+A}, q^+_A] = 0\,, \quad E^{+3} :=  \nabla^{++} q^+_A = 0\,,\lb{Fmodif}
\ee
are satisfied.

A direct calculation shows that $(\delta_2\delta_1 - \delta_1\delta_2)V^{++}$ amounts to
\be
(\delta_2\delta_1 - \delta_1\delta_2)V^{++} = -\nabla^{++}\Lambda +
 i\varepsilon^{abcd}f_{21 [ab]}\partial_{cd} V^{++}\,, \quad
f_{21 [ab]} := \epsilon^A_{2[a}\epsilon_{1b]A}\,,
\ee
where
\be
\Lambda = (D^+)^4\left[(\epsilon_1^{-A}\epsilon^-_{2A})V^{--}\right]
\ee
is a gauge transformation superfield parameter.
Thus, the hidden supersymmetry has the correct off-shell closure on $V^{++}$. This is not the case for $q^{+}_A$. The same bracket
yields
\be
(\delta_2\delta_1 - \delta_1\delta_2)q^{+}_A = [\Lambda, q^+_A] + i\varepsilon^{abcd}f_{21 [ab]}\partial_{cd} q^{+}_A - (D^+)^4
[\epsilon^-_{2 A}\tilde\delta_1 V^{--} - [\epsilon^-_{1 A}\tilde\delta_2 V^{--}]\,,
\ee
where $\tilde{\delta}V^{--}$ is defined from the relation
\be
\delta_0 V^{--} = \epsilon^{-A} \nabla^{--} q^+_A + \tilde{\delta}V^{--}\,,
\ee
and involves the terms vanishing on the
hypermultiplet equations of motion \p{Urq1}, \p{Urq2}.
Thus, in the full analogy
with \p{Lie_bracket},  $(\delta_2\delta_1 - \delta_1\delta_2)q^{+A}$ involves a nontrivial extra term
which vanishes only on mass shell.

Let us now consider the commutators of the hidden supersymmetry with the manifest one, {\it i.e.} with
\be
\hat{\delta} (q^{+}_A, V^{++}) = \eta^{+ a}Q^-_a (q^{+}_A, V^{++})\, \quad Q^{-}_{a} = \frac{\partial}{\partial \theta^{+ a}} + \ldots\,, \quad
\eta^{+ a}= \eta^{ia}u^+_i\,.
\ee
We find
\be
(\hat{\delta} \delta_0 - \delta_0\hat{\delta})V^{++}  = (\eta^{ia}\epsilon_a^B u^+_i) q^+_B := f^{Bi}u^+_i\,q^+_B\,.
\ee
This variation can be identically rewritten as
\be
f^{Bi}u^+_i\,q^+_B= \nabla^{++}( f^{Bi}u^-_i\,q^+_B) - f^{Bi}u^-_i \nabla^{++}q^+_B\,,
\ee
{\it i.e.}, once again, it is reduced to some analytic gauge transformation of $V^{++}$ only on the hypermultiplet mass shell, {\it i.e.}, with
$\nabla^{++}q^{+ A} = 0\,$.

Analogously,
\be
(\hat{\delta} \delta_0 - \delta_0\hat{\delta})q^{+}_A =
-[f^{Bi}u^-_i q^{+}_B, q^+_A] - f_A^iu^{-}_i \left(F^{++} +  \frac12 [ q^{+C}, q^+_C]\right),
\ee
{\it i.e.}, it is reduced to the pure gauge transformation on the mass shell for $V^{++}\,$.

We conclude that the correct ${\cal N}=(1,1)$ closure of the transformations \p{Hidden} with themselves
and with the manifest ${\cal N}=(1,0)$ transformations is achieved only on the mass shell for both superfields $V^{++}$ and $q^{+ A}\,$.
To avoid a possible confusion, we point out that the action \p{Summ} is invariant under the transformations \p{Hidden} {\it off shell},
with $V^{++}$ and $q^{+ A}$
being unconstrained analytic superfields. The mass-shell conditions are required to ensure the correct ${\cal N}=(1,1)$
closure for these transformations.

It is instructive to see how the  superfield equations of motion \p{Fmodif}
are transformed into each other under the transformations \p{Hidden} . Using the properties \p{Vazhnoe},
we find
\be
\delta_0 E^{+3}_{A} = -\epsilon^+_A \,E^{++}\,.
\ee
It is a little more complicated to see that the variation of $E^{++}$ is actually expressed through $E^{+3}$ and the
alternative form \p{Urq2} of the hypermultiplet equation of motion \p{Urq1}. It is easy to show that
\be
\delta_0 V^{--} = \epsilon^{-A} \nabla^{--} q^+_A + \mbox{terms containing $(\nabla^{--})^2 q^{+ B}$ and  $\nabla^{++}q^{+ B}$}\,. \lb{varVmin}
\ee
Then it can be shown that the contribution of the first ``dangerous'' term in \p{varVmin} to $\delta_0 F^{++}$ is exactly canceled by the
variation of the second term in $E^{++}$, so $\delta_0 E^{++}$ is expressed through the terms containing the equivalent forms \p{Urq2}
of the hypermultiplet equation of motion.

In what follows we will meet the situation when the superfields involved in the ${\cal N}=(0,1)$ transformations
above satisfy themselves the mass-shell equations \p{Fmodif}, \p{Urq1} or \p{Urq2}. The various superfields defined earlier are transformed on shell as
\be
&&\delta q^{\pm A} = \hat{\delta} q^{\pm A} - [\epsilon^{- B}q^+_B,\, q^{\pm A}]\,,
 \qquad \hat{\delta} q^{\pm A} = \epsilon^A_a\,W^{\pm a}\,,\lb{qtransfOn1} \\
&&\delta W^{\pm a} = \hat{\delta} W^{\pm a}-[\epsilon^{- B}q^+_B,\, W^{\pm a}]\,, \qquad
\hat{\delta} W^{\pm a}= -i\varepsilon^{abcd}\epsilon^A_b \nabla_{cd}q^{\pm}_A\,. \lb{WtransfOn}
\ee
Note that the transformation \p{qtransfOn1} immediately follows from \p{deltaq} upon using
the equation of motion $E^{++} = 0$ from the set \p{Fmodif}.

 We will need also the on-shell transformation rules for the spinor and vector derivatives of $q^{\pm A}$:
\be
&&\delta(\nabla_{bc}q^{\pm}_A) = \epsilon_{aA}\nabla_{bc} W^{\pm a} +
\sfrac{i}2 \big( \epsilon^B_b [\nabla^-_c q^+_B, \, q^\pm_A] -
\epsilon^B_c [\nabla^-_b q^+_B, \, q^\pm_A] \big)
- [\epsilon^{-B}q^+_B,\,\nabla_{bc}q^{\pm}_A]\,, \lb{transfnabla1} \\
&&\delta(D^+_a q^{-}_A) = -\delta(\nabla^-_a q^{+}_A) =
-\epsilon_{bA}\,D^+_aW^{- b} +\epsilon^B_a\,[q^+_B,\,q^-_A]
-[\epsilon^{-B}q^+_B,\,D^+_a q^{-}_A]\,. \lb{transDq1}
\ee
Note that the last terms in the variations \p{qtransfOn1} - \p{transDq1} are some field-dependent gauge transformations
and
they do not contribute to the variation of the gauge-invariant Lagrangian involving the traces
over the ``color'' indices.

\section{Higher-dimensional ${\cal N}=(1,0)$ and ${\cal N}=(1,1)$ invariants}
\setcounter{equation}0
Now we turn to the discussion of higher-dimensional invariants. As was mentioned in the very beginning,
the pure $6D$ gauge theories are {\it chiral} theories, they involve
only the left-handed  gaugino field and hence are plagued by the chiral anomaly \cite{6Danomaly}. In other words,
gauge symmetry is broken there by quantum effects, which restricts their physical interest. Anomaly can be canceled and
the gauge symmetry  kept intact in the theories involving, besides the left-handed gauginos, also right-handed fermions belonging
to the matter hypermultiplet. This condition is obviously satisfied in the ${\cal N}=(1,1)$ gauge theory.
We will be mainly interested in this section in the on-shell ${\cal N} = (1,1)$  invariant gauge theory, which is written in terms
of ${\cal N} = (1,0)$  superfields and  may or may not possess the full off-shell ${\cal N} = (1,0)$  supersymmetry.

For a higher-dimensional operator to be a  counterterm
  giving a logarithmically divergent contribution
 to the
scattering amplitudes $\sim \ln \Lambda_{UV}$, it must not vanish on mass shell, but its
supersymmetric variation under the on-shell ${\cal N} = (1,1)$
transformations should be reduced
 to a total derivative.   We first discuss  the operators of canonical dimension 6.

\subsection{$d= 6$}

It is very easy to write down the  superfield gauge-invariant action of canonical physical dimension 6 in the
gauge field sector. It has the following unique form \cite{ISZ}:
 \be
\label{dejstvie}
S_{SYM}^{(6)}\  = \frac{1}{2g^2}{\rm Tr}\int  d\zeta^{(-4)}du \,
 \left(F^\pp\right)^2.    \lb{hactan}
\ee
Here the coupling constant $g$ is dimensionless \footnote{Indeed, the superfields $V^{++}, V^{--}$ are dimensionless, and
it follows that $F^{++}$ defined in (\ref{Fpp}) has canonical
dimension 2.}. The component expression for \p{dejstvie} involves extra derivatives,
 \be
\lb{L_d=6_comp}
S = - \frac{1}{2g^2}{\rm Tr} \int \Big[ (\nabla^M F_{ML} )^2 + \frac 12 (\nabla_M {\cal D}_{jk} )^2
+ {\cal D}_{lk} {\cal D}^{kj} {\cal D}_j^{\ l} + \ {\rm fermion \ terms} \Big].
 \ee
We see that the auxiliary fields of the Lagrangian \p{Lagr-comp-d4} enter the $d=6$
Lagrangian with derivatives --- the same phenomenon that we observed in Sect. 2 in a toy SQM
model; the higher-dimensional Lagrangian \p{L_d=6_comp}  is related to the Lagrangian
\p{Lagr-comp-d4} in the same way as the higher-dimensional Lagrangian \p{d2tcomp} to the Witten Lagrangian
\p{LSQMcom}.

We observe that the integrand  in \p{dejstvie} is just the square of the equation of motion
\p{Fpp=0} and therefore this $d=6$ action  {\it vanishes} on mass shell  modulo possible hypermultiplet terms.
Now we are going to show that the same remains true for the $d=6$ actions taking into account the hypermultiplet terms.

 One can write a series of new hypermultiplet
$d=6$ actions $S_n$ representing full  superspace  integrals
\cite{IShyper},
 \be \lb{Snhyp}
 S_n \ \sim \ {\rm Tr} \int du dZ q^{+A} (\nabla^{--} )^n (\nabla^{++} )^{n-1}  q^+_A \, .
  \ee
All the terms with $n > 1$ vanish on mass shell. Using \p{Fpp}, it is convenient to represent the non-vanishing action $S_1$
as an integral over the analytic superspace,
  \be
{\rm Tr} \int dZ  q^{+ A}\nabla^{--} q^+_A =  {\rm Tr}  \int
d\zeta^{(-4)}\, F^{++} [q^+_A, q^{+ A}] \,. \ee

There is one more
 $d=6$ interaction involving the hypermultiplet field. It
does not contain harmonic derivatives and is given by the  analytic superspace integral,
\be S_{\rm quart}  \sim {\rm Tr} \int du d\zeta^{(-4)} [q^{+ A},
q^+_A]^2\,.\lb{quart} \ee

Thus, if disregarding the terms vanishing on the mass shell, a generic
${\cal N} = (1,0)$ invariant  Lagrangian reads
  \be
{\cal L}^{d=6} = \frac{1}{2g^2} {\rm Tr} \int du d\zeta^{(-4)}\,  \Big[
(F^{++})^2  + \alpha  F^{++} [ q^{+ A},
q^+_A]  +\beta [q^{+ A}, q^+_A]^2 \Big].
\lb{L6Ansatz}
\ee
The requirement that its ${\cal N} = (0,1)$
variation vanishes on mass shell imposes the restriction $\alpha =
2\beta + 1/2$ such that the Lagrangian acquires the form
  \be
{\cal L}^{d=6} = \frac{1}{2g^2} {\rm Tr} \int du d\zeta^{(-4)}\, \left( F^{++} + \frac  12 [ q^{+ A}, q^+_A] \right) \left(
F^{++} + 2\beta [ q^{+ A}, q^+_A] \right),
\lb{L6Ansatz-1}
\ee
  which vanishes on shell due to \p{Fmodif}.

We have thus shown that the non-vanishing on-mass-shell counterterms
of canonical dimension 6 are absent, and this proves the one-loop
finiteness of the theory \p{Summ}.

The fact that the algebra of extended supertransformations does not
close off shell suggests that an action corresponding to
\p{L6Ansatz-1} with some fixed $\beta$,
 to which a series of the actions $S_n$ in \p{Snhyp} with arbitrary
coefficients is added, cannot be invariant off shell. Indeed, when
one tries to construct such an invariant (the corresponding
calculations are presented in Appendix B), one meets obstacles that
seem to be unsurmountable. It is easy to see that, in order to ensure the
cancellation of the terms $\propto (q^+)^3$ in the variation, the
coefficient $\beta$ in \p{L6Ansatz-1} should be fixed to $\beta =
1/4$. But the linear in $q^{+A}$ terms do not want to be canceled among themselves, no
matter what you try.

We thus conjecture that  a $d=6$ off-shell  ${\cal N} = (1,1)$
supersymmetric invariant does not exist \footnote{ We did not
rigorously {\it prove} it, however --- it is always difficult to
prove the {\it absence} of something.}.

\subsection{$d=8$}

Once again, we begin with the gauge field sector and write appropriate off-shell
${\cal N}= (1,0)$ supersymmetric gauge-invariant Lagrangians of canonical dimension $d=8$, having in mind to extend them to
the ${\cal N}=(1,1)$ invariants by adding some hypermultiplet terms. It turns out that all such purely gauge field
terms vanish on the mass shell of \eqref{Summ}, in agreement with  \cite{HoweStelle}. Then we
write the full list of different possible $d=8\,$,  ${\cal N}=(1,0)$ superfield terms involving the hypermultiplet contributions and demonstrate
that, on the equations of motion corresponding to the total ${\cal N}=(1,1)$  action \p{Summ}, they are all reduced to a single expression,
which is not invariant under the hidden supersymmetry \p{Hidden}, \p{qtransfOn1} - \p{transDq1} (and there is no way to make it invariant).

We consider first the $d=8$ terms in the pure gauge field sector.  The SYM equations of motion are  $F^\pp = 0$.
The vanishing of some structures (like $ {\rm Tr} \int dZ \,
\nabla^-_a W^{-a}
F^\pp$),  is obvious. We consider now a couple of less trivial examples.
   \begin{itemize}
  \item Let
  \be
\lb{S81}
S^{(8)}_1 \ =\ {\rm Tr}  \int dZ \, (\nabla^-_a W^{+a} ) (D^+_a W^{-a} ) \, ,
 \ee
where $W^{-a} = \nabla^\m W^{+a}$.
We use the identity
  \be
\lb{DW+-}
\nabla^-_a W^{+a} = D_a^+ W^{-a} \,,
 \ee
which is derived  in Appendix A as a corollary of certain Bianchi identities. We obtain
  \be
S^{(8)}_1 \ =\ {\rm Tr} \int dZ \, (\nabla^-_a W^{+a} )^2
=
{\rm Tr} \int d\zeta^{(-4)} (D^+)^4 (\nabla^-_a W^{+a} )^2\, .
 \ee
The terms involving $(D^+)^3 (\nabla^-_a W^{+a} )$ and  $(D^+)^4 (\nabla^-_a W^{+a} )$ contain $F^\pp$ and
vanish on mass shell. We are left with the structure
$$ \propto \epsilon^{cdef} D^+_c D^+_d (\nabla^-_a W^{+a} )  D^+_e D^+_f (\nabla^-_b W^{+b} )\, . $$
On mass shell, it is equivalent to
$$ \propto \epsilon^{cdef} \epsilon_{cdam} \{ W^{+m}, W^{+a} \} \epsilon_{efbn} \{ W^{+n}, W^{+b} \}
\sim \epsilon_{manb} \{ W^{+m}, W^{+a} \}  \{ W^{+n}, W^{+b} \} \, ,
$$
  which {\it vanishes} as the anticommutator $\{ W^{+m}, W^{+a} \}$ is symmetric under $m \leftrightarrow a\,$.

\item Let
 \be
\lb{S82}
S^{(8)}_2 \ =\ {\rm Tr} \int dZ \, (\nabla^-_a W^{+b} ) (D^+_b W^{-a} ) \, .
 \ee
Integrating by parts with respect to $\nabla^-_a$, using the commutation relation
\be
\{D^+_b, \nabla^-_a\}  \ =\ 2i \nabla_{ba} \ =\ -  2i \nabla_{ab} \ =\  -\{D^+_a, \nabla^-_b\} \, ,
 \ee
disregarding the terms involving $F^\pp$, and integrating by parts once again, we reduce
  \p{S82} to \p{S81}.

 \end{itemize}

Now we turn to the general proof that there exist no ${\cal N} = (1,0)$ supersymmetric off-shell invariants of the dimension 8 which could
respect the on-shell ${\cal N} = (1,1)$ invariance.

To this end, we construct
the full set of the superfield Lagrangians of
dimension 4 in the full ${\cal N}=(1,0)$ harmonic superspace (they correspond to the dimension 8 component
Lagrangians)\footnote{For brevity, we omit here the ${\rm Tr}$ symbol with
respect to ``color'' indices, but we will always have it in mind.}:
\be
&& L^{(1)}_W = \nabla^-_a
W^{-a} D^+_bW^{+ b}\,, \; L^{(2)}_W = \nabla^-_a W^{+a} D^+_bW^{-
b}\,, \; L^{(3)}_W = \nabla^-_a W^{+b} D^+_bW^{- a}\,, \nn &&
L^{(4)}_W = \nabla^-_a W^{+b}\nabla^-_b W^{+ a}\,,\; L^{(5)}_W =
D^+_a W^{-b}D^+_b W^{- a}\,,
\lb{Wlagr4} \\
&& L^{(1)}_q = iS^A_a \nabla_{bc} S_{d A}\varepsilon^{abcd}\,,
\;L^{(2)}_q =
 [q^{+(A}, \,q^{-B)}] [q^{+}_A, \,q^{-}_B]\,, \;
L^{(3)}_q = [q^{-A}, \,q^{-}_A][q^{+B}, \,q^{+}_B]\,.\lb{qlagr4}
\ee
Note that a conceivable  term $\sim W^{+ a}\nabla_{ab}W^{-b}$ is
reduced to the other structures in the list \p{Wlagr4}, \p{qlagr4} by integrating
by parts with respect to the spinor derivatives
under the (undisplayed) trace.

Using the off-shell relations \p{DW++}, \p{DWsum} and \p{DW+-} and
also bearing in mind that $\{D^+_a, \,D^+_b\} = \{\nabla^-_a,
\,\nabla^-_b\} =0$, it is easy to show that all Lagrangians in the
set \p{Wlagr4} are reduced to $L^{(2)}_W$ or to $L^{(1)}_W\,$, which in turn are related
to each other by integrating by parts with respect to $\nabla^{--}$.
This proof is valid off shell and does not require
 passing to the analytic subspace at any intermediate step.

 Next, using the on-shell relations \p{DWq1}, it is straightforward to show that
\be
L^{(1)}_W (\mbox{on-shell})\; \Rightarrow \; 4\,L^{(3)}_q\,.
\ee
Also, using simple algebraic manipulations and integrating by parts
with respect to harmonic derivatives, one can show that
\be
L^{(2)}_q (\mbox{on-shell})\; \Rightarrow \;
\sfrac34\,L^{(3)}_q\,.
\ee
It remains to work out $L^{(1)}_q$.
Integrating by parts, it can be reduced to
\be
L^{(1)}_q \;
\Rightarrow \;
-i\varepsilon^{abcd}q^{-A}D^+_a\nabla_{bc}D^+_dq^-_A\,.
\ee
Using
the on-shell relation \p{7Id} and, once again, integrating the term
$q^{- A} \{W^{+ a},\, \nabla^-_a q^+_A \}$ by parts with respect to
$\nabla_a^-$, one reduces  $L^{(1)}_q$, up to a total harmonic
derivative, to $2 L^{(3)}_q\,$.

Thus,  all possible superfield Lagrangians of the
dimension 4 are reduced on  mass shell to the single non-vanishing  structure
\be
L^{(3)}_q =
[q^{-A}, \,q^{-}_A][q^{+B}, \,q^{+}_B]\,.\lb{finLq}
\ee
Bearing in mind the overall trace, the variation of $L^{(3)}_q$ under the hidden ${\cal N}=(0,1)$
supersymmetry \p{qtransfOn1}
is given by
\be
\delta_{\epsilon}L^{(3)}_q \sim \epsilon_{a A}[q^{-B},
\,q^{-}_B][q^{+A}, \, W^{+ a}]\,. \lb{varL3}
\ee
It is non-vanishing, and no terms can be invented to cancel \p{varL3}. Thus, no ${\cal N}=(1,1)$
invariant terms of the dimension 8 can be constructed out of the ${\cal N}=(1,0)$ superfields.

It is worth noting that in the hypermultiplet sector one can contemplate
${\cal N}=(1,0)$ invariants which are not reduced to the product of
``color'' anticommutators as in \p{finLq}, e.g.,
\be
\sim q^{+
A}q^-_A q^{+ B} q^-_B\,, \quad \mbox{or} \sim q^{+ A}q^{-B} q^{+}_A
q^-_B\,.
\ee
Nevertheless, it is impossible to ensure  the mutual
cancellations of the ${\cal N}=(0,1)$ variations of such
terms, while keeping the requirement for the Lagrangian not to vanish on mass shell.
To check this, we wrote down all the independent
terms of this kind, calculated their variations (reducing
$\delta q^{-A}=\nabla^{--}\delta q^{+A}$ to $\delta q^{+ A}$ through
integrating by parts) and found the unique combination of such
terms, $\sim {\rm Tr} \left(q^{+A}q^{-}_A  q^{+B}q^{-}_B +
q^{+A}q^{- B}  q^{+}_{B}q^{-}_A\right)$, the variation of which is
zero up to a total harmonic derivative. However, it is easy to show
that, on the mass shell of  $q^{+A}$, this combination is a total
harmonic derivative on its own.

Surprisingly, the $d=8$ superfield expression which is non-vanishing on shell and respects the on-shell
${\cal N}=(1,1)$ supersymmetry can be constructed if we {\it give up} the requirement of {\it off-shell} ${\cal N}=(1,0)$ supersymmetry.

\subsection{On-shell ${\cal N}=(1,0)$ and ${\cal N} = (1,1)$ invariants}

As the complete off-shell  ${\cal N} = (1,1)$ superfield formalism is absent, it is not
possible to write down operators   of
a fixed canonical dimension $d>4$ which would be  invariant off shell under
 the   ${\cal N} = (1,1)$ transformations. This concerns  the operators of dimension $d=8$ and $d=10$.
However,  in contrast to the case $d=6$, the {\it on-shell} $d=8$ and $d=10$ invariants  exist,
 and it is possible to find them. The basic idea is to seek for the invariants, in which not only
 the hidden ${\cal N} =(0,1)$ supersymmetry is realized on shell, but which are ${\cal N} =(1,0)$ supersymmetric
 also only on shell.

Once again, we start our consideration from the simple example in the gauge field sector.
If we lift the requirement of off-shell ${\cal N}=(1,0)$ supersymmetry, we can define
the non-vanishing $d=8$ operators that are  supersymmetric only on mass shell. One of them reads
   \be
{\tilde S}^{(8)}_{1}  =  \frac14 {\rm Tr} \int d\zeta^{(-4)}\,
\varepsilon_{abcd}W^{+ a} W^{+ b} W^{+ c} W^{+ d}\,, \lb{dim8anal-1}
  \ee
where the factor $\frac14$ was introduced for further convenience. Indeed,
eq. \p{DW0} tells us that $D^+_a W^{+b} = \delta_a^b F^{++}$,
which vanishes on mass shell. Thus, when disregarding the terms proportional to the equations
of motion, $W^{+a}$ is a G-analytic superfield and so the action \p{dim8anal-1} respects ${\cal N}=(1,0)$ supersymmetry on shell
\footnote{Note that a similar on-shell invariant
 appears as a one-loop contribution to the quantum effective action of the ${\cal N}=(1,0)$ gauge theory
in the $6D$ harmonic superspace
in a special background \cite{BuPl2}.}.
Being expressed through components, the bosonic part of \p{dim8anal-1}
gives the known $ F^4$ structure \cite{Gross:1986iv},
\be
\label{F4}
 [{\cal L}^{(8)}_{an}]_{\rm bos} &=& \frac1{2\cdot 81} {\rm Tr}_{(s)} \Big[ 2
F_{MN} F^{MN} F_{PQ} F^{PQ}  +
F_{MN} F_{PQ} F^{MN} F^{PQ} \nn
&&{\;\;\;} -\, 4F^{NM} F_{MR} F^{RS} F_{SN}  - 8
F^{NM} F_{MQ} F_{NR} F^{RQ}\Big].
 \ee
This expression can be derived using the component representation for $W^{+ a}$ \cite{ISZ,BuPl},
  \be
W^{+a} =  \frac i6 F_{MN} (\sigma^{MN})^a_{\ b} \theta^{+b} +  {\rm fermion\  terms} \
+\ {\rm terms\ vanishing\ on\ shell}  + O[(\theta^{+})^2],
 \ee
as well as the identities \p{Trgam_epsilon}, \p{gamgam_epsilon}. Note the presence of the  symmetrized
color traces ${\rm Tr}_{(s)} \sim {\rm Tr} X(Y
Z U  + UYZ + ZUY) $ in \p{F4}.

This tensor structure  reproduces indeed the so-called $t_8$ tensor obtained in the tree-level
four-gluon scattering amplitude \cite{Gross:1986iv}. The complete component form of the associated supersymmetry invariant in six dimensions
was first obtained in \cite{Bergshoeff:1986jm}.

It is also possible to write down an on-shell ${\cal N}= (1,0)$ supersymmetric invariant involving
the product of two color traces,
 \be
{\tilde S}^{(8)}_2 =  \frac14\int d\zeta^{(-4)} \varepsilon_{abcd} {\rm Tr}
(W^{+ a} W^{+ b})\,{\rm Tr} (W^{+ c} W^{+ d})\,. \lb{dim8anal-2}
  \ee

The next step is to seek for the  on-shell  ${\cal N}= (1,1)$ completion of the $d=8$ terms \p{dim8anal-1} and \p{dim8anal-2}.
Clearly, it should be a collection of terms containing the hypermultiplet superfield $q^{+ A}$. First one should construct the full list of
the dimension $d=8$ operators which are G-analytic on the shell of the full set of equations of motion  following from the action
\p{Summ}, {\it i.e.} eqs. \p{Fmodif} and \p{Urq1} [or \p{Urq2}].  Next one needs to select the  ${\cal N}= (0,1)$  invariant combination
of such operators (if it exists).

The minimal on-shell G-analytic extension of \p{dim8anal-1} [{\it i.e.} the expression analytic as a consequence of the full set of equations \p{Fmodif}, \p{Urq1}]
is given by the following expression:
\be
{L}^{+ 4}_0 &=& {\rm Tr}\,\Big\{\frac14 \varepsilon_{abcd} W^{+ a}W^{+ b}W^{+ c}W^{+ d} -i\nabla_{ab}q^{+ A}\big(W^{+ a}q_A^+ W^{+b}
+ 2q_A^+ W^{+a}W^{+b} \big) \nn
&&-\,  W^{+ a}D^+_aq^{-A}\big[ q^+_A (q^+)^{2} + \frac12 (q^+)^{2}q^+_A\big]
+ (q^+)^{2}D^+_aq^{-A}\big( q^+_A W^{+ a} +\frac12W^{+ a}q^+_A\big) \nn
&& -\, 2 (q^+)^{2}\big[q^{- A}q^+_A (q^+)^{2} + \frac12q^{- A} (q^+)^{2}q^+_A \big]\Big\}, \lb{L0}
\ee
where $(q^+)^2 := q^{+A} q^+_A = \frac12 [q^{+A}, q^+_A]\,$. The full list of other possible  $d=8$ superfield G-analytic
terms involving the single trace is given
in Appendix C. It is shown there that, by integrating by parts, they all can be reduced to the two independent structures,
${L}^{+ 4}_2$ and ${L}^{+ 4}_3$ [eqs. \p{L2} and \p{L3}]. Then the ${\cal N}=(1,1)$ supersymmetric combination is uniquely determined to be
\be
{\cal L}^{+ 4}_{(1,1)} = {L}^{+ 4}_0 + {L}^{+ 4}_3\,.\lb{Inv11}
\ee

It is instructive to see how the proof goes on in the abelian case. Passing to the abelian limit in \p{L0} and \p{L3}, we write
\be
 {\cal L}^{+ 4}_{(1,1)} &=&   \frac{1}{4} \varepsilon_{abcd} W^{+a} W^{+b} W^{+c} W^{+d}  + 3i q^{+A} \partial_{ab} q_A^+ W^{+a} W^{+b}
-  q^{+A} \partial_{ab} q_A^+ \, q^{+B} \partial^{ab} q_B^+ \nn
&\equiv&  {\cal L}_{(I)}^{+4} + {\cal L}_{(II)}^{+ 4} + {\cal L}_{(III)}^{+ 4} \, .
  \ee
Our task is to prove that it is invariant on mass shell under the transformations
\be
\delta q^{+A} \ =\ \epsilon^A_a W^{+a}, \ \ \ \ \ \ \ \ \ \ \delta W^{+a} = -2i \epsilon^A_b \partial^{ab} q^+_A \, .
\ee

It is easy to see that the linear in $q$ terms in the sum of $\delta {\cal L}_{(I)}^{+4}$ and $\delta {\cal L}_{(II)}^{+4}$ vanish.
We are left with
\be
\lb{Delta}
\Delta := \delta {\cal L}_{(I)}^{+4} + \delta {\cal L}_{(II)}^{+4}= \ 6 \epsilon^B_c  \varepsilon^{acde}  q^{+A} \partial_{ab} q_A^+  \,
\partial_{de} q^+_B \, W^{+b} \, ,
 \ee
The variation of ${\cal L}_{(III)}^{+4}$ is
  \be
\lb{deltaL3}
\delta {\cal L}_{(III)}^{+4} \ =\ -2 \epsilon^B_c\,\varepsilon^{abed}   q^{+A} \partial_{ab} q_A^+ \, \partial_{ed} q^+_B \, W^{+ c} \, .
 \ee
To see the cancellation of \p{Delta} and \p{deltaL3}, one should use the  cyclic identities
  \be
\lb{cyclic}
  \varepsilon^{abcd}  \delta^e_f + \varepsilon^{bcde}  \delta^a_f + \varepsilon^{cdea}  \delta^b_f
+\varepsilon^{deab}  \delta^c_f + \varepsilon^{eabc}  \delta^d_f  \ = 0 \, , \nn
\varepsilon^{AB}  \delta^C_D +  \varepsilon^{BC}  \delta^A_D + \varepsilon^{CA}  \delta^B_D \ =\ 0
 \ee
{\it and} the equations of motion $\partial_{ab} W^{+b} = 0$, $\Box q^{+A} = 0$.

Namely, we represent
\be
\Delta = -6 \epsilon^B_cq^{+A} \partial_{ab} q^+_A  \, \partial_{de} q^+_B \, W^{+f} \left[  \varepsilon^{cdeb}  \delta^a_f +
\varepsilon^{deba}  \delta^c_f + \varepsilon^{ebac}  \delta^d_f + \varepsilon^{bacd}  \delta^e_f \right] \equiv
{\cal A} + {\cal B} + {\cal C} + {\cal D}
 \ee
and then observe that ${\cal A} = -\Delta$,  ${\cal B} = -3\delta {\cal L}_{(III)}^{+4}$ and
 \be
{\cal C} = {\cal D} = -6  \epsilon^B_c\,\varepsilon^{abce}   q^{+A} \partial_{ab} q^+_A  \, \partial_{de} q^+_B \, W^{+ d} \, .
  \ee
Next, using the second identity in \p{cyclic} and integrating by parts, we derive that ({\it on shell !})
${\cal C} = -\Delta - {\cal C}$ and hence ${\cal C} = -\Delta/2$. This gives
$$\Delta = -\Delta - 3\delta {\cal L}_{(III)}^{+4} - \frac 12 \Delta - \frac 12 \Delta \, ,$$
and, finally, $\Delta = -\delta {\cal L}^{+ 4}_{(III)}$.

The proof in the non-abelian case is much more complicated since there is a lot of various terms coming from different sources.
Nevertheless, we checked that \p{Inv11} is still invariant up to a total derivative. However, this direct method is very cumbersome and
it is natural to seek for another more universal and easier approach. It will be developed in the next sections.
As the important preparatory step, we
note here that \p{Inv11} admits the following equivalent representation through the symmetrized trace:
\be
 \label{L++++fin} {\cal L}^{+4}_{(1,1)}  &=& {\rm Tr}_{(S)}\Big\{  \frac{1}{4} \varepsilon_{abcd}
 W^{+a} W^{+b} W^{+c} W^{+d}  + 3i q^{+A} \nabla_{ab} q_A^+ W^{+a} W^{+b}
-  q^{+A} \nabla_{ab} q_A^+ \, q^{+B} \nabla^{ab} q_B^+ \nn
&& - \, W^{+a} [ D_a^+ q_A^- , q_B^+] q^{+A} q^{+B}
-\frac{1}{2} [ q^{+C} , q^+_C] [ q_A^-,q^+_B] q^{+A} q^{+B}\Big\}.
\ee
The subscript S stands for the symmetrization, meaning that the expression is
symmetrized with respect to the permutation of the four arguments,
\begin{eqnarray} &&  {\rm Tr}_{(S)} \Big( A_1 A_2 A_3 A_4 \Big) \nn
&=& \frac{1}{6} {\rm Tr} \Big( A_1 A_2 A_3 A_4 +
 A_2 A_3 A_1 A_4+A_3 A_1 A_2 A_4+A_3 A_2 A_1 A_4+A_2 A_1 A_3 A_4+A_1 A_3 A_2 A_4 \Big),\lb{SymmTr}
\nonumber
\end{eqnarray}
any commutator being understood as one argument. One can now directly verify, in particular,  that
\begin{eqnarray}  D_a^+ {\cal L}^{+4}_{(1,1)}
= {\rm Tr}_{(S)}\Big\{ E^{++}
 \Bigl( \varepsilon_{abcd} W^{+b} W^{+c} W^{+d} + 6 i q^{+B} \nabla_{ab} q_B^+\, W^{+b} -
[ D_a^+ q_A^- , q_B^+] q^{+A} q^{+B}
 \Bigr) \Big\},  \nonumber
 \end{eqnarray}
 where $E^{++} = F^{++} + \frac12 [q^{+ A}, q^+_A] = 0\,$. This vanishes on mass shell.

Our final comment in this section is that  the double-trace invariant \p{dim8anal-2}
also  admits an ${\cal N} = (1,1)$  completion. Here we present only the minimal $G$-analytic extension [analogous to the extension \p{L0}]. It reads
\be
\tilde{L}^{+ 4}_0 &=& \sfrac14 \varepsilon_{abcd}{\rm Tr}\, (W^{+ a} W^{+ b})\, {\rm Tr}\,(W^{+ c} W^{+ d})
-i{\rm Tr}\,(\nabla_{ab}q^{+A}q^+_A)\,{\rm Tr}\,(W^{+ a} W^{+ b})\nn
&& +\,{\rm Tr}\,(D^+_{a}q^{-A}q^+_A)\,{\rm Tr} [W^{+ a} (q^+)^2]
- {\rm Tr}\,(q^{-A}q^+_A)\,{\rm Tr} \,[(q^+)^2 (q^+)^2]\,.
\ee
It is straightforward to check that this expression is indeed annihilated by $D^{+}_a$ on the mass shell. There exists a freedom of adding
other on-shell analytic Lagrangians, like in the single-trace case. They all vanish in the limit of vanishing $q^{+ A}$.

The double-trace analog of the on-shell ${\cal N}=(1,1)$ invariant \p{Inv11}, \p{L++++fin}
will be derived in the next sections, based on the universal method we are going to expose now.

\section{On-shell ${\cal N}=(1,1)$ harmonic superfields}
\setcounter{equation}0

The most convenient way to construct on-shell ${\cal N}=(1,1)$ invariants of the type we discussed in the subsection 5.3
is to define the on-shell superfields living in extended  harmonic ${\cal N} = (1,1)$
superspace. This and the next two sections are devoted to this subject. Extended on-shell superfields of the similar kind
 were first discussed in \cite{HST,HoweStelle}, but not in the framework of harmonic superspace.
We will see that ``harmonization'', introduced first in \cite{Bossard:2009sy},
  helps a lot. In particular, it allowed us to  resolve explicitly
a set of constraints which the on-shell ${\cal N} = (1,1)$ SYM superfields must obey.

\subsection{The standard and harmonic ${\cal N}=(1,1)$ superspaces}
 We introduce the extended superspace involving, in addition to the odd pseudoreal
left-handed variables $\theta^a_i$, also the odd pseudoreal right-handed
variables $\hat{\theta}^A_a$ ($A=1,2$), which  belong to another
spinor representation,
\be
z = (x^{ab}, \theta^a_i) \; \Rightarrow \; \hat{z} = (x^{ab}, \theta^a_i, \hat{\theta}_{Aa}. \lb{hatCentr}
\ee
We then consider the covariant spinor derivatives,
  \be
\lb{spincovder}
\nabla^i_a \ =\ \frac \partial {\partial \theta^a_i} - i \theta^{bi} \partial_{ab} + {\cal A}^i_a \, ,\nn
\hat{\nabla}^{aA} \ =\ \frac \partial {\partial \hat{\theta}_{Aa}} - i \hat{\theta}_b^A \partial^{ab}  +
\hat{{\cal A}}^{aA} \, ,
  \ee
where ${\cal A}^i_a$ and $ \hat{{\cal A}}^{aA}$
 are the spinor connections and the convention
 $\nabla^{ab} = \tfrac{1}{2} \varepsilon^{abcd}
 \nabla_{cd}$ is assumed.  The superfields
${\cal A}^i_a$, $ \hat{{\cal A}}^{aA}$
 are not arbitrary, but satisfy the constraints
\be
\lb{spincomm0}
\{\nabla^{(i}_a, \nabla^{j)}_b \} \ =\ \{\hat{\nabla}^{a(A}, \hat{\nabla}^{bB)} \} = 0 \, ,
 \ee
 \be
\lb{spincommphi}
\{\nabla^i_a, \hat{\nabla}^{bA} \} \ =\ \delta_a^b \phi^{iA} \, .
 \ee
Bearing in  mind the Bianchi identities, the constraints \p{spincomm0} and \p{spincommphi} imply
 \be
\lb{const-phiiA}
 \nabla^{(i}_a \phi^{j)A}  \ =\   \hat{\nabla}^{a(A}   \phi^{B)i} \ =\ 0 \, .
 \ee
The  constraints \p{spincomm0}, \p{spincommphi},  written in   \cite{HST,HoweStelle},
 define the ${\cal N}=(1,1), \ 6D$ supersymmetric
Yang-Mills theory.
They are known to imply the equations of motion for the  superfields involved.
Below we will show how this property comes about
in the harmonic 
superfield formalism.

We introduce now the harmonics $u^{\hat{\pm}}_A$ which parametrize the second $SU(2)$ automorphism group acting
on the indices $A$ and have the same properties as $ u_i^\pm$. Note, in particular, the identities
 \be
\lb{ident-uhat}
u^{\hat{+}A} u^{\hat{-}}_A = 1, \ \ \ \
u^{\hat{+}}_A u^{\hat{-}}_B  - u^{\hat{-}}_A u^{\hat{+}}_B = \epsilon_{AB} \, .
 \ee
 Respectively, we extend the ${\cal N}=(1,0)$
harmonic superspace \p{hss6} to the ${\cal N}=(1,1)$ harmonic superspace
\be
Z = (x^{ab}, \theta^a_i, u^\pm_k) \; \Rightarrow \; \hat{Z} = (x^{ab}, \theta^a_i, \hat{\theta}_{Aa}, u^\pm_k, u^{\hat{\pm}}_A)\,. \lb{hathss6}
\ee
The analytic basis of this extended harmonic superspace is defined as the set of coordinates
\be
\hat{Z}_{({\rm an})} = (x^{ab}_{({\rm an})}, \theta^{\pm a}, {\theta}^{\hat{\pm}}_a, u^\pm_k, u^{\hat{\pm}}_A),
\ee
where ${\theta}^{\hat{\pm}}_a :=\hat{\theta}^A_a u^{\hat{\pm}}_A$ and
   \be
x^{ab}_{({\rm an})} \ =\ x^{ab} + \frac i2 (\theta^{+a} \theta^{-b} - \theta^{+b} \theta^{-a} )
+ \frac i2 \varepsilon^{abcd} \theta_c^{\hat{+}} \theta_d^{\hat{-}} \, .
   \ee

Next we define the harmonic projection $\phi^{+ \hat{+}} = \phi^{iA} u_i^+ u_A^{\hat{+}}$.
It is clear from \p{const-phiiA} and from the fact that $\phi^{+\hat{+}}$ does not depend on $u_i^-$ and
$u^{\hat{-}}_A$ that $\phi^{+\hat{+}}$ satisfies the constraints
  \be
\lb{const-harm}
 \nabla^+_a \phi^{+\hat{+}} \ =\ \nabla^{a \hat{+}} \phi^{+\hat{+}} \ = \
\partial^{++} \phi^{+\hat{+}} \ =\  \partial^{\hat{+} \hat{+}} \phi^{+\hat{+}} \ =\ 0 \, ,
  \ee
where
 \be
\lb{covder-harm}
\nabla^+_a = \nabla^i_a u^+_i, \ \ \ \nabla^{\hat{+}a} = \hat{\nabla}^{aA} u^{\hat{+}}_A, \ \ \
  \partial^{++}  = u^{+i}  \frac \partial {\partial u^{-i}}, \ \ \partial^{\hat{+}\hat{+}}  =
u^{\hat{+}A} \frac \partial {\partial u^{\hat{-}A}} \, .
  \ee
The spinor covariant derivatives obviously commute with $\partial^{++}$ and $ \partial^{\hat{+}\hat{+}}$.
The full set of defining  (anti)commutators of the gauge ${\cal N}=(1,1), 6D$ theory
in the central basis of the considered bi-harmonic  superspace are
  \be
&& \{\nabla^+_a, \nabla^+_b \} \ =\ \{\nabla^{\hat{+}a}, \nabla^{\hat{+}b} \} = 0, \lb{zerocomm}\\
&& \{\nabla^+_a, \nabla^{\hat{+}b} \} \ =\ \delta_a^b \phi^{+\hat{+}}  \, , \lb{spincomm-harm} \\
&& [ \partial^{++}, \nabla^+_a] = [\partial^{\hat{+}\hat{+}}, \nabla^+_a] = [ \partial^{++}, \nabla^{a\hat{+}}] =
[\partial^{\hat{+}\hat{+}}, \nabla^{a\hat{+}}] = 0\,. \lb{harmcomm-harm}
\ee
Note that, having defined this set, we do not longer need to assume in advance that the $+$ and
$\hat +$ components of the spinor derivatives are as in \p{covder-harm}. It is the relations \p{harmcomm-harm}
 which force them to be linear in harmonics.  Thus, the extended set of constraints \p{zerocomm}
 - \p{harmcomm-harm} is fully equivalent to the  original constraints \p{spincomm0},
\p{spincommphi} without any additional assumptions. The constraints \p{const-harm}
naturally arise as a consequence of Bianchi identities for \p{zerocomm} - \p{harmcomm-harm}.

\subsection{From the central basis to the analytic basis}
 As usual in the harmonic superspace approach,  at the next steps we should pass
to the analytic basis in order to solve the above constraints in terms of the appropriate
 analytic superfield prepotentials and, in particular, to find the explicit form
of the basic superfield strength $\phi^{+ \hat{+}}$. Due to the relation
\p{spincomm-harm}, the analyticities associated with the harmonic sets
 $u^\pm_i$ and $u^{\hat{\pm}}_A$ cannot be made manifest simultaneously.
 In what follows, we will choose the basis in which the spinor derivative
 $\nabla^{\hat{+}a}$ is short,  $\nabla^{\hat{+}a} =
 {\partial}/ { \partial \theta^{\hat{-}}_a}\,$, so that the ``hat'' analyticity is manifest.

Consider first the abelian case, which is much simpler.
Our task is to  find a field $\phi^{+\hat{+}}$
that satisfies the constraints \p{const-harm}.  In the abelian case, the field $\phi^{\hat{+}+}$ does not carry
a charge with respect to the gauge $U(1)$ group, and, as a result, the constraints $\nabla_a^+ \phi^{\hat{+}+} = \nabla^{\hat{+}a} \phi^{\hat{+}+} = 0$ amount
to $D_a^+ \phi^{\hat{+}+} = D^{\hat{+}a} \phi^{\hat{+}+} = 0$ with flat spinor derivatives.
The anticommutator $\{D^+_a, D^{\hat{+}b} \}$ vanishes, so these derivatives can be made ``short'' by passing
to the double analytic basis, where
 $D^+_a =  {\partial}/ { \partial \theta^{-a}}$ and  $D^{\hat{+}}_a =
 {\partial}/ { \partial \theta^{\hat{-}}_a}$. On the contrary, the harmonic derivatives in this basis are lengthened:
 \be
\lb{flat-harm-der}
&&D^{++} = \partial^{++} +i\theta^{+a}\theta^{+b}\pa_{ab}^{({\rm an})}+ \th^{+a}\frac {\partial}{\partial \theta^{-a}}, \nn
&& D^{\hat{+}\hat{+}} = \partial^{\hat{+}\hat{+}} +i\theta^{\hat{+}}_a \theta^{\hat{+}}_b \pa^{ab({\rm an})} +
\th^{\hat{+}}_a \frac {\partial}{\partial \theta^{\hat{-}}_a} \,.
 \ee
It is not difficult now to resolve the abelian constraints \p{const-harm}. The solution reads
   \be
\lb{phi++Ab}
\phi^{+\hat{+}} &=&
\varphi^{+\hat{+}} - \theta^{+a} \psi_a^{\hat{+}} -
\theta_a^{\hat{+}} \lambda^{+a} + \frac i6  \theta_a^{\hat{+}}
\theta^{+b} F^a_{\ b} - i \theta^{+a} \theta^{+b} \partial_{ab} \varphi^{-\hat{+}}\nn
&&  - \, i\theta_a^{\hat{+}}
 \theta_b^{\hat{+}} \partial^{ab} \varphi^{+\hat{-}}
+ i  \theta_a^{\hat{+}}   \theta^{+b} \theta^{+c} \partial_{bc} \lambda^{-a} + i  \theta^{+a}  \theta_b^{\hat{+}}
 \theta_c^{\hat{+}} \partial^{bc} \psi^{\hat{-}}_a \nn
&& -\,  \theta_a^{\hat{+}}
 \theta_b^{\hat{+}}   \theta^{+c} \theta^{+d}  \partial^{ab} \partial_{cd}  \varphi^{- \hat{-}} \,.
 \ee
Here,  the fermionic fields satisfy the Dirac equations $\partial_{ab} \lambda^a = \partial^{ab} \psi_a = 0$, the scalar
field satisfies $\Box \varphi = 0$ and $ F^a_{\ b} = (\sigma^{MN})^a_{\ b} F_{MN}$.
We see that the superfield $\phi^{+\hat{+}}$ satisfying our constraints  automatically satisfies also the equations of motion, {\it i.e.} it is an
on-shell superfield.

For sure, this should not come as a surprise. The same is true for the free hypermultiplet superfield $q^+$
in the usual ${\cal N}=(1,0)$ superspace. In the abelian case, this superfield satisfies the constraints
$D^+ q^+ = D^{++} q^+ = 0$. Its component expansion in the analytic basis
is given by  \p{q+free},
with scalar and fermionic fields
 satisfying the free equations of motion.
The component expansion \p{phi++Ab} of the
free superfield $\phi^{+\hat{+}}$ represents an obvious generalization of \p{q+free}
\footnote{It is also possible to define the  off-shell harmonic ${\cal N}=(1,0)$ superfield $q^+$
whose expansion into harmonics
gives an infinite number of degrees of freedom. For the superfield $\phi^{+\hat{+}}$, this seems to be impossible.}.

Now we come back  to the general non-abelian case. Consider the constraint $\{ \nabla^{\hat{+} a},  \nabla^{\hat{+} b} \} = 0$. Its
generic solution is
  \be
\nabla^{\hat{+} a} \ =\ e^{iV} D^{\hat{+} a} e^{-iV} \, ,
 \ee
where $V$ is a general bi-harmonic superfield (often called {\it bridge}). It is convenient now to perform the similarity transformation
  \be
\lb{conjug}
\nabla^{\hat{+} a} \to D^{\hat{+} a}, \ \ \  \phi^{+\hat{+}} \to e^{-iV} \phi^{+\hat{+}} e^{iV} , \ \ \
 \nabla^+_a \to e^{-iV} \nabla^+_a e^{iV}
 \ee
and define
  \be
&& \nabla^{++} \ = \  e^{-iV} \partial^{++} e^{iV} =  \partial^{++} + V^{++}, \ \ \ \  \nabla^{\hat{+}\hat{+}} \ = \
e^{-iV} \partial^{\hat{+}\hat{+}} e^{iV}  = \partial^{\hat{+}\hat{+}} + V^{\hat{+}\hat{+}}\,,\lb{harmder-conj} \\
&& V^{++} := e^{-iV} \left(\partial^{++} e^{iV}\right)\,, \quad V^{\hat{+}\hat{+}} := e^{-iV} \left(\partial^{\hat{+}\hat{+}} e^{iV}\right). \lb{defVV}
 \ee
The transformed spinor derivatives still satisfy the algebra \p{zerocomm}-\p{harmcomm-harm} and
commute with the transformed harmonic derivatives (which involve now nontrivial harmonic connections $V^{++}$ and $V^{\hat{+}\hat{+}}$).
As was anticipated, to resolve the constraints, we go to the ``hat-analytic'' basis
\footnote{By performing the similarity transformation and going to this basis, we can get rid only of one of the spinor connections,
which we have chosen to be ${\cal A}^{\hat{+}a}$. Alternatively, one could suppress ${\cal A}^+_i$.},
where $D^{\hat{+} a}
=  {\partial} / {\partial \theta^{\hat{-} a} }$ and $\partial^{\hat{+}\hat{+}}$ goes over to $D^{\hat{+}\hat{+}}$ defined in  \p{flat-harm-der}.

In the next section we will solve the system of constraints
 \be
\lb{const-harm-nAb}
 \nabla^+_a \phi^{+\hat{+}} \ =\ D^{a \hat{+}} \phi^{+\hat{+}} \ = \
\nabla^{++} \phi^{+\hat{+}} \ =\  \nabla^{\hat{+} \hat{+}} \phi^{+\hat{+}} \ =\ 0 \, ,
  \ee
with the spinor and harmonic covariant derivatives given in the analytic basis and frame
and satisfying the algebra
  \be
&& \{\nabla^+_a, \nabla^+_b \} \ =\ \{D^{\hat{+}a}, D^{\hat{+}b} \} = 0, \lb{svjazi}\\
&& \{\nabla^+_a, D^{\hat{+}b} \} \ =\ \delta_a^b \phi^{+\hat{+}}  \, , \lb{svjazi10} \\
&& [ \nabla^{++}, \nabla^+_a] = [\nabla^{\hat{+}\hat{+}}, \nabla^+_a] = [ \nabla^{++}, D^{a\hat{+}}] =
[\nabla^{\hat{+}\hat{+}}, D^{a\hat{+}}] = 0\,, \lb{svjazi11} \\
&& [\nabla^{++}, \nabla^{\hat{+}\hat{+}}] = 0\,, \lb{svjazi12}
\ee
which directly follows  from the constraints \p{zerocomm}-\p{harmcomm-harm} written in the central basis.

One can now verify that
an explicit solution of the system of  equations \p{svjazi} - \p{svjazi12}  is
 \be
\lb{nabla+viaq}
\nabla^+_a = D^+_a - \theta^{\hat{+}}_a q^{+\hat{-}} +  \theta^{\hat{-}}_a \phi^{+\hat{+}} \, ,
 \ee

\be
\lb{V++_expan}
V^{\hat{+}\hat{+}} \ =\  i \theta_a^{\hat{+}} \theta_b^{\hat{+}} \, {\cal A}^{ab} -
 \frac 13 \epsilon^{abcd}
 \theta^{\hat{+}}_a \theta^{\hat{+}}_b
\theta^{\hat{+}}_c \, D^+_d q^{- \hat{-}}  + \frac 18  \epsilon^{abcd}
 \theta^{\hat{+}}_a \theta^{\hat{+}}_b
\theta^{\hat{+}}_c  \theta^{\hat{+}}_d \, [q^{+\hat{-}}, q^{-\hat{-}} ]
  \ee
and

 \begin{eqnarray}
\lb{phi++}
\phi^{+\hat{+}} &=& q^{+ \hat{+}} - \theta^{\hat{+}}_a W^{+a} - i
 \theta^{\hat{+}}_a \theta^{\hat{+}}_b \nabla^{ab} q^{+\hat{-}}  + \frac{1}{6}  \varepsilon^{abcd}
 \theta^{\hat{+}}_a \theta^{\hat{+}}_b  \theta^{\hat{+}}_c [ D^+_d q^{-\hat{-}}  , q^{+\hat{-}}] \nonumber \\
&& \hspace{20mm} +  \frac{1}{24}  \varepsilon^{abcd}  \theta^{\hat{+}}_a \theta^{\hat{+}}_b
 \theta^{\hat{+}}_c\theta^{\hat{+}}_d  [ q^{+\hat{-}}, [ q^{+\hat{-}} , q^{-\hat{-}}] ] \,.
 \end{eqnarray}
Here the objects $q^{+\hat{\pm}} = q^{+A}u^{\hat{\pm}}_A$ and $q^{+A}$, ${\cal A}^{ab}$, $W^{+a}$, as well as
 $V^{++}$ entering the covariant derivative $\nabla^{++}$, are the ${\cal N}=(1,0)$ superfields
discussed in the previous sections. For self-consistency, they should satisfy their equations of motion, e.g., $\nabla^{++}q^{+ A} =0\,$.
In the next section, we will present an accurate {\it derivation} of this solution from the set of constraints \p{svjazi} - \p{svjazi12}
and show thereby that the solution \p{nabla+viaq}-\p{phi++} is {\it unique}.  We will also derive the variations of the on-shell
superfields  $\phi^{+\hat{+}}$ and $V^{\hat{+}\hat{+}}$ under the  $\cN=(0,1)$
 supersymmetry transformations and demonstrate that the particular representation \p{V++_expan} for the gauge
 superfield $V^{\hat{+}\hat{+}}$ in \p{V++_expan} is none other than  the appropriate  Wess-Zumino gauge choice for it.

In the abelian case, the commutators vanish, the covariant derivative $\nabla^{ab}$ is replaced by the
ordinary one, and the  superfield \p{phi++} is reduced to the abelian superfield \p{phi++Ab} ($\lambda^a$ being the
lowest component of $W^{+a}$). Note that the non-abelian expression for $\phi^{+\hat{+}}$
does not enjoy anymore the symmetry
under interchange $\theta \leftrightarrow \hat{\theta}$.
That is due to our choice to work in the frame, where
the {\it hatted} spin connection vanishes.

To close this section, we write the variations of the superfields \p{V++_expan}, \p{phi++} under
the ${\cal N} = (0,1)$ supertransformations, just anticipating their derivation in the next section:
\be
&& \delta V^{\hat{+}\hat{+}} = -\epsilon^{\hat{+}}_a \frac{ \partial\, }
{\partial \theta_a^{\hat{+}}}V^{\hat{+}\hat{+}}  - 2 i \epsilon_a^{\hat{-}}
 \theta_b^{\hat{+}} \partial^{ab} \phi^{+\hat{+}}  + \nabla^{\hat{+} \hat{+}}\Lambda^{(comp)}\,,\lb{TranVhat} \\
&&
\delta \phi^{+\hat{+}} = - \epsilon^{\hat{+}}_a \frac{ \partial\, }
{\partial \theta_a^{\hat{+}}} \phi^{+\hat{+}} - 2 i \epsilon_a^{\hat{-}}
 \theta_b^{\hat{+}} \partial^{ab} \phi^{+\hat{+}}  - [\Lambda^{(comp)}, \phi^{+\hat{+}}]\,,\lb{variation2}
\ee
where the field-dependent compensating gauge parameter $\Lambda^{(comp)}$ is given by the expression
\be
\Lambda^{(comp)} &=& (\epsilon^{-B}q^+_B) + 2i \epsilon^{\hat{-}}_a \theta^{\hat{+}}_b{\cal A}^{ab}
-\frac12 \varepsilon^{abcd}\epsilon^{\hat{-}}_a \theta^{\hat{+}}_b\theta^{\hat{+}}_c D^+_d q^{-\hat{-}} \nn
&& +\frac16 \varepsilon^{abcd}\epsilon^{\hat{-}}_a \theta^{\hat{+}}_b\theta^{\hat{+}}_c\theta^{\hat{+}}_d
[q^{+\hat{-}}, q^{-\hat{-}}]\,.\lb{Compen}
\ee
The first two terms in \p{variation2} and \p{TranVhat}  are induced by the supersymmetric
variations of $\theta^{\hat{+}}_a$ and $x^{ab}$. The third term is an extra
 gauge transformation needed to preserve the Wess-Zumino form of the superfield $V^{\hat{+}\hat{+}}$ after the
supersymmetry transformation. It is worth pointing out that the simple form \p{variation2} and \p{TranVhat}
of the hidden ${\cal N}=(0,1)$ transformations is obtained, provided that the involved   ${\cal N}=(1,0)$ superfields are
subject to their equations of motion. At the same time, under the manifest ${\cal N}=(1,0)$ supersymmetry the expressions
\p{phi++} and \p{V++_expan} behave as the standard off-shell ${\cal N}=(1,0)$ harmonic superfields.

In fact, the transformations  \p{variation2} and \p{TranVhat} can be derived directly from the on-shell
transformation laws \p{qtransfOn1} - \p{transDq1}   of the involved ${\cal N} = (1,0)$ superfields, using the identities
listed in Appendix A.

\setcounter{equation}{0}
\section{Solving the ${\cal N}=(1,1)$ SYM  constraints in terms of ${\cal N}=(1,0)$ superfields}

In this section, we solve the constraints in the analytic basis and frame and show that
their general solution is given by eqs. \p{nabla+viaq}, \p{V++_expan}, \p{phi++}.

\subsection{Input and gauge-fixing}
We start with the whole set of constraints \p{zerocomm} - \p{harmcomm-harm} written in a more detailed form,
 \be
({\rm a})\;\, \{\nabla^+_a, \nabla^+_{b} \}  = 0\,, \qquad ({\rm b})\;\, \{D^{\hat{+}a},
D^{\hat{+}b} \}  = 0\,, \qquad ({\rm c})\;\,
\{\nabla^+_a, D^{\hat{+}b} \}  = \delta_a^b \phi^{+\hat{+}}\,, \lb{antic12}
\ee
\be
({\rm a})\;\, [ \nabla^{\hat{+}\hat{+}}, \nabla^+_a]  = 0\,, \;\;  ({\rm b})\;\,[\tilde{\nabla}^{{+}{+}}, \nabla^+_a]  = 0\,, \;\;
({\rm c})\;\,[\nabla^{\hat{+}\hat{+}}, D^{a\hat{+}}]  = 0\,, \;\;
({\rm d})\;\, [ \tilde{\nabla}^{++}, D^{a\hat{+}}]  = 0\,, \lb{harmspin}
\ee
\be
[\tilde{\nabla}^{++}, \nabla^{\hat{+}\hat{+}}]  = 0\,. \lb{HarmHarm}
\ee
Here
\be
\nabla^+_a = D^+_a + {\cal A}^+_a(\hat{Z})\,, \lb{nabla+}
\ee
and the ``hatted'' spinor derivatives were chosen to be short,
$\nabla^{\hat{+}a} = D^{\hat{+}a} = \partial/\partial \theta_a^{\hat{-}}\,$ \footnote{One can always get rid of the
spinor
connection $\hat{\cal A}$ in  the covariant derivatives $\nabla^{\hat{+}a}$,
capitalizing on their anticommutativity in any basis and frame.}. Thus, in the chosen basis,
 the ``hatted'' G-analyticity is manifest\footnote{In the general non-abelian case,
one cannot make simultaneously manifest both the hatted and unhatted G-analyticities
 because of the non-vanishing anticommutator (\ref{antic12}c).} and the constraints
(\ref{harmspin}c) and (\ref{harmspin}d)
imply that both harmonic gauge connections
in the harmonic derivatives $\tilde{\nabla}^{++}$ and $\nabla^{\hat{+}\hat{+}}$ are independent
of the coordinates $\theta^{\hat{-}}_b$:
\be
\tilde{\nabla}^{++} = D^{++} + \tilde{V}^{++}(\hat{\zeta})\,, \quad \nabla^{\hat{+}\hat{+}} =
D^{\hat{+}\hat{+}} + V^{\hat{+}\hat{+}}(\hat{\zeta})\,, \lb{nablas1}
\ee
where $\hat{\zeta} = (x^{ab}_{({\rm an})}, \theta^{\pm a}, \,\theta^{\hat{+}}_a,\, u^\pm_i, \,u^{\hat{\pm}}_A)$.
In what follows, we omit the index
``$({\rm an})$'' for the analytic coordinate  $x$.
We use the notation $\tilde{\nabla}^{++}$ in order to distinguish this harmonic derivative acting in the
full ${\cal N}=(1,1)$ superspace from its  ${\cal N}=(1,0)$ counterpart.

At this step, both harmonic connections are arbitrary functions of the hatted analytic coordinates $\theta^{\hat{+}}_a$
and the harmonics $u^{\hat{\pm}}_A$, as well as of the full set of the  ${\cal N}=(1,0)$ harmonic superspace coordinates.
They are transformed with the hat-analytic superfield parameter $\Lambda(\tilde{\zeta})$:
\be
&&\delta \tilde{V}^{++} = \tilde{\nabla}^{++}\Lambda(\hat{\zeta})\,, \lb{gaugeTransf++}\\
&&\delta V^{\hat{+}\hat{+}} = \nabla^{\hat{+}\hat{+}}\Lambda(\hat{\zeta}).\lb{gaugeTransf++hat}
\ee
The constraints (\ref{harmspin}c) and (\ref{harmspin}d) imply no other consequences.

As the next steps,
we wish to show that the dependence of the harmonic
connections $V^{\hat{+}\hat{+}}$ and $\tilde{V}^{++}$  on the coordinates $\theta^{\hat{+}}_a\,, u^{\hat{\pm}}_A$
can be drastically simplified {\bf (i)} by choosing the Wess-Zumino- type gauge for $V^{\hat{+}\hat{+}}$ and {\bf (ii)}
by exploiting the constraint \p{HarmHarm} for  $\tilde{V}^{++}$ (see the next subsection).

It is straightforward to see that the gauge freedom associated with the superfield transformation
parameter $\Lambda(\hat\zeta)$ can be partially fixed by putting $V^{\hat{+}\hat{+}}$
in the ``short'' form,\footnote{For further convenience, we
use the  abbreviations $\Psi^{\hat{+}3d} := \varepsilon^{abcd}\theta^{\hat{+}}_a\theta^{\hat{+}}_b\theta^{\hat{+}}_c\,$,
$\Psi^{\hat{+}4} :=\varepsilon^{abcd}\theta^{\hat{+}}_a\theta^{\hat{+}}_b\theta^{\hat{+}}_c\theta^{\hat{+}}_d\,$. The
identities
 $\theta^{\hat{+}}_a\theta^{\hat{+}}_b\theta^{\hat{+}}_c  = \frac16\varepsilon_{abcd}\Psi^{\hat{+}3d}\,, \quad
\theta^{\hat{+}}_a\theta^{\hat{+}}_b\theta^{\hat{+}}_c\theta^{\hat{+}}_d = \frac1{24}\varepsilon_{abcd}\Psi^{\hat{+}4}\,, \quad
\theta^{\hat{+}}_a\Psi^{\hat{+}3b} =-\frac14 \delta^b_a\Psi^{\hat{+}4}$ hold.}
\be
V^{\hat{+}\hat{+}} = i\theta^{\hat{+}}_a \theta^{\hat{+}}_b\hat{\cal A}^{ab} + \Psi^{\hat{+}3\,d}\varphi_d^{\hat{-}} +
\Psi^{\hat{+}4}{\cal D}^{\hat{-}2}\,, \quad \varphi_d^{\hat{-}} = \varphi_d^A u^{\hat{-}}_A\,,\quad
{\cal D}^{\hat{-}2} ={\cal D}^{(AB)}u^{\hat{-}}_A u^{\hat{-}}_B\,, \lb{WZhat}
\ee
where $\hat{\cal A}^{ab}, \varphi_d^A$ and ${\cal D}^{(AB)}$ are some ${\cal N}=(1,0)$ harmonic superfields, still arbitrary at this step.
While passing to \p{WZhat}, the $(\theta^{\hat{+}}_a, u^{\hat{\pm}}_A)$ dependence of $\Lambda(\hat{\zeta})$ was fully used up,  so the residual gauge
freedom is associated with the gauge function $\Lambda_{{\rm int}}(x, u^{\pm}_i, \theta^{\pm a})\,,\; \Lambda \rightarrow \Lambda_{{\rm int}}$.
Note that this gauge parameter still depends on $\theta^{-a}$.
Now we are going to show that this dependence can be fixed by a further
 gauge choice.

To this end, we need to inspect the structure of the spinor derivative $\nabla_a^+ = D^{+}_a + {\cal A}^+_a$.
First of all, the Bianchi identities, following from the full set \p{antic12}, imply
the  G-analyticity conditions for $\phi^{+\hat{+}}$,
\be
({\rm a})\;\, D^{\hat{+} a}\phi^{+\hat{+}} = 0\,, \qquad  ({\rm b})\;\,\nabla_a^+\phi^{+\hat{+}} = 0\,.   \lb{GanalVhat}
\ee
Postponing the discussion of the condition (\ref{GanalVhat}b) to the next subsection, we focus  here on
the constraint (\ref{GanalVhat}a).
Due to the ``shortness''  of $D^{\hat{+}a}$, 
it implies that $\phi^{+\hat{+}}$ does not depend on $\theta^{\hat{-}}_a$.
 In addition, this constraint together with (\ref{antic12}c)
 uniquely fixes the spinor connection
${\cal A}^+_a$ to be
\be
{\cal A}^+_a = \tilde{\cal A}^+_a + \theta^{\hat{-}}_a \phi^{+\hat{+}}\,, \lb{A+}
\ee
where
\be
\tilde{\cal A}^+_a = f^+_a + \theta^{\hat{+}}_b f_a^{+\hat{-} b} + \theta^{\hat{+}}_b \theta^{\hat{+}}_c f^{+ \hat{-}\hat{-}bc}_a +
\Psi^{\hat{+}3\,d}f_d^{+\hat{-}3} + \Psi^{\hat{+}4} f^{+ \hat{-}4}_a\,. \lb{tildeA+}
\ee
The component superfields in this expansion depend on both the ${\cal N}=(1,0)$ coordinates (including the
harmonics $u^\pm_i$)
and the extra harmonics $u^{\hat{\pm}}_A$.

One of the consequences of the constraint (\ref{antic12}a) is
\be
D^+_a f^+_b + D^+_b f^+_a + \{f^+_a, f^+_b \} = 0\,,
\ee
whence $f^+_b = e^{i\tilde{v}}(D^+_be^{-i\tilde{v}})$, where $\tilde{v}$ is an additional ``bridge'' which does not depend on
$\theta^{\hat{+}}_a$ (because $f^+_b$ does not). Using this bridge, we can pass to the frame where $f^+_b=0$ and the residual
gauge group is  represented by the standard
analytic superfield parameter $\Lambda(\zeta)$ of the ${\cal N}=(1,0)$ gauge theory.
Indeed, the residual gauge transformations preserving
the condition $f^+_b=0$  commute with $D^+_a$, whence $D^{+}_a \Lambda = 0\,$.

Hereafter, we will use the spinor connection ${\cal A}^+_a$ in the form \p{A+}, \p{tildeA+} with the condition
     \be
f^+_a =0\,,\lb{f+a0}
\ee
and the following $\theta^{\hat{+}}_b$ expansions for the hat-analytic superfields $\phi^{+\hat{+}}$ and $\tilde{V}^{++}$,
\be
&& \phi^{+\hat{+}} = q^{+\hat{+}} - \theta^{\hat{+}}_a W^{+ a} + \theta^{\hat{+}}_a\theta^{\hat{+}}_b\beta^{+\hat{-}ab}
+ \Psi^{\hat{+}3 d}G^{+\hat{-}\hat{-}}_d + \Psi^{\hat{+}4} G^{+\hat{-}3}\,, \lb{expphi} \\
&& \tilde{V}^{++} = V^{++} + \theta^{\hat{+}}_a v^{++\hat{-}a} + \theta^{\hat{+}}_a\theta^{\hat{+}}_b v^{++\hat{-}\hat{-}ab}
+ \Psi^{\hat{+}3 d}v^{++\hat{-}3}_d + \Psi^{\hat{+}4}v^{++\hat{-}4}\,. \lb{tildeV}
\ee
In \p{expphi}, \p{tildeV} we introduced the notation $q^{+\hat{+}}, W^{+ a}$ and $V^{++}$, having in mind that these quantities will be finally identified with
the ${\cal N}=(1,0)$ superfields considered before. However, at the present stage,
all the coefficients in the expansions \p{expphi}, \p{tildeV}
are still generic ${\cal N}=(1,0)$ superfields involving  an extra dependence on the  harmonics $u^{\hat{\pm}}_A$.

Now we are ready to explore all the consequences of the constraints
\p{antic12} - \p{HarmHarm}.

\subsection{Harmonic equations}

We start by showing that  $\tilde{V}^{++}$ does not actually depend
on the coordinates $\theta^{\hat{+}}_a$ and $u^{\hat{\pm}}_A$, if fixing the gauge as in \p{WZhat}.
This follows from
 the constraint \p{HarmHarm}, which amounts to the mixed ``harmonic flatness'' condition
\be
D^{++} V^{\hat{+}\hat{+}} - D^{\hat{+}\hat{+}} \tilde{V}^{++} +
[\tilde{V}^{++},  V^{\hat{+}\hat{+}}] = 0\,.\lb{VVhat}
\ee
Substituting the WZ expression \p{WZhat} for $V^{\hat{+}\hat{+}}$ 
and equating to zero the coefficients in the $\theta^{\hat{-}}_a$ expansion of
the  l.h.s. of \p{VVhat}, we find the set of equations
\be
&& \partial^{\hat{+}\hat{+}}\tilde{V}^{++} =0\,, \quad \partial^{\hat{+}\hat{+}}v^{++\hat{-}a} = 0\,, \lb{zero1}\\
&& \partial^{\hat{+}\hat{+}}v^{++\hat{-}\hat{-}ab} - i(\nabla^{++}\hat{A}^{ab} - \partial^{ab}
\tilde{V}^{++}) = 0\,, \lb{sec} \\
&& \partial^{\hat{+}\hat{+}}v^{++\hat{-}3}_d - \nabla^{++} \varphi^A_d u^{\hat{-}}_A = 0\,, \lb{third} \\
&& \partial^{\hat{+}\hat{+}}v^{++\hat{-}4} - \nabla^{++} {\cal D}^{AB}u^{\hat{-}}_A u^{\hat{-}}_B= 0\,. \lb{fourth}
\ee
Eqs. \p{zero1} imply the independence of $V^{++}$ of the harmonics $u^{\hat{\pm}}_A$ and, bearing in mind
the  {\underline {\bf Lemma}} \p{lemma}, also the condition
\be
v^{++\hat{-}a} = 0\,. \lb{01}
\ee
Already at this step we can identify $\tilde{V}^{++}$ with the
familiar from the previous sections harmonic ${\cal N}=(1,0)$ gauge potential, since
the  $\theta^{\hat{\pm}}_a$ - independent part of
the constraint (\ref{harmspin}b) is equivalent to the ${\cal N}=(1,0)$ G-analyticity
condition, (\ref{harmspin}b) $\rightarrow D^+_a \tilde{V}^{++} = 0\,$.

Eq. \p{sec} is equivalent to two separate equations, the one for $v^{++\hat{-}\hat{-}ab}$, which implies
\be
v^{++\hat{-}\hat{-}ab} =0\,,\lb{02}
\ee
and another independent condition arising in the zero order in $u^{\hat{\pm}}$,
\be
\nabla^{++}\hat{A}^{ab} - \partial^{ab} \tilde{V}^{++} = 0\,. \lb{AV++}
\ee

Analogously, the remaining equations \p{third} and \p{fourth} imply
\be
v^{++\hat{-}3}_d  = v^{++\hat{-}4} = 0\,,\lb{034}
\ee
as well as
\be
\nabla^{++}\varphi^A_d = 0\,, \quad \nabla^{++}{\cal D}^{AB} = 0\,.\lb{++AD}
\ee

Thus, we derived that
\be
\tilde{V}^{++} \equiv  V^{++}, \qquad \tilde{\nabla}^{++} \equiv \nabla^{++}\,.
\ee
We have also obtained
 the harmonic constraints \p{AV++} and \p{++AD}.
Note that \p{AV++} is equivalent to the vanishing of the commutator
\be
[\nabla^{++}, \hat{\nabla}^{ab}] = 0\,, \quad \hat{\nabla}^{ab} :=\partial^{ab} + \hat{A}^{ab}\,. \lb{AV++1}
\ee
The constraint \p{HarmHarm} has thereby  been fully used and solved.

Our next task is to further fix the spinor connection \p{A+}. It involves the superfield  $\phi^{+\hat{+}}$.
 Consider it in more details. Besides the G-analyticity conditions (\ref{GanalVhat}), it
satisfies the harmonic constraints
\be
({\rm a})\; \nabla^{\hat{+}\hat{+}}\phi^{+\hat{+}} = 0; \qquad ({\rm b})\; \nabla^{++}\phi^{+\hat{+}} = 0\,, \lb{Harmanal1}
\ee
which also come out as the Bianchi identities [they are derived by commuting both sides of (\ref{antic12}c) with
$\nabla^{\hat{+}\hat{+}}$ and $\tilde{\nabla}^{++} = \nabla^{++}$
and taking into account the constraints \p{harmspin}]. Eq. (\ref{Harmanal1}a)
 amounts to the following set of equations for the ${\cal N}=(1,0)$
components in the expansion \p{expphi}:
\be
&& \partial^{ \hat{+} \hat{+}}q^{+\hat{+}} = 0 \;\Rightarrow \; q^{+\hat{+}} = q^{+ A}u^{\hat{+}}_A\,,\lb{qA} \\
&& \partial^{ \hat{+} \hat{+}}\beta^{+\hat{-}ab} +  i\hat{\nabla}^{ab}q^{+\hat{+}} = 0 \; \Rightarrow \; \beta^{+\hat{-}ab} =
-i\hat{\nabla}^{ab}q^{+\hat{-}}\,, \quad q^{+\hat{-}} := q^{+ A}u^{\hat{-}}_A\,, \lb{beta} \\
&& \partial^{ \hat{+} \hat{+}} G^{+\hat{-}2}_d + [\varphi^{\hat{-}}_d, q^{+\hat{+}}] +
\frac{i}{6}\varepsilon_{dabc}\hat{\nabla}^{ab}W^{+c} = 0\,,\lb{Geq} \\
&& \partial^{ \hat{+} \hat{+}} G^{+\hat{-}3} + \frac{1}{24}\varepsilon_{abcd}\hat{\nabla}^{ab}\hat{\nabla}^{cd}q^{+\hat{-}} +
[{\cal D}^{\hat{-}2}, q^{+\hat{+}}] + \frac14 \{\varphi^{\hat{-}}_a,  W^{+ a}\} = 0\,, \lb{G2eq}
\ee
where $\hat{\nabla}^{ab} = \partial^{ab} + \hat{\cal A}^{ab}$. Eqs. \p{Geq} and \p{G2eq} amount to the
equations for defining
the superfields $G^{+\hat{-}2}_d, G^{+\hat{-}3}$ and to the additional self-consistency conditions which appear
in the  zero order in harmonics $u^{\hat{\pm}}_A\,,$
\be
&& \varepsilon_{dabc}\hat{\nabla}^{ab}W^{+ c} - 3i [\varphi_{dA},q^{+A}] =0\,, \lb{CondW1} \\
&& \varepsilon_{abcd}\hat{\nabla}^{ab}\hat{\nabla}^{cd}q^{+A} + 6\{\varphi_{a}^A, W^{+ a}\} -
 16 [{\cal D}^{AB}, q^+_B] = 0\,.
\ee
These self-consistency conditions can be shown to be satisfied on the final solution of the constraints.
The harmonic equations for $G^{+\hat{-}2}_d$ and $G^{+\hat{-}3}$ uniquely fix these superfields as
\be
&&G^{+\hat{-}2}_d = G^{+ (AB)}_d u^{\hat{-}}_Au^{\hat{-}}_B\,, \qquad G^{+ (AB)}_d = -\frac12
 [\varphi_{d}^{(A},q^{+B)}]\,, \nn
&& G^{+\hat{-}3} = G^{+ (ABC)}u^{\hat{-}}_A u^{\hat{-}}_B u^{\hat{-}}_C\,, \qquad  G^{+ (ABC)} =
 -\frac13 [{\cal D}^{(AB}, q^{+C)}]\,.
\ee
When deducing these solutions, we made use of the reduction relations
$$
u^{\hat{+}}_A u^{\hat{-}}_B = u^{\hat{+}}_{(A} u^{\hat{+}}_{B)} + \frac12 \varepsilon_{AB}\,, \quad
u^{\hat{-}}_A u^{\hat{-}}_B u^{\hat{+}}_C = u^{\hat{-}}_{(A} u^{\hat{-}}_B u^{\hat{+}}_{C)} +
\frac13 (\varepsilon_{CA}u^{\hat{-}}_B +
\varepsilon_{CB}u^{\hat{-}}_A)\,.
$$

Eq. (\ref{Harmanal1}a) also implies
\be
\partial^{ \hat{+} \hat{+}} W^{+ a} = 0\,,
\ee
which means independence of $W^{+ a}$ of the hatted harmonics.

Thus, we have fully fixed the $u^{\hat{+}}_A, u^{\hat{-}}_B$ dependence in the $\theta^{\hat{+}}$ expansion \p{expphi} of
$\phi^{+\hat{+}}$.
At this stage,
it is instructive to write $\phi^{+\hat{+}}$ in the form which takes into account the explicit solutions given above,
\be
\phi^{+\hat{+}} &=& q^{+ A}u^{\hat{+}}_A - \theta^{\hat{+}}_a W^{+ a} -
i \theta^{\hat{+}}_a\theta^{\hat{+}}_b\hat{\nabla}^{ab}q^{+ A}u^{\hat{-}}_A
-\frac12 \Psi^{\hat{+}3 d}[\varphi^A_d,q^{+ B}]u^{\hat{-}}_A u^{\hat{-}}_B \nn
&&-\, \frac13 \Psi^{\hat{+}4} [{\cal D}^{AB}, q^{+ C}]u^{\hat{-}}_A u^{\hat{-}}_Bu^{\hat{-}}_C\,.\lb{explphi}
\ee

Now we are ready to explore the conditions imposed by the second harmonic constraint (\ref{Harmanal1}b). It implies
\be
\nabla^{++}q^{+\hat{+}} = 0\,, \; \nabla^{++}W^{+ a} = 0\,, \;  \nabla^{++}\hat{\nabla}^{ab}q^{+\hat{-}} = 0\,, \;  \nabla^{++}G^{+(AB)}_d =
 \nabla^{++} G^{+(ABC)} = 0\,.\lb{standHarmanal}
\ee
The first of these equations is recognized as the equation of motion for the hypermultiplet, so already at this step we can identify
$q^{+ A}$ with the  ${\cal N}=(1,0)$ hypermultiplet
superfield of the previous sections. Its analyticity follows from the
G-analyticity condition (\ref{GanalVhat}b) (see below).  The second constraint coincides with \p{HarmW}. The third harmonic equation in \p{standHarmanal}
is satisfied as a consequence
of the first one and \p{AV++1}. The last two equations are satisfied as a consequence of the first equation
and the constraints \p{AV++} and \p{++AD}.

Now we can come back to the problem of the ultimate fixing of the spinor connection ${\cal A}^+_a\,$.
This fixing is accomplished by the constraint (\ref{harmspin}a). Like in the case of $\tilde{V}^{++}$ and the constraint \p{HarmHarm}, eq. (\ref{harmspin}a)
eliminates all the negatively charged components in the expansion \p{tildeA+} [with the condition \p{f+a0}],
except for the first term  $f^{+\hat{-}b}_a$,
\be
f^{+ \hat{-}\hat{-}bc}_a = f_d^{+\hat{-}3} = f^{+ \hat{-}4}_a = 0\,, \lb{zeroA+}
\ee
whereas $f^{+\hat{-}b}_a$ is fixed as
\be
f^{+\hat{-}b}_a = -\delta^b_a q^{+A}u^{\hat{-}}_A. \lb{1stExpr}
\ee
Simultaneously we obtain a few differential conditions relating the ${\cal N}=(1,0)$ components of $\phi^{+\hat{+}}$ to those of
$V^{\hat{+}\hat{+}}$ defined in \p{WZhat}. These are as follows:
\be
&& D^+_a \hat{\cal A}^{bc} = \frac{i}{2}\Big(\delta^b_a W^{+c}- \delta^c_a W^{+b} \Big)\,, \lb{hathat+1} \\
&& D^+_a \varphi^{\hat{-}}_d = \frac{i}{3}\varepsilon_{adbc}\hat{\nabla}^{bc}q^{+\hat{-}}\,. \lb{varphieq}\\
&& D^{+}_a{\cal D}^{AB} -\frac{3}{8} [\varphi^{(A}_a, q^{+B)}] = 0\,. \lb{Deq1}
\ee
Like in the previous cases, these extra equations are self-consistency conditions which are identically satisfied for the general solution
of all constraints. As we will see, eq. \p{hathat+1} plays the especially important role, giving rise to the expression of $W^{+ a}$ in terms
of the ${\cal N}=(1,0)$ analytic potential $V^{++}$.

 The final form for the spinor connection ${\cal A}^+_a\,$ that takes into account
 the solutions \p{zeroA+}, \p{1stExpr} is
\be
{\cal A}^+_a = -\theta^{\hat{+}}_a q^{+ A}u^{\hat{-}}_A + \theta^{\hat{-}}_a\phi^{+\hat{+}}\,.\lb{finA}
\ee

It remains to work out the conditions following from the G-analyticity constraint  (\ref{GanalVhat}b).
Using the explicit expressions \p{finA}, \p{explphi}, we find that (\ref{GanalVhat}b) amounts to the following set of equations:
\be
&& D^+_a q^{+ \hat{+}} =0\,, \quad D^{+}_a W^{+ b} = \delta^b_a[ q^{+ \hat{-}}, q^{+ \hat{+}}] =
-\frac12 \delta^b_a[ q^{+ A}, q^{+}_A]\,, \lb{eqsmot} \\
&& D^{+}_a \hat{\nabla}^{cd} q^{+\hat{-}} - \frac{i}{2}\Big(\delta^c_a[q^{+\hat{-}}, W^{+d}] -
\delta^d_a[q^{+\hat{-}}, W^{+c}]\Big)= 0\,, \lb{conseq} \\
&& [D^+_a\varphi^{-(A}_d, q^{+B)}] + \frac{i}{3}\varepsilon_{adcf}[q^{+(A}, \hat{\nabla}^{cf}q^{+B)}] = 0\,, \lb{conseq2} \\
&& [D^+_a{\cal D}^{(AB}, q^{+C)}] + \frac{3}{8}[q^{+(A}, [\varphi^B_a, q^{+C)}]] = 0\,.
\ee
The first equation in \p{eqsmot}
provides the standard analyticity condition for the hypermultiplet $q^{+ A}$, while the second equation is going
to become the
equation of motion for  the ${\cal N}=(1, 0)$ analytic potential $V^{++}$. The remaining equations prove to be satisfied as a consequence
of the basic equations of motion.

At last, it is straightforward to check that the constraint (\ref{antic12}a) does not result in any new restrictions and
is identically satisfied as a consequence of G-analyticity of $q^{+\hat{-}}$ and the condition (\ref{GanalVhat}b).

Let us discuss the peculiarities of the realization of the hidden supersymmetry in the considered frame.
As usual, to preserve the Wess-Zumino gauge \p{WZhat},  one needs to make a compensating gauge transformation.
 The appropriate
gauge parameter is easily found to be
\be
\Lambda_{(1)}^{(comp)} = 2i \epsilon^{\hat{-}}_a\theta^{\hat{+}}_b{\cal A}^{ab} +
\frac{3}{2}\varepsilon^{abcd}\epsilon^{\hat{-}}_a \theta^{\hat{+}}_b\theta^{\hat{+}}_c \varphi^{\hat{-}}_d +
\frac{4}{3}\varepsilon^{abcd}\epsilon^{\hat{-}}_a\theta^{\hat{+}}_b\theta^{\hat{+}}_c\theta^{\hat{+}}_d{\cal D}^{\hat{-}2}\,.\lb{WZlambda1}
\ee
Besides this, one needs to preserve the ``short'' form of the spinor connection \p{finA}. The appropriate
 compensating gauge parameter is
\be
\Lambda_{(2)}^{(comp)} = \epsilon^{-A}q^+_A\,, \lb{WZlambda2}
\ee
such that the total compensating gauge parameter is
\be
\Lambda^{(comp)} = \Lambda_{(1)}^{(comp)} + \Lambda_{(2)}^{(comp)}\,.
\ee
Correspondingly, the hidden supersymmetry transformations of $V^{\hat{+}\hat{+}}$ and ${\cal A}^+_a$ are
\be
&&\delta V^{\hat{+}\hat{+}} = -\epsilon^{\hat{+}}_a \frac{ \partial\, }
{\partial \theta_a^{\hat{+}}}V^{\hat{+}\hat{+}}  - 2 i \epsilon_a^{\hat{-}}
 \theta_b^{\hat{+}} \partial^{ab} V^{\hat{+}\hat{+}} + \nabla^{\hat{+}\hat{+}}\Lambda^{(comp)}, \lb{modSUSY1} \\
&& \delta {\cal A}^+_a = -\epsilon^{\hat{+}}_b \frac{ \partial\, }
{\partial \theta_b^{\hat{+}}}{\cal A}^+_a - \epsilon^{\hat{-}}_b \frac{ \partial\, }
{\partial \theta_b^{\hat{-}}}{\cal A}^+_a - 2 i \epsilon_c^{\hat{-}}
 \theta_b^{\hat{+}} \partial^{cb} {\cal A}^+_a + \nabla^{+}_a\Lambda^{(comp)}\,. \lb{modSUSY2}
\ee
Note that $\Lambda_{(2)}^{(comp)}$ does not contribute to \p{modSUSY1}.

Since all superfields should undergo the same compensating gauge transformation under the hidden supersymmetry,
one can wonder what happens in the case of $V^{++}$. Its transformation law looks as
\be
\delta V^{++} = -2 i \epsilon_a^{\hat{-}}
 \theta_b^{\hat{+}} \partial^{ab} V^{++} + \nabla^{+ +}\Lambda^{(comp)} \lb{SUSYV++}
 \ee
 and seemingly contradicts the fact that $V^{++}$ should not depend on the hatted coordinates.  However, let us
 look at $\nabla^{+ +}\Lambda^{(comp)}$. Using the constraints \p{++AD} and \p{AV++}, we find
 $$
 \nabla^{+ +}\Lambda^{(comp)} = 2i \epsilon^{\hat{-}}_a\theta^{\hat{+}}_b \partial^{ab}V^{++} +
\nabla^{++}(\epsilon^{-A}q^+_A)\,.
 $$
The first term cancels the unwanted term in \p{SUSYV++}, while the second term, with taking into account the on-shell
condition $\nabla^{++}q^{+ A} = 0$, yields the already known transformation law of $V^{++}$ under the hidden supersymmetry,
\be
\delta V^{++} = \epsilon^{+A}q^+_A\,.
\ee

In a similar way, by considering the transformation of the superfield $\phi^{+\hat{+}}$ under the hatted supersymmetry,
one can derive the
hidden supersymmetry transformations of its ${\cal N}=(1,0)$ superfield components $q^{+ A}$ and $W^{+ a}$.

At this stage, we succeeded to express all the involved geometric
quantities of the ${\cal N}=(1,1)$ gauge theory in terms of the ${\cal N}=(1,0)$ superfields appearing in the
$\theta^{\hat{+}}$ expansion of $V^{\hat{+}\hat{+}}$ in the WZ gauge
\p{WZhat}: the hypermultiplet $q^{+A}$ and
the ${\cal N}=(1,0)$ superfield $W^{+a}$,  which is going to become the covariant
${\cal N}=(1,0)$ superfield strength considered in the previous
sections. It remains to relate the superfields in  \p{WZhat} to the
known ${\cal N}=(1,0)$ superfields in a pure algebraic way, without
solving various differential conditions deduced above. This can be
achieved by requiring for the  vector superfield connections derived in
the hatted and unhatted sectors to coincide (our superspace involves
hatted and unhatted odd coordinates, but only one set of bosonic coordinates
$x^M$).

\subsection{Identifying vector connections}
Let us now proceed to the vector connections.

 We consider first the unhatted sector. Since $\nabla_a^+$ includes $\theta^{\hat{\pm}}_a$, its counterpart $\nabla_a^-$
should also include now such a dependence\footnote{It is thus not the same as
$\nabla_a^-$ in  \p{AAgauge}. We have chosen, however, not to invent other notation and hope
that this  will not lead to confusion.},
 and the same concerns the full ${\cal N}=(1,1)$ superfield vector connection. We define
$\nabla_a^-$ in the standard way:
\be
\nabla_a^- := D^-_a + {\cal A}^-_a = [\nabla^{--}, \nabla_a^+]\,, \quad {\cal A}^-_a = {\cal A}^{-(0)}_a  - \theta^{\hat{+}}_a q^{-\hat{-}} +
\theta^{\hat{-}}_a\nabla^{--}\phi^{+\hat{+}}\,,\lb{nablaFull-}
\ee
where
\be
\nabla^{--} = D^{--} + V^{--}\,,
\ee
$V^{--}$ is the same as in the previous sections [it is constructed from $V^{++}$ by the harmonic zero curvature equation \p{nabcomm}] and
${\cal A}^{-(0)}_a = -D^+_a V^{--}$. The relevant
full superfield vector connection is defined in the standard way:
\be
&& \{\nabla_a^+, \nabla_b^- \} = 2i( \partial_{ab} + {\cal V}_{ab})\,, \quad {\cal V}_{ab} =
\frac{1}{2i} (\nabla_a^+ {\cal A}^-_b + D_b^-{\cal A}^+_b)\,,   \lb{Vconndef} \\
&& {\cal V}_{ab} = {\cal A}_{ab} + \frac{1}{2i}\Big(\theta^{\hat{+}}_b D^+_a q^{-\hat{-}} +
\theta^{\hat{+}}_a \nabla^-_b q^{+\hat{-}} - \theta^{\hat{-}}_b D^+_a\nabla^{--} \phi^{+\hat{+}}
- \theta^{\hat{-}}_a \nabla^-_b\phi^{+\hat{+}} \nn
&&{\quad\quad\quad} +\, \theta^{\hat{+}}_a \theta^{\hat{+}}_b [q^{+\hat{-}}, q^{-\hat{-}}] - \theta^{\hat{+}}_a \theta^{\hat{-}}_b
[q^{+\hat{-}}, \nabla^{--}\phi^{+\hat{+}}] + \theta^{\hat{+}}_b \theta^{\hat{-}}_a[\phi^{+\hat{+}}, q^{-\hat{-}}] \nn
&& {\quad\quad\quad}+ \,\theta^{\hat{-}}_a \theta^{\hat{-}}_b [\phi^{+\hat{+}}, \nabla^{--}\phi^{+\hat{+}}]\Big), \quad  {\cal A}_{ab} =
\frac{1}{2i}\,D_a^+ {\cal A}^{-(0)}_b\, .\lb{Vab}
\ee
It has the restricted $\theta^{\hat{-}}_a$ dependence (only the terms of the first and second order in $\theta^{\hat{-}}_a$),
but includes all  $\theta^{\hat{+}}_a$ monomials.

On the other hand, one can perform an analogous construction for the derivatives with hatted indices. We define the relevant second harmonic
connection $V^{\hat{-}\hat{-}}$  from the hatted flatness relation
\be
D^{\hat{+}\hat{+}}V^{\hat{-}\hat{-}} -  D^{\hat{-}\hat{-}}V^{\hat{+}\hat{+}} + [V^{\hat{+}\hat{+}},  V^{\hat{-}\hat{-}} ] =0 \lb{hatFlat}
\ee
and then introduce the hatted spinor and vector connections  as
\be
&& [\nabla^{\hat{-}\hat{-}}, D^{\hat{+} a}] := \nabla^{\hat{-}a} = D^{\hat{-}a} + {\cal A}^{\hat{-}a}\,, \quad
{\cal A}^{\hat{-}a} = -\frac{\partial}{\partial \theta^{\hat{-}}_a} V^{\hat{-}\hat{-}}\,, \lb{hatAspin} \\
&& \{D^{\hat{+} a}, \nabla^{\hat{-}b} \} = 2i(\partial^{ab} + \hat{\cal V}^{ab})\,, \quad \hat{\cal V}^{ab} =
\frac{i}{2}\frac{\partial}{\partial \theta^{\hat{-}}_a} \frac{\partial}{\partial \theta^{\hat{-}}_b}V^{\hat{-}\hat{-}}\,, \lb{hatVvect}
\ee
where $\nabla^{\hat{-}\hat{-}} = D^{\hat{-}\hat{-}} + V^{\hat{-}\hat{-}}$.

The calculation of $V^{\hat{-}\hat{-}}$ is the most boring part of the whole story. We parametrize  the $\theta^{\hat{-}}_a$ expansion of
$V^{\hat{-}\hat{-}}$ as
\be
V^{\hat{-}\hat{-}} = i\theta^{\hat{-}}_a \theta^{\hat{-}}_b v^{ab} + \Psi^{\hat{-}3d}v^{\hat{+}}_d + \Psi^{\hat{-}4} v^{\hat{+}2}\,.\lb{hatV--}
\ee
All coefficients here are hat-analytic ${\cal N}=(1,0)$ superfields, the $u^{\hat{\pm}}_A$ and $\theta^{\hat{+}}_a$ dependence of which
will be strictly fixed by the corresponding hat-harmonic equations following from \p{hatFlat}. The possible terms of the zeroth and first orders
in $\theta^{\hat{-}}_a$ can be shown to vanish by the same mechanism as in the previous examples: their $\theta^{\hat{+}}$ expansions contain only
components with negative ``hat'' charges and these components  are killed by the equations like
$\partial^{\hat{+}\hat{+}} \omega^{\hat{-}n} = 0\, \rightarrow\, \omega^{\hat{-}n} = 0$, following from \p{hatFlat}.

The $\theta^{\hat{-}}_a$ expansion of the l.h.s. of the constraint \p{hatFlat} contains the Grassmann monomials of the first, second, third
and fourth degrees. Equating the corresponding coefficients to zero, we obtain the following set of equations:
\be
&& 2i\theta^{\hat{+}}_b(\hat{\cal A}^{ba} - v^{ba}) -3 \varepsilon^{abcd}\theta^{\hat{+}}_b\theta^{\hat{+}}_c\varphi^{\hat{-}}_d
- 4\varepsilon^{abcd}\theta^{\hat{+}}_b\theta^{\hat{+}}_c\theta^{\hat{+}}_d{\cal D}^{\hat{-}2} = 0\,, \lb{1storder} \\
&& \nabla^{\hat{+}\hat{+}} v^{ab} -3i\varepsilon^{abcd} \theta^{\hat{+}}_c v^{\hat{+}}_d
-i \theta^{\hat{+}}_c\theta^{\hat{+}}_d\partial^{ab}\hat{\cal A}^{cd} - \Psi^{\hat{+}3 d}\partial^{ab}\varphi^{\hat{-}}_d
- \Psi^{\hat{+}4}\partial^{ab}{\cal D}^{\hat{-}2} = 0\,, \lb{2ndorder} \\
&& \nabla^{\hat{+}\hat{+}}v^{\hat{+}}_d + 4 \theta^{\hat{+}}_d v^{\hat{+}2} = 0\,, \lb{3dorder} \\
&& \nabla^{\hat{+}\hat{+}}v^{\hat{+}2} = 0\,. \lb{4thorder}
\ee
To solve eqs. \p{1storder} - \p{4thorder}, one expands the corresponding unknowns over $\theta^{\hat{+}}_a$ and
then fix the $u^{\hat{\pm}}_A$ dependence of the coefficients by these equations. For instance, we write
\be
v^{\hat{+}2} = v_{(0)}^{\hat{+}2} + \theta^{\hat{+}}_a v^{\hat{+}a} + \theta^{\hat{+}}_a\theta^{\hat{+}}_b w^{ab} +
\Psi^{\hat{+}3 d}v_{d}^{\hat{-}} + \Psi^{\hat{+}4}v^{\hat{-}2}\,, \lb{decom1}
\ee
and obtain from \p{4thorder} the following equations and their solutions
\be
&& \partial^{\hat{+}\hat{+}}v_{(0)}^{\hat{+}2} = 0 \quad \Rightarrow \quad v_{(0)}^{\hat{+}2} = v^{(AB)}u^{\hat{+}}_A u^{\hat{+}}_B\,, \lb{0reduce}\\
&& \partial^{\hat{+}\hat{+}}v^{\hat{+}a} = 0 \quad \Rightarrow \quad v^{\hat{+}a} = v^{aA}u^{\hat{+}}_A\,, \lb{1streduce} \\
&&  \partial^{\hat{+}\hat{+}}w^{ab} +i \hat{\nabla}^{ab}v_{(0)}^{\hat{+}2}= 0 \quad \Rightarrow \quad w^{ab} = w^{ab}_0 -i\hat{\nabla}^{ab}
v^{(AB)}u^{\hat{+}}_A u^{\hat{-}}_B \,, \lb{2ndreduce} \\
&& \partial^{\hat{+}\hat{+}}v^{\hat{-}}_d - \frac{i}{6}\varepsilon_{dabc}\hat{\nabla}^{ab}v^{\hat{+}c} + [\varphi^{\hat{-}}_d, v^{\hat{+}2}_{(0)}] = 0\,,
\lb{3ndreduce} \\
&& \partial^{\hat{+}\hat{+}}v^{\hat{-}2} + \frac{i}{24}\varepsilon_{abcd}\hat{\nabla}^{ab}w^{cd} + [{\cal D}^{\hat{-}2}, v_{(0)}^{\hat{+}2}]
-\frac14 \{\varphi^{\hat{-}}_a, v^{\hat{+}a}\} = 0\,. \lb{4threduce}
\ee
Eq. \p{3ndreduce} has the following solution,
\be
v^{\hat{-}}_d = \frac{i}{6}\varepsilon_{dabc}\hat{\nabla}^{ab}v^{cA}u^{\hat{-}}_A -\frac23 [\varphi_{dA}, v^{(AB)}]u^{\hat{-}}_B
-\frac12 [\varphi_{d}^{(A}, v^{BC)}]u^{\hat{-}}_Au^{\hat{-}}_Bu^{\hat{+}}_C\,. \lb{2ndsol}
\ee
Eq. \p{4threduce} yields both the solution for $v^{\hat{-}2}$,
\be
v^{\hat{-}2} &=& \frac12\Big([v^{B(A}, D_B^{D)}]  - \frac{1}{24}\varepsilon_{abcd}\hat{\nabla}^{ab}\hat{\nabla}^{cd} v^{AD}
+ \frac18 \{\varphi^{(A}_d, v^{d D)} \}\Big) u^{\hat{-}}_Au^{\hat{-}}_D \nn
&&-\,\frac13[D^{(AB}, v^{CD)}] u^{\hat{-}}_Au^{\hat{-}}_Bu^{\hat{-}}_Cu^{\hat{+}}_D\; ,  \lb{v-2sol}
\ee
and the additional self-consistency condition
\be
[D^{AB}, v_{AB}] + \frac{i}{8}\varepsilon_{abcd}\hat{\nabla}^{ab} w_0^{cd} - \frac{3}{8}\{\varphi_{dB}, v^{dB}\} = 0\,.\lb{v-2scons}
\ee

The remaining equations can be solved analogously. Instead of writing the analogs of the equations \p{0reduce} - \p{4threduce},
we will present their solutions, omitting various self-consistency constraints which are identically
satisfied on the final full solution.

We start with the equation \p{3dorder}. We have
\be
v^{\hat{+}}_d = v^{\hat{+}}_{(0)d}  + \theta^{\hat{+}}_b v^b_d + \theta^{\hat{+}}_a\theta^{\hat{+}}_b v^{\hat{-}[ab]}_d +
\Psi^{\hat{+}3 d}v^{\hat{-}2}_{db} + \Psi^{\hat{+}4}v^{\hat{-}3}_{d}\,.
\ee
The solution is
\be
&& v^{\hat{+}}_{(0)d} = v^{A}_{d}u^{\hat{+}}_A\,, \quad  v^b_d  = v^b_{0d} - 4 \delta^b_d v^{(AB)}u^{\hat{+}}_Au^{\hat{-}}_B\,, \lb{solut1} \\
&& v^{\hat{-}[ab]}_d = -\Big[i\hat{\nabla}^{ab}v^A_d + 2(\delta^a_d v^{bA} - \delta^b_d v^{aA})\Big]u^{\hat{-}}_A\,, \lb{solut2} \\
&&  v^{\hat{-}2}_{db} = v^{(AB)}_{db}u^{\hat{-}}_Au^{\hat{-}}_B\,, \quad v^{(AB)}_{db} = -\frac12 \{\varphi^{(A}_b, v_d^{B)}\} +
i\frac23 \varepsilon_{dbcf}\hat{\nabla}^{cf} v^{(AB)}\,, \lb{solut3} \\
&& v^{\hat{-}3}_{d} = v^{(ABC)}_du^{\hat{-}}_A u^{\hat{-}}_B u^{\hat{-}}_C\,, \quad v^{(ABC)}_d = \frac12 [v^{(AB}, \varphi^{C)}_d]
+ \frac13 [ v^{(A}_d, {\cal D}^{BC)}]\,.\lb{solut4}
\ee
One more important relation following from \p{3dorder} is
\be
\varepsilon_{abcd}w_0^{cd} + \frac{i}{4}\varepsilon_{fbcd}\hat{\nabla}^{cd}v_{0a}^f + \frac{1}{2}\{\varphi_{bD}, v^D_a\} = 0\,. \lb{Eqw0}
\ee

We now turn to  \p{2ndorder}. Once again, we expand
\be
v^{[ab]}= v^{[ab]}_{(0)}  + \theta^{\hat{+}}_d v^{\hat{-} d[ab]} + \theta^{\hat{+}}_c\theta^{\hat{+}}_d v^{\hat{-}2[cd][ab]} +
\Psi^{\hat{+}3 d}v^{\hat{-}3[ab]}_{d} + \Psi^{\hat{+}4}v^{\hat{-}4[ab]}\,.
\ee
The solution is
\be
&& v^{[ab]}_{(0)} = v^{[ab]}_{0}\,, \quad v^{\hat{-}3[ab]}_{d} = v^{\hat{-}4[ab]} = 0\,,\lb{solut1a} \\
&& v^{\hat{-} d[ab]} = 3i \varepsilon^{dabc}v^A_c u^{\hat{-}}_A\,, \quad v^{\hat{-}2[cd][ab]} =
-6i\varepsilon^{cdab}v^{AB}u^{\hat{-}}_Au^{\hat{-}}_B\,, \lb{solut2a}\\
&& \varepsilon^{ab[cf}v^{d]}_{0g} = \frac13 \Big(\hat{\nabla}^{cd}v_0^{[ab]} - \partial^{ab}\hat{\cal A}^{cd}\Big). \lb{solut4a}
\ee
An important consequence of   \p{2ndorder} is also the relation
\be
\partial^{ab}\varphi^{A}_d + [v_0^{[ab]}, \varphi^{A}_d] + \frac12\varepsilon^{abcf} \varepsilon_{cdug}\Big[\hat{\nabla}^{ug}v^A_f
-2i(\delta^u_f v_0^{gA} - \delta^g_f v_0^{uA})\Big] = 0\,.  \lb{solut5a}
\ee

The most crucial is eq. \p{1storder}. It gives
\be
v_0^{[ab]} = \hat{\cal A}^{[ab]}\,, \quad v^A_b = -\frac12 \varphi^A_b\,, \quad v^{(AB)} = \frac13 {\cal D}^{(AB)}\,. \lb{crucial}
\ee
Now, substituting all this into  \p{solut5a}, we can determine $v_0^{bA}$,
\be
v_0^{bA} = \frac{i}{4}\hat{\nabla}^{ba}\varphi^A_a\,.
\ee
Using \p{solut4a} and \p{Eqw0}, we can also express $w_0^{[ab]}$ and $\,v^d_{0b}$ through the basic superfields $\hat{\cal A}^{[ab]}$,  $\varphi^A_b$
and ${\cal D}^{(AB)}$. Thus, we obtain the full solution for $V^{\hat{-}\hat{-}}$.

Now we are ready to explicitly construct the full superfield vector connection $\hat{\cal V}^{ab}$. Using the definition \p{hatVvect}, we obtain
\be
\hat{\cal V}^{ab} = v^{[ab]} + 3i \epsilon^{abcd}\theta^{\hat{-}}_c v^{\hat{+}}_d +
6i \epsilon^{abcd}\theta^{\hat{-}}_c\theta^{\hat{-}}_d v^{\hat{+}2}\,.\lb{hatVvectexpl}
\ee
The crucial requirement now is that this connection is related to the connection ${\cal V}^{ab}$ in
the sector of ``unhatted'' spinor derivatives as
\be
{\cal V}_{ab} = \frac12\epsilon_{abcd}\hat{\cal V}^{cd}\,. \lb{constrVect}
\ee
Comparing two expressions in the zeroth order in $\theta^{\hat{-}}_c$, we immediately find
\be
\hat{\cal A}^{ab} = \frac12\epsilon^{abcd}{\cal A}_{cd}\,, \quad \varphi^{\hat{-}}_a = -\frac13 D^+_a q^{-\hat{-}}\,, \quad
{\cal D}^{AB} = \frac18 [q^{+(A}, q^{-B)}]\,, \lb{identif}
\ee
which, being substituted into the basic superfields $V^{\hat{+}\hat{+}}$ and $\phi^{+\hat{+}}$ \p{WZhat}, \p{explphi},  precisely
reproduce the  solution \p{V++_expan} and \p{phi++} that we have presented in the previous Section. Comparing the
coefficients of the next terms of expansion in $\theta^{\hat{-}}_c$
in the equation \p{constrVect} gives relations that are identically satisfied, when taking into account
the ${\cal N}=(1,0)$  equations of motion and the G-analyticity conditions, and so does not produce new constraints.

Note that after the identification \p{identif}, the relation
\p{hathat+1} becomes equivalent to \p{Da,nabcb}.  It gives  the expression \p{W+Def} of $W^{+ a}$ in terms of $V^{--}\,$.

\section{On-shell ${\cal N}=(1,1)$ supersymmetric actions}
\setcounter{equation}0

 \subsection{Invariant actions: $d=8$}

We can now use the techniques developed in  Sect. 6 and 7
to write down the actions invariant under the extended ${\cal N}=(1,1)$ on-shell supersymmetry.
The original $d=4$ action \p{Summ} is off-shell invariant.
For $d=6$, the invariants non-vanishing on shell are absent. Dimension 8
is the first nontrivial case.

Consider the density
\be
\lb{L++++}
{\cal L}^{+4}_{(1,1)}   = -{\rm Tr}\int d\hat{\zeta}^{(-4)}  \,\frac{1}{4}
(\phi^{+\hat{+}} )^4\ ,
\ee
where the hatted analytic superspace measure $d\hat{\zeta}^{(-4)}$
is defined as $d\hat{\zeta}^{(-4)} = d\hat{u}(\hat{D}^-)^4$ (in contrast to
 $d\zeta^{(-4)}$, it does not involves $d^6 x$).
Bearing in mind the property (\ref{GanalVhat}b),
  \be
\nabla^{+}_{a}\phi^{+\hat{+}} = D^{+}_{ a}\phi^{+\hat{+}} + [{\cal A}^{+}_{ a}, \phi^{+\hat{+}}] = 0\,,
 \ee
where ${\cal A}^+_a$ is given in \p{finA}, and the fact that the commutator term does not contribute under the trace,
we derive that the Lagrangian \p{L++++} is ${\cal N}=(1,0)$ analytic:
\be
D^+_a{\cal L}^{+4}_{(1,1)} = 0\,.
\ee
This analyticity holds only on shell since the constraint (\ref{GanalVhat}b)
necessarily implies the second equation of motion in \p{eqsmot}.

It is easy to see that the integral \p{L++++} is shifted by a total derivative under the ${\cal N}=(0,1)$ transformations \p{variation2}.
 Indeed, the integrand  transforms as
\begin{eqnarray}
\delta \,{\rm Tr}\,\Big[\frac{1}{4} (\phi^{+\hat{+}} )^4\Big]  =
 - \epsilon^{\hat{+}}_a \frac{ \partial\, }{\partial \theta_a^{\hat{+}}} {\rm Tr}\Big[
 \frac{1}{4} (\phi^{+\hat{+}} )^4 \Big] -  2 i \epsilon_a^{\hat{-}} \theta_b^{\hat{+}}
 \partial^{ab} {\rm Tr} \Big[  \frac{1}{4} (\phi^{+\hat{+}} )^4 \Big], 
 \end{eqnarray}
and hence
\be \delta {\cal L}^{+4}_{(1,1)} &=&  - 2 i \partial^{ab} {\rm Tr}  \int d\hat{\zeta}^{(-4)}  \,
\Bigl( \epsilon_a^{\hat{-}} \theta_b^{\hat{+}}    \frac{1}{4} (\phi^{+\hat{+}} )^4
 \Big). \ee
The  action
\be
S_{(1,1)} = \int d\zeta^{(-4)}{\cal L}^{+4}_{(1,1)}
\ee
is clearly invariant.

To express \p{L++++} in terms of ${\cal N}=(1,0)$ superfields, we need to substitute there the  explicit
expression \p{phi++} for $\phi^{+\hat{+}}$ and
to integrate over $d\hat{\zeta}^{(-4)} d\hat{u}$, using \p{harmint} (with the capital $SU(2)$ indices
$A,B, \ldots$ instead of $i, j,\ldots$). Doing this, we reproduce the result \p{L++++fin} quoted above.

Though it is not at all seen in the expression \p{L++++fin}, we expect that the Lagrangian expressed
{\it in components} is invariant under the permutation $\lambda \leftrightarrow \psi$.

The Lagrangian \p{L++++fin} represents a ${\cal N} = (1,1)$ generalization of the ${\cal N} = (1,0)$
supersymmetric  single-trace Lagrangian \p{dim8anal-1} involving the vector supermultiplet.
 It is trivial to generalize in a similar way the
double-trace Lagrangian \p{dim8anal-2}. It is sufficient to consider the density
 \be
\lb{L++++double}
\hat{\cal L}^{+4}_{(1,1)} = -\frac14 \int  d\hat{\zeta}^{(-4)} \,
{\rm Tr}\,(\phi^{+\hat{+}} )^2\, {\rm Tr}\,(\phi^{+\hat{+}} )^2 \ ,
\ee
and perform the integrals over the hatted variables. The result is expressed
 in terms of the ${\cal N}=(1,0)$ superfields as follows
\be
\hat{\cal L}^{+4}_{(1,1)} &=&
\frac14\varepsilon_{abcd} {\rm Tr}\,(W^{+a}W^{+b})\, {\rm Tr}\,\Big( W^{+c}W^{+d} + 2i q^{+ A}\nabla^{cd}q^+_{\,A}\Big)\nn
&& -\,\frac12 {\rm Tr}\,(q^{+ A}\nabla^{ab}q^+_{\,A}) \,{\rm Tr}\,(q^{+ B}\nabla_{ab}q^+_{\,B}) + \frac{1}{12}\partial^{ab} {\rm Tr}\,(q^{+ A}q^{+ B})
\,\partial_{ab} {\rm Tr}\,(q^{+}_{\,A}q^{+}_{\, B}) \nn
&& +\,{\rm Tr}\,(q^{+ A}W^{+ a})\,{\rm Tr}\,\Big\{D^+_a q^-_{\,A}(q^+)^2  - 2i W^{+b}\nabla_{ba} q^+_{\,A}\Big\}\nn
&& +\,\frac13 {\rm Tr}\,(q^{+ A}q^{+ B})\,{\rm Tr}\,\Big\{ (q^+)^2\, [q^+_{\,(A}, q^-_{\,B)}]- \nabla^{ab}q^+_{\, A} \nabla_{ab} q^+_{\,B}
- W^{+ a}[D^+_aq^-_{\,(A}, q^+_{\,B)}]\Big\}. \lb{L++++double1}
\ee

Thus, the nontrivial on-shell $d=8$ invariants exist. Still the perturbative
expansion for the amplitudes in the theory \p{Summ} does not involve divergences at the two-loop level.
The matter is that these invariants do {\it not} possess the  full {\it off-shell} ${\cal N} = (1,0)$ supersymmetry,
 which the physically relevant counterterms should obey. Indeed, we have in our disposal the  off-shell
${\cal N} = (1,0)$ harmonic superfield description, which implies the existence of the  gauge-covariant ${\cal N} = (1,0)$
supergraph techniques, such that all the relevant counterterms enjoy this off-shell symmetry.

\subsubsection{Gauge non-invariant off-shell supersymmetric realization}
Our remark is that one still {\it can} write an
off-shell supersymmetric $d=8$ action, if renouncing the requirement
of gauge invariance. The corresponding density reads:
  \be
 \label{L++++OffShell} \hat{{\cal L}}^{+4} &=& {\rm Tr}_{(S)}\Big\{  \frac{1}{4} \varepsilon_{abcd}
 W^{+a} W^{+b} W^{+c} W^{+d}  + 3i q^{+A} \nabla_{ab} q_A^+ W^{+a} W^{+b}
-  q^{+A} \nabla_{ab} q_A^+ \, q^{+B} \nabla^{ab} q_B^+ \nn \
&& \hspace{15mm} - W^{+a} [ D_a^+ q_A^- , q_B^+] q^{+A} q^{+B}
-\frac{1}{2} [ q^{+C} , q^+_C] [ q_A^-,q^+_B] q^{+A} q^{+B} \nn
&& \quad  + \Bigl( F^{++} + \frac{1}{2} [ q^{+A} , q^+_A]\Bigr) \Big( 2 i {\cal A}_{ab} W^{+a} W^{+b} - 2 F^{++} {\cal A}_{ab}
{\cal A}^{ab} \nn  &&
\hspace{15mm}  - 3 q^{+B} \nabla^{-}_a q_B^+ \, W^{+a} + 3 q^{+B} \nabla^{--} q_B^+ \, F^{++} + [ q^-_B, q^+_C] q^{+B} q^{+C} \Big)
 \Big\}\,.  \ee
Indeed, it is not difficult to check that the expression \p{L++++OffShell}
is G-analytic off  mass shell. One further notices that
the density \p{L++++OffShell} coincides with  \eqref{L++++fin} modulo the terms
proportional to the equations of motion. In other words, \p{L++++OffShell}
represents {\it the same} counterterm as  \eqref{L++++fin}; one expression is
obtained from another by a field redefinition.

Consider now a deformation of the action \eqref{Summ} involving
not the density \p{L++++fin}, but the gauge non-invariant off-shell supersymmetric density
 \p{L++++OffShell},
\be
 \lb{Sdeformed}
 S = S^{Vq^+}  + f^2 \int d\zeta^{(-4)} {\hat{\cal L}}^{+4}  + f^{6} S_2 + \dots\,.
 \ee
To  order $f^2$, the standard
gauge transformation of the complete action reads
\be
\delta S = \frac{1}{f^2}{\rm Tr}_{(s)}  \int d\zeta^{(-4)} \Big[\Big
( F^{++} + \frac{1}{2} [ q^{+A} , q^+_A]\Big) \Big(  \delta V^{++} +
2 i f^4 \partial_{ab} \Lambda \bigl(  W^{+a} W^{+b} -
2 F^{++} {\cal A}^{ab}\bigr)  \Big) \Big],
\ee
which vanishes on shell, but not off shell.
On the other hand, one can notice that  the  action \p{Sdeformed}
is invariant under the {\it modified} gauge transformation
     \be \lb{modif_gauge} \delta V^{++} = \nabla^{++} \Lambda -
 2 i f^4 T\Bigl(  \partial_{ab} \Lambda \bigl(  W^{+a} W^{+b} -
2 F^{++} A^{ab}\bigr)\Bigr)  + \mathcal{O}(f^8) \,,
    \ee
where $T(\cdots)$ stands for the symmetrized product projected on the Lie algebra,
\be T( X_1 X_2 X_3  ) =  \ \frac 16 \, T^p \, {\rm Tr}  \,  \left[ T^p \bigl( X_1 \{X_2 ,X_3 \}+ X_2 \{X_3 ,X_1 \}+ X_3 \{X_1 ,X_2 \}
\bigr)  \right], \ee
with the generators $T^p$ normalized by Tr$\,(T^p T^q)  \ =\ \delta^{pq}$.

This gauge transformation preserves the G-analyticity of $V^{++}$. The algebra
of the modified gauge transformations closes,
\be  \bigl( \delta(\Lambda_1)  \delta(\Lambda_2)-\delta(\Lambda_2)  \delta(\Lambda_2)\bigr)
  V^{++}  = \delta([\Lambda_1,\Lambda_2])  V^{++}  + \mathcal{O}(f^8) \, . \ee

The situation when the action representing an infinite series \p{Sdeformed} is invariant
under the modified gauge transformations \p{modif_gauge}, which  also are given by an infinite
series, is exactly the same as what happens for off-shell supersymmetry,
when choosing the gauge-invariant realization, cf. eqs. \p{LSQMeff}, \p{trans_eff} and their discussion in
Sect. 2.

Alternatively, one can restore
the standard realization of gauge transformations by
redefining
the superfield $V^{++}$ in such a way that
\be V^{++} \ \to \  V^{++} +  2 i f^4 T\Bigl( {\cal A}_{ab}W^{+a} W^{+b} -
F^{++}  {\cal A}_{ab}  {\cal A}^{ab}\bigr)\Bigr)  + \mathcal{O}(f^8) \, .
 \ee
The modified $V^{++}$ is not analytic anymore,
\be \lb{harm_curv}
D^+_a V^{++} \ =\ [  D_a^+, \nabla^{++}] = - f^4 \varepsilon_{abcd} T\bigl( W^{+b} W^{+c} W^{+d} \bigr)
 + \mathcal{O}(f^8) \, . \ee
The nonzero commutator \p{harm_curv} is related to the non-zero curvature constraints derived in
 \cite{Bergshoeff:1986jm} in the ordinary ${\cal N} = (1,0)$ superspace formalism.

\subsection{Invariant actions: $d=10$}
We again start with the invariants in the pure gauge sector.

One can write two different off-shell ${\cal N}=(1,0)$ supersymmetric and gauge
invariant Lagrangians of canonical dimension 10.
One of them is known as the {\it single-trace} invariant; the corresponding action reads
  \be
\lb{1trace}
S^{(10)}_1 = \int dZ\,  \varepsilon_{abcd} {\rm Tr}\,\big(W^{+a} W^{-b} W^{+c} W^{-d}\big).
 \ee
Any $d=10$ invariant with a different ordering of the covariant superfield strengths is
reduced to \p{1trace} by integrating by parts with respect to
the harmonic derivatives [using the relations \p{HarmW}].
One can derive in this way
the following convenient representation for \p{1trace},
 \be
\lb{1traceA}
S^{(10)}_1 = \frac13 \int dZ  \varepsilon_{abcd} {\rm Tr} \,\big(\{W^{+a},  W^{-b}\}\,\{ W^{+c},  W^{-d}\}\big).
 \ee
This form of the $d=10$ invariant implies, in particular,
that all possible $q^{+ A}$-dependent terms completing this off-shell
${\cal N}=(1,0)$ invariant
to an on-shell ${\cal N}=(1,1)$ invariant should
represent a trace of the product either of two anticommutators or of two commutators.
Thus, they should vanish in the abelian limit together with the term \p{1traceA}.

There is also the {\it double trace} invariant,
  \be
\lb{2trace}
S^{(10)}_2  =  \int dZ \,\epsilon_{abcd}\,{\rm Tr}\,
(W^{+a} W^{-b})\,{\rm Tr}\,(W^{+c} W^{-d})\,. 
 \ee
Its uniqueness can be as well  proved via integrating by parts and taking advantage of the relations
\p{HarmW}.
In the abelian limit,  \p{2trace} vanishes by the same token as
 \p{1trace}.

The difference of the invariants \p{1trace} and \p{2trace} from  \p{dim8anal-1} and \p{dim8anal-2} is that the harmonic charge of the
integrand in the former is zero, and the integral now goes over  the whole superspace rather than its analytic subspace. This brings
about two additional
powers of mass in the component Lagrangians. Another crucial difference is that \p{1trace} and \p{2trace} are
${\cal N}=(1,0)$ supersymmetric {\it off shell},
whereas \p{dim8anal-1} and \p{dim8anal-2}  are supersymmetric only {\it on shell}.

To construct the possible on-shell ${\cal N}=(1,1)$ completion of \p{1trace} and \p{2trace} one can proceed in the spirit of
Sect. 5. Namely, one can add to these expressions  all possible
${\cal N}=(1,0)$ superfield
invariants of dimension $d=10$ with hypermultiplets and require the sum to be invariant
 up to a total derivative   under the ${\cal N}=(0,1)$ transformations \p{qtransfOn1} - \p{transDq1}
 on the mass shell
\p{Fmodif}, \p{Urq1}.
In Sect. 5, we managed to carry out this program for
the single-trace $d=8$ invariant, but, for $d=10$, this turns out
to be an extremely difficult task. The calculations are {\it much} more simple, if
 using the on-shell
${\cal N}=(1,1)$ harmonic superspace formalism.

We introduce the superfield
\begin{eqnarray}
\lb{phi-hat+}
\phi^{-\hat{+}} = \nabla^{--} \phi^{+\hat{+}}
&=& q^{- \hat{+}} - \theta^{\hat{+}}_a W^{-a} - i
\theta^{\hat{+}}_a \theta^{\hat{+}}_b \nabla^{ab} q^{-\hat{-}}
 +\frac{1}{6}  \varepsilon^{abcd}  \theta^{\hat{+}}_a \theta^{\hat{+}}_b
\theta^{\hat{+}}_c [ D^+_d q^{-\hat{-}}  , q^{-\hat{-}}] \nonumber \\
&& \hspace{20mm} -  \frac{1}{24}  \varepsilon^{abcd}  \theta^{\hat{+}}_a
 \theta^{\hat{+}}_b  \theta^{\hat{+}}_c\theta^{\hat{+}}_d  [ q^{-\hat{-}},
 [ q^{-\hat{-}} , q^{+\hat{-}}] ] \ . \end{eqnarray}
It satisfies the constraints,
\be
\lb{const-phi-+}
D^{\hat{+}}_a  \phi^{-\hat{+}} \ =\ {\nabla}_a^-  \phi^{-\hat{+}}
= \nabla^{--}  \phi^{-\hat{+}} \ =\ \nabla^{\hat{+}\hat{+}}  \phi^{-\hat{+}}
\ =\  0 \, ,
  \ee
where
 \be
\lb{tilde-}
\nabla_a^-    \ =\ D_a^- - D^+_aV^{--}\ -
\theta^{\hat{+}}_a q^{-\hat{-}} + \theta^{\hat{-}} \phi^{-\hat{+}}
 \ee
[cf. \p{nablaFull-}]. The superfield \p{phi-hat+}  appears in the anticommutator
 \be
\lb{comm_nabla-}
\{ D^{\hat{+}b}, \ \nabla^-_a \}   \ =\
\delta_a^b \phi^{- \hat{+}} \,,
  \ee
which can be obtained by applying $\nabla^{--}$ to both sides of the constraint (\ref{antic12}c).

 In the full analogy with \p{variation2}, the ${\cal N}=(0,1)$ variation of
\p{phi-hat+} is a combination of a total space-time derivative,
total $\theta$ derivative and the commutator term. The latter involves
 the same compensating superfield $\Lambda^{\rm comp}$ as
in \p{variation2}.

With the superfields $\phi^{+\hat{+}}$ and $\phi^{-\hat{+}}$ in hand, it is rather clear how
to define the two $d=10$ ${\cal N}=(1,1)$ invariant actions generalizing \p{1trace} and \p{2trace}. They are:
\be
S_1^{(10)} = {\rm Tr} \int  dZ d\hat{\zeta}^{(-4)}  \,
 (\phi^{+\hat{+}} )^2 (\phi^{-\hat{+}} )^2 \ \lb{1Trace}
 \ee
and
\be
S_2^{(10)} = -\int dZ d\hat{\zeta}^{(-4)}  \,  {\rm Tr} \Big(\phi^{+\hat{+}}
\phi^{-\hat{+}}\Big)\,{\rm Tr} \Big(\phi^{+\hat{+}}  \phi^{-\hat{+}}\Big)\, ,\lb{2Trace}
\ee
where the minus sign in \p{2Trace} was chosen for further convenience.

In contrast to \p{L++++}, these invariants {\it vanish} in the abelian limit (in agreement with the fact that \p{1trace}
and \p{2trace} vanish in this limit).
This property can be made manifest for \p{1Trace} by rewriting it as
\be
S_1^{(10)} = -\frac16 {\rm Tr} \int  dZ d\hat{\zeta}^{(-4)}  \,
 [\phi^{+\hat{+}}, \phi^{-\hat{+}}]\, [\phi^{+\hat{+}}, \phi^{-\hat{+}}]\ .\lb{11Trace}
\ee

The single trace invariant can also be written as a full superspace integral
\be
\label{fullint}
 S_1^{(10)}  \ \sim \ {\rm Tr} \int dZ d\hat{Z} \ \phi^{+ \hat{+}} \phi^{- \hat{-}}\,,
 \quad \phi^{- \hat{-}} = \nabla^{\hat{-}\hat{-}} \phi^{- \hat{+}}\,.\lb{Zform}
 \ee
To show this, we represent
\be
d\hat{Z} = d\hat{\zeta}^{(-4)}(D^{\hat{+}})^4\,, \quad (D^{\hat{+}})^4 :=
-\frac{1}{24}\varepsilon_{abcd}D^{a\hat{+}}D^{b\hat{+}}D^{c\hat{+}}D^{d\hat{+}}
\ee
and use the hat-analyticity of $\phi^{+\hat{+}}$ and $\phi^{-\hat{+}}$, as well as the relations
\be
D^{d\hat{+}}\phi^{- \hat{-}} = -\nabla^{d\hat{-}}\phi^{- \hat{+}}\,, \quad D^{c\hat{+}}D^{d\hat{+}} \phi^{- \hat{-}} =
- 2i \nabla^{cd}\phi^{- \hat{+}}
\ee
to bring \p{Zform} in the form
\be
\sim \frac{i}{12} \int d\hat{\zeta}^{(-4)}{\rm Tr} \Big(\phi^{+
\hat{+}}\varepsilon_{abcd}D^{a\hat{+}}D^{b\hat{+}} \nabla^{cd}\phi^{- \hat{+}}\Big).
\ee
After that we rewrite $\nabla^{cd}$ through the unhatted covariant derivatives as
\be
\nabla^{cd} = \frac12\varepsilon^{cdab}\nabla_{ab}\,, \quad \nabla_{ab} = \frac{1}{2i} \{\nabla^+_a, \nabla^-_b\}
 \lb{AntiComm-hat}
\ee
and then pull out the remaining two hatted derivatives $D^{a\hat{+}}D^{b\hat{+}}$ to the right
through the anticommutator
\p{AntiComm-hat},
using the constraint (\ref{antic12}c) and its corollary \p{comm_nabla-}. The result is
\be
\int d\hat{Z} \ {\rm Tr}\,\Big( \phi^{+ \hat{+}} \phi^{- \hat{-}}\Big) \ =\
\int d\hat{\zeta}^{(-4)} \,{\rm Tr}\,\Big([\phi^{+ \hat{+}}, \phi^{- \hat{+}}]\,[\phi^{+ \hat{+}},
\phi^{- \hat{+}}]\Big).
\ee

 On the other hand,
the double trace invariant cannot be written as a full-superspace integral and can be considered
as a 1/4 BPS protected operator.
This allows one to explain the absence of the associated logarithmic divergence in the pure spinor
formalism \cite{Berkovits:2009aw:2009aw,Bjornsson:2010wm,Bjornsson:2010wu}. However, it is not yet
sufficient to prove the non-renormalization theorem in the standard quantum field theory framework.
Eq. \eqref{2Trace} is a full-superspace integral over ${\cal N}=(1,0)$ harmonic superspace, and is {\it \'a priori} allowed by
the harmonic superspace Feynman rules.
One may anticipate nonetheless that the Ward identities for the non-linearly realized extra supersymmetry
 would permit
to rule it out as an allowed counterterm. The integrand in  \eqref{2Trace} is invariant with respect to
the transformations \eqref{Hidden} modulo a total derivative in ${\cal N}=(1,0)$ harmonic superspace
and taking into account the equations of motion.
The variation of a total derivative  with respect to \eqref{Hidden} gives again a total
 derivative,
and one gets in this way a chain of co-forms associated to a given supersymmetry invariant
(see Sect. 5.3 of \cite{Bossard:2013rza}).
One shows then in the framework of algebraic renormalization \cite{Piguet:1995er} that the cohomology class
associated to this chain of
co-forms must be compatible with the cohomology class associated to the classical (dimension 4)
 Lagrangian. In this way,
one would combine the constraints following from the ${\cal N}=(1,0)$ harmonic superspace
 Feynman rules and the constraints following
from the full ${\cal N}=(1,1)$ on-shell supersymmetry of the action in the framework of algebraic renormalization.
 One knows that neither of these methods, taken separately, is powerful
 enough to explain the absence of non-planar divergences at three loops
 \cite{Bossard:2010pk}. But we hope that,  being combined in this way,
they may allow to prove the required non-renormalization
theorem. We will not, however,
 investigate this issue further in this paper.

Let us come back to the explicit form of the $d=10$  invariants in the ${\cal N}=(1,0)$ harmonic superspace.
It is rather straightforward to perform
the integration over $d\hat\zeta^{(-4)}$ in the invariant \p{11Trace} and obtain its ${\cal N}=(1,0)$ superfield form.

The result of integration can be written as a sum of the three terms
\be
\tilde{S}_1^{(10)} =  \int  dZ \Big( {\cal L}_{(1)}^{(10)} +
{\cal L}_{(2)}^{(10)} + {\cal L}_{(3)}^{(10)} \Big),\lb{three}
\ee
where
\be
{\cal L}_{(1)}^{(10)} &=&  \frac 16 \varepsilon_{abcd}\,{\rm Tr}\,\Big( \{W^{+ a},
 W^{-b}\}\{W^{+ c}, W^{-d}\} - 2i \{W^{+ a}, W^{-b}\}[q^{-A}, \nabla^{cd}q^+_A]  \nn
&&\, -\frac 1{2} \nabla^{ab}(q^-)^2\nabla^{cd}(q^+)^2 + \frac 1{6} \nabla^{ab}[q^{-(A}, q^{+ B)}]\,
\nabla^{cd}[q^{-}_{(A}, q^{+}_{ B)}] \nn
&& - \, [q^{+ A}, \nabla^{ab}q^-_A]\, [q^{+ B}, \nabla^{cd}q^-_B]\Big), \lb{1st} \\
{\cal L}_{(2)}^{(10)} &=& \frac 23  {\rm Tr}\,\Big\{\Big( [q^{+ A}, W^{-a}] - [q^{- A}, W^{+a}]\Big)
 \Big( i[W^{+b}, \nabla_{ab}q^-_A]
 +  \frac13\,[q^{+ B}, [ q^-_{(B}, D^+_aq^-_{A)}]] \Big)\Big\}, \lb{2st} \\
{\cal L}_{(3)}^{(10)} &=& \frac 1{18} {\rm Tr}
 \Big\{[q^{+(A}, q^{-B)}]\Big([q^{+C},[q^-_C,[q^+_{(A}, q^-_{B)}]]]  + 2[q^{+C},[q^-_A,[q^+_{(B}, q^-_{C)}]]] \nn
&& - 4\{W^{+ a}, [D^+_aq^-_A, q^-_B]\} - 4 [\nabla^{ab}q^+_A, \nabla_{ab}q^-_B]\Big)\Big\}. \lb{3st}
\ee
While deriving \p{1st} - \p{3st}, we essentially used the integration by parts and various
 on-shell conditions like $\nabla^{--}q^{- A} = 0$  etc.
Perhaps, these expressions can be further simplified by integrating by parts and using
some $SU(2)$ Fierz identities. Anyway, it would be very difficult
to guess them entirely within the ${\cal N}=(1,0)$ superfield formalism.

A good check of the correctness of \p{1st} - \p{3st} is the verification of the fact that
 the variation of the first term in \p{1st}
under the hidden supersymmetry $\hat{\delta}W^{\pm a} = -2i \epsilon^A_b\nabla^{ab}q^{\pm}_A$
[see \p{WtransfOn}] is canceled (modulo various terms vanishing on-shell)
by the $(W)^3$ part of the variation of the term $\sim W^2$  in \p{three}.
 The latter term is assembled from  the pieces coming from \p{1st} and \p{2st}
and, after some algebra, is represented as
$$
2i {\rm Tr}\,\Big( [\nabla_{ab}q^+_A, W^{-a}] [W^{+b}, q^{-A}]\Big).
$$
Its variation under $\hat{\delta}q^{\pm A} = \epsilon^A_a W^{\pm a}$ [see \p{qtransfOn1}] exactly
 cancels the variation of the first term in \p{1st}.

The double-trace invariant \p{2Trace} can also be straightforwardly cast into the ${\cal N}=(1,0)$ superfield form:
\be
S_2^{(10)} = \int  dZ \Big( \hat{\cal L}_{(1)}^{(10)} + \hat{\cal L}_{(2)}^{(10)} + \hat{\cal L}_{(3)}^{(10)} \Big),
 \lb{threeDouble}
\ee
\be
\hat{\cal L}_{(1)}^{(10)} &=& \varepsilon_{abcd}\Big\{ {\rm Tr}\,(W^{+a}W^{-b})\,{\rm Tr}\,(W^{+c}W^{-d})
-i{\rm Tr}\,(W^{+a}W^{+b})\,{\rm Tr}\,(q^{-A}\nabla^{cd}q^-_{\,A})\Big\} \nn
&& + \,{\rm Tr}\,(q^{+A}\nabla^{ab}q^+_{\, A})\,{\rm Tr}\,(q^{-B}\nabla_{ab}q^-_{\,B}) -
 \frac1{6}\partial^{ab}{\rm Tr}\,(q^{+A}q^{+B})\,\partial_{ab}{\rm Tr}\,
(q^{+}_{\,A} q^{+}_{\,B})\,, \nn
\hat{\cal L}_{(2)}^{(10)} &=& \frac43\,{\rm Tr}\,(q^{+ A}\,W^{+a})\,{\rm Tr}\Big\{q^{-B}\, [D^+_a q^-_{\,(A}, q^-_{\,B)}]
- 3i \nabla_{ab}q^-_{\,A}\, W^{-b} \Big\}, \nn
\hat{\cal L}_{(3)}^{(10)} &=& \frac23\,{\rm Tr}\,(q^{+ A}q^{-B})\,{\rm Tr} \Big\{[q^{+ C},
 q^-_{\,C}]\, [q^+_{\,(A}, q^-_{\,B)}]
- 2\nabla^{ab}q^+_{(A}\,\nabla_{ab}q^-_{B)} \nn
&&+\, 2[D^+_aq^-_{(A}, q^-_{\,B)}]\, W^{+ a}\Big\}.\lb{LagrHat}
\ee
As a good self-consistency check, one can verify that, in the abelian limit,
the ${\cal N}=(1,0)$ superfield Lagrangian in \p{threeDouble} is indeed reduced to
a total derivative. This check is not trivial because not all terms in \p{LagrHat}
contain (anti)commutators under the trace [in contrast to \p{1st} - \p{3st}].

Finally, we want to point out once more that the actions \p{three} and \p{threeDouble}
respect the {\it off-shell} ${\cal N}=(1,0)$ supersymmetry, being written in terms
of the off-shell ${\cal N}=(1,0)$ superfields. They also respect the {\it on-shell}
 ${\cal N}=(0,1)$ invariance because they
admit an equivalent representation as integrals over the ${\cal N}=(1,1)$ harmonic
 superspace and its non-trivial subspaces supporting a linear realization
of both ${\cal N}=(1,0)$ and ${\cal N}=(0,1)$ supersymmetries. The second supersymmetry
 becomes nonlinear, when is realized in terms of the ${\cal N}=(1,0)$ superfields.
To avoid a possible confusion, we note that  the ${\cal N}=(1,1)$ form \p{1Trace} and
\p{2Trace} of the $d=10$ terms already enjoys
the {\it off-shell} ${\cal N}=(1,0)$ supersymmetry. The equations of motion come into play,
 only when checking the ${\cal N}=(0,1)$ invariance of
these expressions.

\section{Summary and outlook}
In this paper, we applied the off-shell ${\cal N} = (1,0)$ and on-shell  ${\cal N} = (1,1)$
harmonic superspace approaches for constructing higher-dimensional invariants in the six-dimensional
${\cal N} = (1,0)$ SYM and ${\cal N} = (1,1)$ SYM theories. The ${\cal N} = (1,1)$ SYM theory
constraints were solved in terms of
 ${\cal N} = (1, 0)$ harmonic superfields. This allowed us to explicitly construct
the full set of on-shell ${\cal N} = (1,1)$
supersymmetry  invariants of canonical dimensions 8 and 10 in ${\cal N} = (1, 0)$ superspace.
All possible  $d = 6$,  ${\cal N} = (1,1)$ invariants were shown
 to vanish on shell, confirming the UV finiteness of ${\cal N} = (1,1)$ SYM at one loop.
We have also shown that there are no  $d =8$
 on-shell ${\cal N} = (1, 1)$ supersymmetric invariants which possess the full off-shell
  ${\cal N} = (1, 0)$ supersymmetry together with off-shell gauge invariance.
Assuming the use of a gauge-invariant regularization scheme for  ${\cal N} = (1, 0)$
supergraphs, this implies the absence of two-loop counterterms.
On the other hand, the on-shell
${\cal N} = (1, 0)$ and ${\cal N} = (1, 1)$ supersymmetric $d=8$ invariants  exist. They  are represented
as the analytic harmonic ${\cal N} = (1, 0)$ superspace integrals of densities which are  both analytic
and gauge invariant only on mass shell,  {\it i.e.} assuming the equations of motion to be satisfied. We show that one can enforce
either off-shell analyticity or off-shell gauge invariance of the relevant density, but not both of them simultaneously.
Structures of this kind appear in the derivative expansion of the supersymmetric Born-Infeld action.

Two $d = 10$ invariants were explicitly constructed as integrals over the
whole ${\cal N} = (1,0)$ harmonic superspace. The single-trace invariant can
be rewritten as an integral over the full ${\cal N} = (1, 1)$ superspace, while the
double-trace invariant cannot. This property, being combined with an additional
reasoning based on the algebraic renormalization ideas adapted to the ${\cal N}=(1,0)$
 harmonic superspace formalism,
could potentially explain why the double-trace invariant is UV protected. However, to prove this,
 we would need
first to compute the chain of ${\cal N}=(1,0)$ harmonic superspace co-forms associated to these two invariants
and then establish that the chain associated to the double-trace invariant is, indeed, incompatible
with the one of the classical action.

The ${\cal N}=(1,1)$ harmonic superspace is also useful to conveniently combine on-shell particle states
into a G-analytic superfield in momentum space. This provides an efficient tool to apply the generalized
unitary method to compute on-shell scattering amplitudes in ${\cal N} = (1,1),\ 6D$ SYM theory \cite{Sieg}
 (see also \cite{KazNew} and references therein).
It would be interesting to clarify the relations between the Feynman rules in ${\cal N}=(1,0)$ harmonic superspace
and this generalized unitary method in (on-shell) momentum harmonic superspace.

We conclude by mentioning some other problems where our on-shell harmonic approach  could be applied.

It could be used, e.g., to construct the invariants of higher dimension $d\geq 12$  in the ${\cal N} = (1, 1), \ 6D$
SYM theory and to inspect whether some kind of the non-renormalization theorems could be formulated.
The same methods could be applied for constructing the Born-Infeld effective action
for coincident D5-branes in type IIB string theory, with the manifest  ${\cal N} = (1,0)$ off-shell and hidden
${\cal N} = (0, 1)$ on-shell supersymmetries. It would be also interesting to develop an analogous on-shell
${\cal N} = 4, \ 4D$
harmonic  superspace approach for the ${\cal N} = 4, \ 4D$ SYM theory in the off-shell
${\cal N} = 2$ superfield formulation and apply it to the problem of constructing the relevant
quantum effective action and proving its identity with the appropriate D3-brane Born-Infeld action.
An intriguing feature of such effective actions is the presence there of Chern-Simons (or Wess-Zumino)
type terms of non-tensorial character \cite{tseyt5, Intril, arg}. It would be interesting to see how
such terms (and their possible $6D$ counterparts) could be identified in the on-shell harmonic superspace approach.

The last (but not least) domain where the on-shell harmonic superspace methods could help in selecting relevant
counterterms and other higher-dimensional invariants is extended supergravities in diverse dimensions.

\section*{Acknowledgements}
E.I. acknowledges the support from the RFBR
grant 15-02-06670 and a grant of the IN2P3-JINR Program. E.I. and G.B would like to thank SUBATECH,
Universit\'{e} de Nantes, and A.S. would like to thank Ecole Polytechnique and JINR for the warm hospitality
in the course of this study. The authors thank Joseph Buchbinder,
Shahin Sheikh Jabbari, Emery Sokatchev and Kelly Stelle for the interest in the work
and useful comments.

\section*{Appendix A. Bianchi identities}
\setcounter{equation}0
\renewcommand\theequation{A.\arabic{equation}}
Many important  identities between the harmonic superfields
in the pure gauge ${\cal N} = (1,0)$ theory
were derived and discussed in Sect. 3.
The Bianchi relations allow one to obtain further interesting and useful identities.

When one includes
the hypermultiplet superfields and imposes the on-shell constraints \p{Urq1}, \p{Urq2}, many
other relations
 relevant to  ${\cal N} = (1,1)$ SYM theory can also be derived.

\subsection*{A1. Off-shell relations}

We first discuss the pure gauge theory off-shell relations.
 Taking the anticommutator of $D^+_b$ with the second of  relation in
 \p{Da,nabcb}, we find
\be
[\nabla_{ab}, \nabla_{cd}] = \sfrac14\left(\varepsilon_{acdf}\nabla^-_b W^{+ f} +
\varepsilon_{bcdf} D^+_a W^{- f} \right). \lb{nabnab}
\ee

An important Bianchi identity is obtained from \p{nabnab} by contracting its both sides with $\varepsilon^{abcd}$ and using the fact that
$[\nabla_{ab}, \nabla_{cd}]\varepsilon^{abcd} \equiv 0$. One obtains
\be
\nabla^-_b W^{+ b} = D^+_b W^{- b}\,. \lb{DW+-2}
\ee
Rewriting the identity \p{DW0} as
\be
D^+_a W^{+ b} = \sfrac14\,\delta^b_a \,D^+_c W^{+ c}\,, \lb{DW++}
\ee
acting on its both sides by $\nabla^{--}$ and using \p{DW+-2}, one also finds
\be
D^+_a W^{- b} + \nabla^{-}_a W^{+ b} = \sfrac12 \delta^b_a\,\nabla^-_c W^{+ c} = \sfrac12 \delta^b_a\,D^+_c W^{- c}\,. \lb{DWsum}
\ee
One of the corollaries of \p{nabnab} is that its right-hand side
 is antisymmetric under
the permutations $a \leftrightarrow b$ and $ab \leftrightarrow cd$,
 as its left-hand side is (the antisymmetry
under $c \leftrightarrow d$ is manifest on both sides).

One more important Bianchi identity following from the basic (anti)commutation
relations can be derived by anticommuting both sides of \p{nabnab} with
$D^+_a$ or $\nabla^-_b$,
\be
\nabla_{ab}W^{+b} = -\sfrac{i}{4} D^+_a \nabla^-_b W^{+b} = -\sfrac{i}{4} D^+_a D^+_b W^{-b}\; \Leftrightarrow \;
D^+_a\, \nabla^-_b W^{+b} +\sfrac12 \nabla^-_a\, D^+_b W^{+b} = 0\,.\lb{nablaW+}
\ee
All other relations obtained in this way are identically satisfied as  a consequence of \p{nablaW+} and the identities derived above. By
commuting \p{nablaW+} with $\nabla^{--}$, one obtains an analogous identity for $W^{-a}$,
\be
\nabla_{ab}W^{-b} = \sfrac{i}{4} \nabla^-_a \nabla^-_b W^{+b}= \sfrac{i}{4} \nabla^-_a D^+_b W^{-b}\;
\Leftrightarrow \; \nabla^-_a\, D^+_b W^{-b} +\sfrac12 D^+_a\, \nabla^-_b W^{-b} = 0\,.\lb{nablaW-}
\ee

\subsection*{A2. On-shell ${\cal N} = (1,1)$ relations}

 The presence of the hypermultiplet does not affect  the off-shell identities derived above.
But the on-shell identities are modified. For example, instead of $ D^{+}_bW^{+a} = 0$, we obtain, using \p{DW0},
  \be
\lb{DW1}
D^{+}_b W^{+a} \ =\ -\sfrac 12 \delta^a_b [q^{+A}, q^+_A] \, .
  \ee

Bearing in mind  \p{DW1},  one obtains
\be
&& D^{+}_aW^{-a} = \nabla^-_a W^{+a} = -2 [q^{- A},\, q^+_A]\,, \quad \nabla^-_a W^{-a} =  -2 [q^{- A}, \,q^-_A]\,, \lb{DWq1} \\
&& \nabla_{ab}W^{+b} = -\sfrac{i}{2}[\nabla^-_aq^{+ C}, \,q^+_C] = \sfrac{i}{2}[D^+_aq^{- C}, \,q^+_C]\,, \nn
&& \nabla_{ab}W^{-b} = \sfrac{i}{2}[D^+_aq^{- C},\, q^-_C] = -\sfrac{i}{2}[\nabla^-_aq^{+ C},\, q^-_C]\,. \lb{DWq2}
\ee

Starting from the evident identity
$$
\nabla_{ab}D^+_c q^{-A} = \sfrac{1}{2i} \{D^+_a, \,\nabla^-_b\} D^+_c q^{- A}\,,
$$
and repeatedly using \p{Da,nabcb} together with the on-shell relations \p{onsh1},
it is straightforward to obtain the following cyclic on-shell identity,
\be
\left(\nabla_{ab} D^+_c + \nabla_{ca} D^+_b + \nabla_{bc} D^+_a\right)q^{-A} = \sfrac{i}{2}\varepsilon_{abcd}\left([W^{-d},\, q^{+ A}] - [W^{+d}, \,q^{- A}] \right),
\lb{cycl1}
\ee
which, in virtue of the analyticity condition $D^+_a q^{+ A} =0\,$,  also implies
\be
\left(\nabla_{ab} \nabla^-_c + \nabla_{ca} \nabla^-_b + \nabla_{bc} \nabla^-_a\right)q^{+A} = -\sfrac{i}{2}\varepsilon_{abcd}\left([W^{-d},\, q^{+ A}]
- [W^{+d},\, q^{- A}] \right).
\lb{cycl11}
\ee

Some useful consequences of these identities are
\be
&& \nabla^{ab}D^+_b q^{-}_A = \sfrac i2 \left([W^{+ a},\, q^{-}_A] - [W^{- a},\, q^{+}_A]\right), \lb{1Id} \\
&& D^+_b\nabla^{ab} q^{-}_A = - \sfrac i2 \left([W^{- a},\, q^{+}_A] + 2[W^{+a},\, q^{-}_A]\right),\lb{2Id} \\
&&  \nabla^{ad} \nabla_{d e} q^{+}_A = -\sfrac14 [ D^+_e W^{- a},   q^{+}_A] - \sfrac14 \delta^a_e\big( \{W^{+f},  D^+_fq^{-}_A\} -
\sfrac12 \left[q^{-}_A, \,[q^{+C}, q^+_C]\right]\big),\lb{3Id+} \\
&& \nabla^{ad} \nabla_{de} q^{-}_A = \sfrac14 [ \nabla^-_e W^{+ a},   q^{-}_A] - \sfrac14 \delta^a_e\big( \{W^{-f},  D^+_fq^{-}_A\} +
\sfrac12 \left[q^{+}_A, \,[q^{-C}, q^-_C]\right]\big),\lb{3Id} \\
&&\nabla^{cd} \nabla_{cd} q^{-}_A = \{W^{-e}, D^+_e q^{-}_A\} + \sfrac 12 \left[q^{+}_A, \,[q^{-C}, q^-_C]\right] -
\sfrac 12 \left[q^{-}_A, \,[q^{-C}, q^+_C]\right],\lb{4Id} \\
&&\nabla^{cd} \nabla_{cd} q^{+}_A = \{W^{+e}, D^+_e q^{-}_A\} + \sfrac 12 \left[q^{+}_A, \,[q^{-C}, q^+_C]\right] -
\sfrac 12 \left[q^{-}_A, \,[q^{+C}, q^+_C]\right],\lb{5Id} \\
&& D^+_a D^+_b \nabla^{ab} q^{-}_A = i\big(\{W^{+e}, D^+_e q^{-}_A\} -
\left[q^{+}_A, \,[q^{-C}, q^+_C]\right]
- 2 \left[q^{-}_A, \,[q^{+C}, q^+_C]\right]\big),\lb{6Id} \\
&&D^+_a \nabla^{ad} D^+_d q^{-}_A = -\sfrac i2 \big(\{W^{+e}, D^+_e q^{-}_A\} +2 \left[q^{+}_A, \,[q^{-C}, q^+_C]\right]
- 2\left[q^{-}_A, \,[q^{+C}, q^+_C]\right]\big),\lb{7Id}
\ee
where we took advantage of the on-shell relations \p{DWq1}. Note that \p{1Id}, \p{4Id} and
\p{5Id} are none other than the covariant superfield form
of the $6D$ Dirac and Klein-Fock-Gordon equations for the physical fermionic and bosonic components of $q^{+}_A$.

One more useful on-shell consequence of the Bianchi identities is the following cyclic identity:
\be
\nabla^{ab}W^{+ c} + \nabla^{ca}W^{+ b} + \nabla^{bc}W^{+ a} = \sfrac i2 \varepsilon^{abcd}\nabla^-_d  [q^{+A}, q^+_A] \, , \lb{nablW}
\ee
from which one can derive
\be
\nabla^{bc}\nabla_{bc} W^{+ a} = [\nabla^-_d W^{+a}, W^{+ d}] +
 2i \nabla^{ab}\nabla^-_b (q^{+})^2 + \sfrac 12 [[q^{-A}, q^{+}_A], W^{+ a}]\,.\lb{nabl2W}
\ee

\section*{Appendix B. In quest of an off-shell ${\cal N}=(1,1)$ $d=6$ invariant}
\setcounter{equation}0
\renewcommand\theequation{B.\arabic{equation}}

We continue here the discussion of Subsection 5.1 and study the symmetries of  the higher derivative actions
  of canonical dimension 6. A generic ${\cal N} = (1,0)$ supersymmetric action of this kind is a linear combination of  the supergauge action
\p{dejstvie} and the hypermultiplet actions \p{S0hyp}, \p{Snhyp} and \p{quart}.
The question is whether one can define a specific linear combination $S^{(6)}$ which would be invariant off shell under the variations  \p{Hidden} of the
hidden  ${\cal N} = (0,1)$ supersymmetry.

Requiring the
cancellation of the terms generated by the variations \p{Hidden} in the first order in $q^{+ A}$ and using the formula
\be
\delta S_{SYM}^{(6)} = \sfrac{1}{2g^2}  {\rm Tr} \int dZ \, \delta V^{++} (\nabla^{--})^2 F^{++}  = \sfrac{1}{2g^2} {\rm Tr} \int
d\zeta^{(-4)} \,\delta V^{++}  (D^+)^4 (\nabla^{--})^2(D^+)^4 V^{--} \,,
\ee
we deduce the following form for the candidate action
\be
S^{(6)} &=& \sfrac{1}{2g^2} {\rm Tr} \Big[ \int  d\zeta^{(-4)}\,
(F^{++})^2  - \int dZ \, q^{+ A}\nabla^{--} q^+_A
- \sfrac12 \int d Z \, q^{+ A}(\nabla^{--})^2\nabla^{++} q^+_A   \Big]    \nn
&&+\, \sum_{n \geq 3} \alpha_n S_n + \beta S_{\rm quart} \, ,
\lb{S2}
\ee
where $\alpha_n$ and $\beta$ are arbitrary coefficients; $S_n$ and $S_{\rm quart}$ were defined in \p{Snhyp}, \p{quart}.
The variation of the second line in \p{S2}
does not involve the linear in $q$ terms (this is obvious for $S_{\rm quart}$ and one can also show this to be true
for all $S_n$ with $n \geq 3$). These terms are also cancelled in the variation
of the first line.

The variation of the first
line in \p{S2} can be represented as
\be
\delta S_{\rm first\ line}^{(6)} =
 \frac{1}{4g^2} {\rm Tr} \int d\zeta^{(-4)}\, [(D^+)^4 C^{-3 A}]
\nabla^{++}q^+_A \nn
 -\, \frac{1}{2g^2}  {\rm Tr} \int d\zeta^{(-4)}\, [ (D^+)^4 (\epsilon^{-A}
V^\m, q^+_A] [q^+_B, q^{+B} ]
\, , \lb{varS2prime}
\ee
where
\be
C^{-3 A} &=& \epsilon^{-B}[q^+_B, (\nabla^{--})^2 q^{+ A}] + \epsilon^{-A}[\nabla^{--}q^+_B, \nabla^{--} q^{+ B}] \nn
&& +\, [\nabla^{--}\delta_0V^{--}, q^{+ A}] - 2[\delta_0V^{--}, \nabla^{--}q^{+ A}]\,,
\ee
or, in the equivalent form,
\be
C^{-3 A} &=& \epsilon^{-B}[q^+_B, (\nabla^{--})^2 q^{+ A}] - \epsilon^{+B}[(\nabla^{--})^2q^+_B, \nabla^{--} q^{+ A}] \nn
&& +\, [\nabla^{--}\delta_0V^{--}, q^{+ A}] + [\nabla^{++}(\nabla^{--}\delta_0V^{--}), \nabla^{--}q^{+ A}]\,.
\ee
While calculating this variation, we took advantage of the general formula \p{Rel1} for the variation  of $V^{--}$.

The second term in the variation \p{varS2prime}  cancels the variation of the quartic term $S_{\rm quart}$ in $S^{(6)}$ under
a particular choice $\beta = \frac 1{8g^2}$ that we adopt.
Next, we note that the second term in \p{S2} can be rewritten as an integral over the analytic subspace,
$$
{\rm Tr}\, \int dZ \, q^{+ A}\nabla^{--} q^+_A = - {\rm Tr}\,\int d\zeta^{(-4)} F^{++}[q^{+ A}, q^+_A]\,.
$$

Then the first two terms in \p{S2} together with the quartic term  can be written as
\be
\frac{1}{2g^2} {\rm Tr} \int d\zeta^{(-4)}   \left(F^{++} + \frac12 [q^{+A}, q^+_A]\right)^2.
\ee
Note that the expression in the parentheses is nothing but the equation of motion for $V^{++}$ for  the standard $d=4$
action $S^{Vq^+}$ of Eq. \p{Summ}. Then the action $S^{(6)}$ can be written as
\be
\lb{candidate}
S^{(6)} = \frac{1}{2g^2} {\rm Tr} \Big[\int d\zeta^{(-4)}  \left(F^{++} + \frac12 [q^{+A}, q^+_A]\right)^2   -
\frac12 \int d Z \, (\nabla^{--})^2q^{+ A}\nabla^{++} q^+_A \Big].\lb{S2prime}
\ee
Similar to the first term in \p{S2prime}, the second term is
 also the product of two equivalent forms of the equation of motion for $q^{+A}$, eqs. \p{Urq1} and \p{Urq2}.

The total variation of \p{candidate} is given by the first term in the variation \p{varS2prime}.
It does not seem to represent a total derivative.
If adding the sum over $S_n$ with nonzero coefficients, the cubic in $q$ terms in the variation are modified, and they would
include the terms
with higher powers of harmonic derivatives. This does not seem to help.

Let us focus on the case of vanishing $\alpha_n$.
  To check that the integrand in the variation of such an action is not a total derivative and hence that
\p{varS2prime} does not vanish, we can use the following trick. If the integrand {\it were} a total derivative, the related
``equation of motion'' obtained by varying \p{varS2prime} with respect to $q^{+A}\,$,
treated as an unconstrained superfield
on the analytic harmonic superspace, should identically vanish. After some algebra, with making use of the analyticity of $q^{+A}\,$,
we find that the variational derivative of \p{varS2prime} with respect to  $q^{+A}\,$ is reduced to the expression
\be
\frac{\delta(\delta_0 S_2')}{\delta q^{+}_A} \sim  (D^+)^4 Y^{-A}\,, \nonumber
\ee
where
\be
\lb{Y-A}
Y^{-A} &=& \epsilon^{+ A} [q^{+D}, (\nabla^{--})^2q^+_D]+ 2[(\nabla^{--}\delta_0V^{--}), \nabla^{++}q^{+ A}] \nn
&& +\,  2[\delta_0V^{--}, (1 + \{\nabla^{++},\nabla^{--}\})q^{+ A}] \nn
&& -\, \frac{\delta(\delta_0 V^{--})}{\delta q^{+}_A} \Big([q^{+ B}, \nabla^{--}\nabla^{++}q^+_B] +
3 [\nabla^{--}q^{+ B}, \nabla^{++}q^+_B]\Big).
\ee
It is easy to show that for the particular class of $q^{+ A}$ subjected
to the dynamical equations \p{Urq1} and \p{Urq2}, the variation $\delta_0V^{--}$ is
reduced to
\be
\delta_0V^{--}| = \epsilon^{-A}\nabla^{--}q^+_A\,,
\ee
and
\be
Y^A| = 4\epsilon^{-B}[\nabla^{--} q^+_B, q^{+ A}]\,.
\ee
Then it immediately follows that $(D^+)^4 (Y^A|) = 0$ because of the analyticity of $q^{+ A}\,$.

However,  we do not  see any reason for \p{Y-A} to vanish in the general case,
 when $q^{+ A}$ is not subject to any dynamical constraint. One of the arguments {\it against} the existence
of an off-shell ${\cal N}=(1,1)$ supersymmetric $d=6$ action is the following. In the $d=4$ action \p{Summ}, the extended supersymmetry implies
among other things the symmetry between the physical fermion of the gauge multiplet and the fermion
of the hypermultiplet. This symmetry is manifest when the action is expressed in  components.
On the other hand, there is no such  symmetry for the $d=6$ action. It was shown in \cite{IShyper}
that an action $S_n$ in \p{Snhyp} involves an infinite number of physical propagating
bosons and fermions associated with the harmonic
expansion of $q^{+A}$. This cannot match with the gluino sector that involves a
finite number (=12) of the fermionic degrees of freedom for each value of the color index.

\section*{Appendix C.
  ${\cal N} = (1,0)$ on-shell $d=8$ invariants}
\renewcommand\theequation{C.\arabic{equation}}
\setcounter{equation}0

We give here the full list of the planar (single-trace) G-analytic $d=8$ invariants
\footnote{The pure ${\cal N}=(1,0)$ SYM invariant \p{dim8anal-1} is G-analytic, and its ${\cal N} = (1,1)$ extension we are
interested in is also G-analytic.}
 involving
the gauge supermultiplet and the hypermultiplet and study their properties.

\subsubsection*{C1. Most general analytic ${\cal L}^{+4}$}

A generic on-shell analytic $d=8$ single-trace Lagrangian ${\cal L}^{+4}$
can be represented as a linear combination of the following six terms,
\be
{\cal L}^{+4}_{gen} = L^{+ 4}_0 + \sum_{i=1}^{5} \alpha_iL^{+ 4}_i\,,\lb{Sum}
\ee
where $L^{+ 4}_0$ is given by \p{L0} and
\be
L^{+ 4}_1 &=& {\rm Tr}\Big\{ i[q^{+ A}, \nabla_{ab}q^+_A] W^{+ a}W^{+ b}
+ \frac12\{W^{+a}, (q^+)^{2}\} [D^+_aq^{-A}, q^+_A] + (q^+)^4\,[q^{- A}, q^+_A]\Big\} ,\lb{L1}\\
L^{+ 4}_2 &=& {\rm Tr}\Big\{ q^{+ A}\nabla_{ab}q^+_A\, q^{+ B}\nabla^{ab}q^+_B + q^{+ A}D^+_aq^-_A \big[ q^{+B}W^{+ a}q^+_B
- (q^+)^{2}W^{+ a}\big] \nn
&& +\, (q^+)^{2}\, q^{+ B} q^{+ A}q^-_A q^+_B + (q^{+})^4\, q^{+ A}q^-_A \Big\}, \lb{L2} \\
L^{+ 4}_3 &=& {\rm Tr}\Big\{q^{+ A}\nabla_{ab}q^+_A\, \nabla^{ab}q^{+ B}q^+_B - \frac12 q^{+ A}D^+_a q^-_A\big[q^{+ B}W^{+ a}q^+_B
- W^{+ a}\,(q^+)^{2} \big] \nn
&& +\frac12 D^+_a q^{- A}q^+_A \big[ q^{+ B}W^{+ a}q^+_B - (q^+)^{2}\,W^{+ a}\big]
+ \frac12(q^+)^4 \big(q^{-A}q^+_A - q^{+A}q^-_A\big) \nn
&& +\,\frac12 (q^+)^{2}\,q^{+A}\big(q^{-B}q^+_B - q^{+B}q^-_B\big)q^+_A\Big\}, \lb{L3}
\ee
\be
L^{+ 4}_4 &=& {\rm Tr}\Big\{\nabla_{ab}q^{+ B}q^+_B\,\nabla^{ab}q^{+ A}q^+_A + D^+_a q^{- A}q^+_A \big[W^{+ a}\,(q^+)^{2} - q^{+B}W^{+ a}q^+_B\big]\nn
&&-\, (q^+)^{2}\big[q^{+A}\,q^{-B}q^+_B q^+_A + (q^+)^{2}q^{-B}q^+_B\big]\Big\}, \lb{L4} \\
L^{+ 4}_5 &=& {\rm Tr}\Big\{\nabla_{ab}q^{+ B}\,\nabla^{ab}q^{+}_B (q^+)^{2} + \frac12 [q^{+ A}, \,W^{+ a}][D^+_aq^{-}_A, \,(q^+)^{2}] \nn
&&+\,\frac12 [q^{+ A}, \,(q^+)^{2}][q^{-}_A, \,(q^+)^{2}]\Big\}.\lb{L5}
\ee
Here $\quad (q^+)^{2} := q^{+ A}q^+_A = \frac12 [q^{+ A},\,q^+_A]\,$.  It is straightforward to check
that each term in the sum \p{Sum} is separately annihilated by $D^+_f\,$.

Taking into account that the analytic superspace integration measure is $\sim d^6 x du (D^{-}_a)^4$, one can integrate
the above Lagrangians by parts  not only with respect to $\nabla_{ab}$ but also with respect to $\nabla^{-}_a$ and $\nabla^{--}$.
Using this opportunity and  making use of the on-shell relation \p{5Id},
one can show that  $L^{+ 4}_1$ is a total derivative on shell and establish
the following on-shell relations,
\be
\lb{Rel}
 L^{+ 4}_2 \ =\  L^{+ 4}_4\,, \quad \quad L^{+ 4}_5  =\  - L^{+ 4}_2 - L^{+ 4}_3 \, .
\ee
 We are left with only two independent Lagrangians,  $L_{2}^{+4}$ and $L_{3}^{+4}$. The representation \p{Sum}  is thus
rewritten, modulo a total derivative, as
\be
{\cal L}^{+4}_{gen} = L^{+ 4}_0 + \alpha_2L^{+ 4}_2 + \alpha_3L^{+ 4}_3\,.\lb{Sum1}
\ee

\subsubsection*{C2. Hidden ${\cal N} =(0,1)$ supersymmetry}

Now we can vary ${\cal L}^{+4}_{gen}$ with respect to the on-shell ${\cal N}=(0,1)$
transformations \p{qtransfOn1} - \p{transDq1} in order to learn at which values of the free parameters $\alpha_i$
it is invariant (up to a total derivative).  A general
variation contains the terms not involving
the superfield strengths $W^{\pm a}$ as well as  the terms of the first, second and third degrees in $W^{\pm a}$.
   It is easy to check
that the cubic term $\sim (W)^3$ in the variation comes only from $L^{+4}_0$ and represents a
 total derivative.
To explore the cancellation of the other terms
[{\it i.e.} terms  $\sim (W)^2, (W)^1$ and $(W)^0$]
in the  variation of the generic Lagrangian \p{Sum1}
 is not an easy problem.

Consider a symmetrized
trace invariant \p{L++++fin},  which, as we showed, is the on-shell ${\cal N}=(1,1)$ invariant.
It can be expressed via the structures
 \p{L0}, \p{L1} - \p{L5} as
\be
{\cal L}^{+4}_{(1,1)} = L_0^{+ 4} + L_1^{+ 4} - \frac1{6}\left(L_2^{+ 4} + L_4^{+ 4} -4 L_3^{+ 4} + 2 L_5^{+ 4}\right).\lb{invSymm}
\ee
This corresponds to the following particular choice of the coefficients in the general formula \p{Sum},
\be
\alpha_1 = 1\,, \quad \alpha_2 = \alpha_4 = -\frac1{6}\,, \quad \alpha_3 = \frac23\,, \quad \alpha_5 = -\frac13\,.
\ee
Taking into account the on-shell equivalence relations \p{Rel}, the invariant \p{invSymm}
can be cast, up to a total derivative and modulo equations of motion, in the simple form \p{Sum1}
with $\alpha_2 =0\,, \; \alpha_3 = 1$:
\be
{\cal L}^{+4}_{(1,1)} = L_0^{+ 4} + L_3^{+ 4}\,.
\ee
We have explicitly checked the cancellation, up to a total derivative,  of the quadratic terms   $\sim (W)^2$
in the variation of this expression under \p{qtransfOn1} - \p{transDq1}.

\end{document}